%% file: TWC_backhaul_paper_arxiv.tex
\documentclass[12pt, draftclsnofoot, onecolumn]{IEEEtran}
\usepackage{array}
\usepackage{color}
\usepackage{epsf}
\usepackage{times}
\usepackage{epsfig}
\usepackage{graphicx}
\usepackage{epstopdf}
\usepackage{hyperref}
\usepackage{amsmath}
\usepackage{amssymb}
\usepackage{amsxtra}
\usepackage{amsthm}
\usepackage{algorithmic}
\usepackage{algorithm}
\usepackage{caption}
\usepackage{subcaption}
\usepackage{authblk}
\usepackage{bbm}
\usepackage{comment}
\usepackage{soul}

\input{Figures/defs_tikzpgf}
\usetikzlibrary{petri}
\usetikzlibrary{shapes}
\usetikzlibrary{positioning,arrows,patterns}
\usepackage[utf8x]{inputenc}
\usepackage{scalefnt}
\usetikzlibrary{calc,decorations.markings}
\pgfplotsset{compat=1.10}
\usetikzlibrary{arrows.meta}
\usepackage[columnwise,switch,mathlines]{lineno} 

\usepackage{pgfplots}
\pgfplotsset{compat=newest}
\usepgfplotslibrary{groupplots}

\usepackage{xcolor,cite,etoolbox}
\makeatletter
\pretocmd\@bibitem{\color{black}\csname keycolor#1\endcsname}{}{\fail}
\newcommand\citecolor[1]{\@namedef{keycolor#1}{\color{blue}}}
\makeatother

\IEEEoverridecommandlockouts
\newtheorem{theorem}{Theorem}
\newtheorem{lemma}{Lemma}

\begin{document}
	
	\title{Analysis of Large Scale Aerial Terrestrial Networks with mmWave Backhauling}
	
	\author{Nour~Kouzayha,~\IEEEmembership{Member,~IEEE,}
		Hesham~ElSawy,~\IEEEmembership{Senior Member,~IEEE,}
		Hayssam~Dahrouj,~\IEEEmembership{Senior Member,~IEEE,}
		Khlod~Alshaikh,
		Tareq Y. Al-Naffouri,~\IEEEmembership{Senior Member,~IEEE,}
		and~Mohamed-Slim Alouini,~\IEEEmembership{Fellow,~IEEE}
		\thanks{This work has been presented in part in~\cite{NourICC} at IEEE International Conference on Communications (ICC'2020), Dublin, Ireland.}
		\thanks{N. Kouzayha, H. Dahrouj, T. Y. Al-Naffouri, and M.-S. Alouini are with the Division of Computer, Electrical and Mathematical Sciences and Engineering, King Abdullah University of Science and Technology, Thuwal, Saudi Arabia (e-mail: nour.kouzayha@kaust.edu.sa; hayssam.dahrouj@kaust.edu.sa; tareq.alnaffouri@kaust.edu.sa; slim.alouini@kaust.edu.sa).}
		\thanks{H. ElSawy is with the Electrical Engineering Department, King Fahd University of Petroleum and Minerals (KFUPM), Dhahran, Saudi Arabia. (e-mail: hesham.elsawy@kfupm.edu.sa).}
		\thanks{K. Alshaikh is an enterprise solutions architect at Udacity (e-mail: khlod.alshaikh@gmail.com).}
	}
	
	
\maketitle
\vspace{-1.5cm}
\begin{abstract}
\vspace{-0.1cm}
Service providers are considering the use of unmanned aerial vehicles (UAVs) to enhance wireless connectivity of cellular networks. To provide connectivity, UAVs have to be backhauled through terrestrial base stations (BSs) to the core network. In particular, we consider millimeter-wave (mmWave) backhauling in the downlink of a hybrid aerial-terrestrial network, where the backhaul links are subject to beamforming misalignment errors. In the proposed model, the user equipment (UE) can connect to either a ground BS or a UAV, where we differentiate between two transmission schemes according to the backhaul status. In one scheme, the UEs are served by the UAVs regardless of whether the backhaul links are good or not. In the other scheme, the UAVs are aware of the backhaul links status, and hence, only the subset of successfully backhauled UAVs can serve the UEs. Using stochastic geometry, the performance of the proposed model is assessed in terms of coverage probability and validated against Monte-Carlo simulations. Several insights are provided for determining some system parameters including the UAVs altitude and required number and the beamforming misalignment error of the backhaul link. The obtained results highlight the impact of the UAVs backhaul link on the UE experience.
\end{abstract}
\vspace{-0.5cm}
	\begin{IEEEkeywords}
		\vspace{-0.3cm}
		UAV, mmWave, terrestrial BS, backhaul unaware transmission, backhaul aware transmission, coverage probability, stochastic geometry.
	\end{IEEEkeywords}
	\IEEEpeerreviewmaketitle
\vspace{-0.6cm}
\section{Introduction}
Unmanned aerial vehicles (UAVs) are receiving significant interest to enhance the wireless coverage of cellular networks~\cite{Mozaffari}. UAVs are used when terrestrial cellular systems get damaged, in hard to reach areas, and wherein there is an occasional need for supplementary coverage. One of the main features of UAVs is the existence of line-of-sight (LOS) links, which can improve the signal quality compared to non-line-of-sight (NLOS) links. The LOS transmission is considered as one of the key factors that can boost capacity in beyond $5$G (B$5$G) ultra dense networks. The placement of UAVs at elevated altitudes can effectively avoid obstacles, thus enabling millimeter wave (mmWave) based technologies such as multiple-input-multiple-output (MIMO) transmissions~\cite{Evans}. In addition, UAVs can be further integrated with emerging network architectures such as Internet of things~\cite{Bushnaq,Bushnaq2} and multiple edge computing~\cite{Amer}. In this work, we consider a hybrid aerial-terrestrial cellular network where UAVs are deployed above a built-up dense urban area to assist terrestrial networks and are backhauled through mmWave links.
\vspace{-0.5cm}
\subsection{Related Work}
\vspace{-0.1cm}
The foreseen potential of UAVs’ to support massive wireless connectivity has attracted significant research efforts to model, analyze, and design UAVs networks. Instances of such research efforts include developing air-to-air and air-to-ground communications channel models~\cite{Khawaja,Chang}, optimal placement and resource allocation for UAVs~\cite{Yang,Cai}, trajectory planning~\cite{Wu, Hu,Bushnaq}, charging ground sensors using UAVs radio frequency (RF) signals~\cite{Saab,Benbuk} and evaluating the performance of UAV-assisted cellular networks~\cite{Wang,Galkin,Qi}.

To best analyze the performance of UAV networks, stochastic geometry becomes indispensable to account for the mutual interference between the active aerial-terrestrial links~\cite{Sultan}. In~\cite{Chetlur}, the authors use stochastic geometry to provide an expression of the coverage probability for a UAV network under guaranteed LOS conditions. However, the model in~\cite{Chetlur} overlooks the effect of LOS/NLOS components of the aerial channel. A probabilistic LOS/NLOS propagation model for aerial channels is adopted in~\cite{Wang} for UAV-assisted cellular networks. Such LOS/NLOS assumption leads to more accurate results and design insights at the expense of more involved analysis. In~\cite{Osama}, the authors use stochastic geometry to model a network where tethered UAVs are deployed to assist traffic offloading under practical challenges.

Terrestrial BSs can be backhauled via wired or wireless links. In contrast, wireless backhauling is mandatory for UAVs to support their mobility. The UAV backhaul links can be provided by terrestrial BSs using sub-$6$~GHz technologies, mmWave technologies~\cite{Xiao}, or free space optics (FSO)~\cite{Pappi}. For instance, the authors in~\cite{Alzenad} consider point-to-point FSO backhaul links for UAVs. Laser beams are used in~\cite{Lahmeri} to boost UAVs operations with both power and backhauling capabilities. Across the RF spectrum, backhauling at the mmWave range is shown to be superior to sub-$6$~GHz backhauling due to the larger bandwidth and the active beam steering capabilities. In this regard, 3GPP envisions an integrated access and backhaul architecture for small cell BSs in which the same infrastructure and spectral resources are used by the macro BSs on the access links and the small BSs on the backhaul links~\cite{zhang2}. This new architecture introduces new design challenges such as the proper partitioning of resources between the access and the backhaul links~\cite{Saha2,Saha1}. As for UAV networks, the authors in~\cite{Xiao} utilize mmWave backhauling, where issues related to beam tracking, LOS blockage and UAV discovery are investigated.

Despite UAV communications multiple promises, modeling and analyzing the impact of UAV backhauling remains relatively unexplored. The main focus in the literature is mostly on the UAV access links while assuming guaranteed UAV backhaul. The work in~\cite{Boris} is one of few exceptions that explicitly accounts for the UAV backhaul. However, this work assumes that backhauling BSs cannot directly serve UEs, which underutilizes the role of terrestrial BSs. Furthermore, the UAV network in~\cite{Boris} is modeled as an infinite Poisson Point Process (PPP) and the backhauling impact is only incorporated via simulations. The backhaul effect is also considered in~\cite{Chen}, where a single UAV is introduced to maximize the throughput in a dense urban environment. Hence, the work in~\cite{Chen} does not account for the possible coexistence of multiple UAVs and the subsequent mutual interference between them. Up to the authors' best knowledge, the joint consideration of the access and mmWave backhaul links in a large-scale hybrid aerial-terrestrial cellular network has not been investigated in literature, and so this paper addresses the details and intricacies of this problem from a stochastic geometry perspective.
\vspace{-0.7cm}
\subsection{Contribution and Organization}
\vspace{-0.2cm}
In this paper, a hybrid aerial-terrestrial network is considered, where UAVs are used to assist BSs in a dense urban environment in providing coverage to UEs. The UAVs are backhauled through mmWave links, which are subject to beamforming misalignment errors. Using stochastic geometry, we aim to evaluate the performance of the network and the effect of its parameters, while highlighting the impact of mmWave backhauling on UE coverage. To the best of the authors' knowledge, this work is the first to jointly consider the access and mmWave backhaul links in a large-scale hybrid aerial-terrestrial network, where the UE can be either served by a BS or by a UAV. The main contributions of this paper can be summarized as follows:
\begin{enumerate}
\item \textbf{\textit{System model:}} We consider a hybrid aerial-terrestrial network with a finite number of backhaul-enabled UAVs. The UAV network is modeled as a binomial point process (BPP), which fits more realistic use cases than infinite PPPs. The developed mathematical framework explicitly accounts for the LOS/NLOS mmWave links along with possible beamforming misalignment errors.
\item \textbf{\textit{Comparison between backhaul aware/unaware transmission models:}} The proposed model differentiates between backhaul aware and backhaul unaware transmission models. For the unaware scenario, the serving UAV transmits data to UE without considering its backhaul link quality. In the backhaul aware scheme, the serving UAV is aware of the backhaul status and does not communicate with the UE in case of backhaul link outage. The obtained results show that prior knowledge of the backhaul link quality relieves parts of the network interference, and hence, improves the overall coverage probability.
\item \textbf{\textit{Design guidelines and insights:}} The analytical results, validated using Monte-Carlo simulations, highlight the fundamental impact of the backhaul link quality on the UE coverage. Based on the proposed framework, we investigate the impact of various parameters including the intensity of terrestrial BSs, the fraction of backhaul-enabled BSs, the altitude and number of UAVs, as well as the beamforming misalignment errors. The results reveal that the number and altitude of UAVs must be chosen carefully to optimize the UE experience.
\end{enumerate}
\vspace{-0.7cm}
\begin{table}[htp!]
	\centering
	\caption{Notations Summary.}
	\vspace{-0.3cm}
	\begin{tabular}{m{0.23\linewidth}|m{0.73\linewidth}}
		\hline\hline
		\textbf{Notation} & \textbf{Description} \\\hline
		$D_{u}$, $r_{u}$, $v_0$ & Disk in which the UAVs are distributed, Radius of $D_{u}$, Distance from the UE to origin\\\hline
		$N_{u}$, $h_{u}$, $\lambda_{g}$, $h_{g}$ & Number and height of UAVs, Density and height of terrestrial BSs with reference to ground\\\hline
		$\delta_{b}$, $\lambda_{b}$ & Fraction and density of terrestrial backhaul-enabled BSs\\\hline
		$\Phi_{u}$, $\Phi_{g}$, $\Phi_{g_l}$, $\Phi_{g_n}$ &Set of UAVs, terrestrial BSs, LOS BSs and NLOS BSs respectively \\\hline
		$\kappa_{u_l}(\cdot)$, $\kappa_{u_n}(\cdot)$, $\kappa_{b_l}(\cdot)$, $\kappa_{b_n}(\cdot)$ & LOS and NLOS probabilities of the access link and backhaul link, respectively\\\hline
		$P_{u}$, $P_{g}$, $P_{b}$ & Transmit power of UAVs, BSs on the access link, BSs on the backhaul link, respectively\\\hline
		$\eta_{g}$, $\eta_{l}$, $\eta_{n}$ & Path-loss exponent parameter for terrestrial BS, LOS UAV, or NLOS UAV, respectively \\\hline
		$m_{l}$, $m_{n}$ & Nakagami-m fading parameter for aerial LOS link, or aerial NLOS link, respectively \\\hline
		$\Omega_{g,x_{i}}$, $\Omega_{u_{l},y_{j}}$, $\Omega_{u_{n},y_{j}}$, $\Omega_{b_{l},x_{i},y_{j}}$, $\Omega_{b_{n},x_{i},y_{j}}$ & Small scale fading gain between UE and $i$-th BS, $j$-th LOS or NLOS UAV on the access link and between $j$-UAV and $i$-th LOS or NLOS BS on the backhaul link, respectively\\\hline
		$s_{g,x_{i}}$, $s_{u_{l}, y_{j}}$, $s_{u_{n}, y_{j}}$, $s_{b_{l},x_{i},y_{j}}$, $s_{b_{n},x_{i},y_{j}}$  & Horizontal distance between the projections of the UE and the $i$-th BS or the $j$-th LOS or NLOS UAV and between the $i$-th LOS or NLOS BS and the $j$-th UAV, respectively \\\hline
		$z_{g,x_{i}}$, $z_{u_{l},y_{j}}$, $z_{u_{l},y_{j}}$, $z_{b_{l},x_{i}, y_{j}}$, $z_{b_{n},x_{i}, y_{j}}$ & Distance between the UE and the $i$-th BS or the $j$-th LOS or NLOS UAV and between the $i$-th LOS or NLOS BS and the $i$-th UAV, respectively\\\hline
		$P^{r}_{g,x_{i}}$, $P^{r}_{u_{l},y_{j}}$, $P^{r}_{u_{n},y_{j}}$, $P^{r}_{b_{l},x_{i}, y_{j}}$, $P^{r}_{b_{n},x_{i}, y_{j}}$ & Received power from the $i$-th BS or the $j$-th LOS/NLOS UAV at the UE on the access link, or from $i$-th LOS/NLOS BS to the $j$-th UAV on the backhaul link, respectively \\\hline
		$x_{g}$, $y_{u_l}$, $y_{u_n}$, $x_{b_l}$, $x_{b_n}$ & Distances between the UE and its serving BS, LOS or NLOS UAV on the access link and between the UAV and its serving LOS or NLOS BS on the backhaul link, respectively\\\hline
		$C_{l}$, $C_{n}$, $\sigma_{b}^2$  & LOS and NLOS near-field path loss and noise power for backhaul connection\\\hline
		$G_{s}^{(\mathrm{max})}$, $G_{s}^{(\mathrm{min})}$, $\theta_{s}$ & Antenna parameters for BSs ($s=g$) and UAVs ($s=u$) for backhaul connection \\\hline
		$\varepsilon_{g}$, $\varepsilon_{u}$, $\sigma_{g}^{2}$, $\sigma_{u}^{2}$ & Additive beam-steering errors and corresponding variances of BS and UAV antennas \\\hline
		$\hat{I}_{g}$, $\hat{I}_{u_l}$, $\hat{I}_{u_n}$ & Interference from all BSs or UAVs except serving BS, LOS or NLOS UAV, respectively\\ \hline
		$I_{u}$, $I_{g_l}$, $I_{g_n}$ & Interference from all UAVs or BSs if UE associates to a BS, a LOS or NLOS UAV, respectively\\\hline
		$A_{g}$, $A_{u_l}$, $A_{u_n}$, $A_{b_l}$, $A_{b_n}$ & Association probabilities between the UE and a BS, a LOS or a NLOS UAV for backhaul unaware and between the UAV and a LOS or a NLOS BS for backhaul, respectively\\\hline
		$S_{l}(\cdot)$, $S_{n}(\cdot)$, $S(\cdot)$ & Conditional backhaul probabilities given that the UAV is associated with a LOS BS, or a NLOS BS and overall backhaul probability, respectively\\\hline
		$P_{cov,g}$, $P_{cov,u_l}$, $P_{cov,u_n}$ & Conditional coverage probabilities given that the UE is associated with a terrestrial BS, a LOS UAV, or a NLOS UAV for backhaul unaware transmission, respectively\\\hline
		$\tilde{A}_{g}$, $\tilde{A}_{u_l}$, $\tilde{A}_{u_n}$, $\tilde{A}_{f}$ & Association probabilities between the UE and a terrestrial BS, a LOS UAV, or a NLOS UAV and service failure probability for backhaul aware transmission, respectively \\\hline
		$\tilde{P}_{cov,g}$, $\tilde{P}_{cov,u_l}$, $\tilde{P}_{cov,u_n}$ & Conditional coverage probabilities given that the UE is associated with a terrestrial BS, a LOS UAV, or a NLOS UAV for backhaul aware transmission, respectively\\\hline
		$\tilde{N_u}$, $I_{\tilde{u}}$ & Number of UAVs with successful backhaul links for backhaul aware transmission, Interference from UAVs with successful backhaul links\\\hline
		$P_{cov}$, $\tilde{P}_{cov}$ & Overall coverage probabilities for backhaul unaware and aware transmission, respectively \\\hline
		$\tau_a$, $\tau_b$& $\mathrm{SIR}$ threshold on the access link, $\mathrm{SINR}$ threshold on the backhaul link\\\hline
		\hline
	\end{tabular}
	\label{table:notations_access}
	\vspace{-0.5cm}
\end{table}
\subsection{Paper Notations and Organization}
\vspace{-0.1cm}
The subscripts $\{\cdot\}_u$, $\{\cdot\}_g$, and $\{\cdot\}_b$ refer to UAV, ground BS and backhaul, respectively. The subscripts $\{\cdot\}_{l}$, $\{\cdot\}_{n}$ refer to LOS and NLOS, respectively. $\tilde{\{\cdot\}}$ denotes the backhaul aware transmission scheme. $\mathbb{P}\{\cdot\}$ denotes probability, $\mathbb{E}[\cdot]$ denotes expectation, $\mathcal{L}_x(\cdot)$ denotes the Laplace transform of a random variable $x$, and $_{2}F_{1}(a,b;c;x)=\sum_{n=0}^{\infty}\frac{(a)_n(b)_n}{(c)_n}\frac{x^n}{n!}$ denotes the Gauss hyper geometric function. Other notations are listed in Table~\ref{table:notations_access}.

The rest of the paper is organized as follows. Section~\ref{sec:system_model} describes the system model. The analysis of the backhaul unaware and aware transmission scenarios are presented in Sections~\ref{sec:unaware} and \ref{sec:aware}, receptively. Numerical results and design insights are discussed in Section~\ref{sec:results} and validated using Monte-Carlo simulations. Finally, the paper is concluded in Section~\ref{sec:conclusion}.
\vspace{-0.6cm}
\section{System Model}\label{sec:system_model}
\vspace{-0.5cm}
\subsection{Network Model}
\begin{figure}[t]
	\centering
	\begin{minipage}[t]{0.5\linewidth}
		\includegraphics[width=\linewidth,height=0.21\textheight]{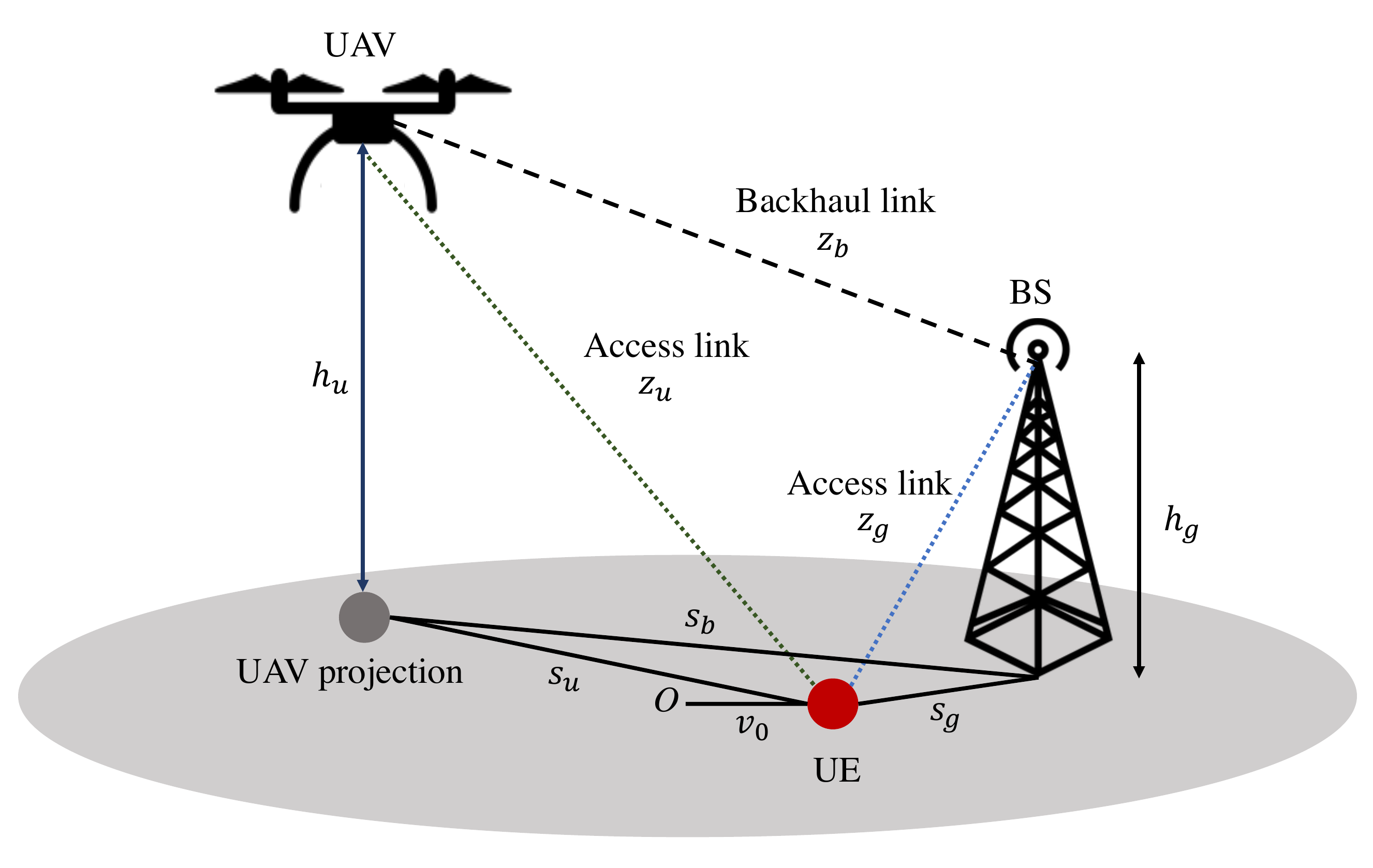}
		\vspace{-0.8cm}
		\caption{Proposed system model}
		\label{fig:system_model}
	\end{minipage}
	\begin{minipage}[t]{0.4\linewidth}
		\includegraphics[width=\linewidth,height=0.12\textheight]{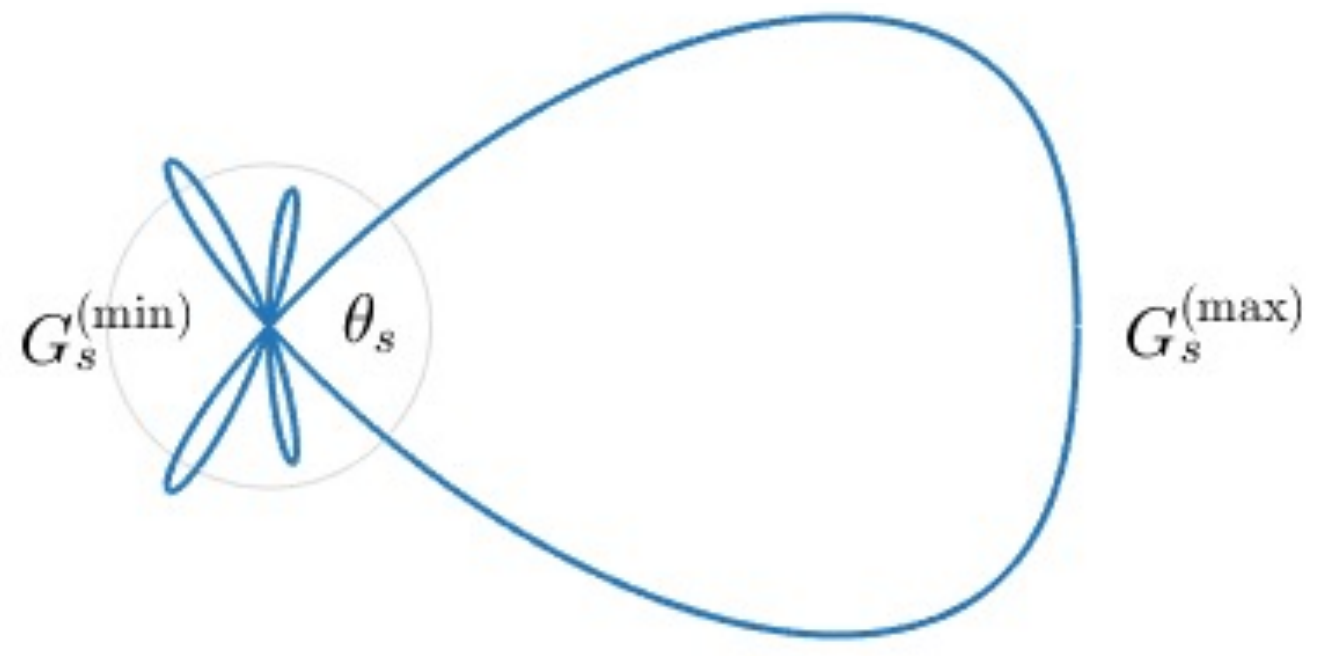}
		\vspace{-0.8cm}
		\caption{Sectored-pattern antenna.}
		\label{fig:antenna}
	\end{minipage}
	\vspace{-1cm}
\end{figure}

\vspace{-0.2cm}
UAVs are deployed to improve the downlink (DL) coverage of a one-tier cellular network, where UEs can be served by either terrestrial BSs or UAVs. The BSs provide wireless access to the UEs, and mmWave backhaul links to UAVs. Fig.~\ref{fig:system_model} presents the proposed network model. Since we consider a DL scheme, the UAVs transmit to UEs and receive from BSs for backhauling purposes. The UAVs are considered as simple relays with no queuing capabilities. On the other hand, the BSs are assumed to have saturated buffers which always have packets for their served UEs, where UEs can be directly served from the BS or through a UAV relay. The analysis of spatio-temporal models with non-saturated BSs buffers and queuing UAVs is beyond the scope of this paper and is left as a future research direction.

The ground BSs are spatially distributed according to a PPP $\Phi_{g}=\{\textbf{x}_{i}\}$, where $\textbf{x}_{i}$ is the location of the $i$-th BS. We assume that the density of the ground BSs is $\lambda_g$ and that they are all at the same height $h_g$ and transmit with the same power $P_{g}$ to UEs. A fraction $\delta_{b}$ of the terrestrial BSs is equipped with backhauling capabilities and transmit to the UAVs with the same power $P_{b}$ for backhaul. Thus, the density of backhaul-enabled BSs $\lambda_{b}$ is equal to $\delta_{b}\lambda_{g}$. We assume that $N_u$ UAVs hover at an altitude $h_u$ and are distributed uniformly in a finite disk $D_u$ with radius $r_u$ forming a BPP $\Phi_u=\{\textbf{y}_{j}\}$, where $\textbf{y}_{j}$ refers to the location of the $j$-th UAV. Without loss of generality, we assume that the BPP is centered around $\textbf{o'}=(0,0,h_{u})$ and that all the UAVs transmit with the same power $P_{u}$ to UEs. For the access link, the BSs and UAVs are equipped with single antennas, so as to serve single antenna UEs on the sub-$6$~GHz band. On the other hand, dedicated antenna arrays at the BSs and UAVs are used to align the mmWave transmissions for backhaul. Using separate antennas for mmWave backhaul and access at the terrestrial BSs has two main benefits. Firstly, it allows the BSs to steer their antennas to align with those of the UAVs, and hence, to maximize the directionality gain. Secondly, it reduces the interference on both access and backhaul links as they operate on different frequency bands. The analysis is conducted for a UE positioned at an arbitrary distance $\textbf{v}_{0}=(v_{0},0,0)$ from the origin $\textbf{o}=(0,0,0)$ on the ground. All frequency resources are universally reused across the network.
\vspace{-0.7cm}
\subsection{Channel Model}
\vspace{-0.3cm}
\subsubsection{Terrestrial Access Channel (BS-UE)}
Due to the nature of terrestrial communication, the fading channel of the BS-UE access link consists of a large-scale fading modeled using a distance-dependent path-loss with path-loss exponent $\eta_g$, and a small-scale Rayleigh fading with exponential distribution and unit mean. The signal power received at the UE from the $i$-th BS located at $x_{i}$ can thus be expressed as $P_{g,x_{i}}^{r}=P_{g}(s_{g,x_{i}}^2+h_{g}^2)^{-\eta_g/2}\Omega_{g,x_{i}}$, where $\Omega_{g,x_{i}}$ is the small-scale fading, $s_{g,x_{i}}=\sqrt{z_{g,x_{i}}^{2}-h_{g}^{2}}$ is the horizontal distance separating the UE and the projection of the $i$-th BS on the ground and $z_{g,x_{i}}=||\textbf{x}_{i}-\textbf{v}_{0}||$ is the actual distance.
\subsubsection{Aerial Access (UAV-UE) and Backhaul (BS-UAV) Channels}\label{sec:aerial_channel}
Both the aerial access and backhaul channels are affected by obstacles in the environment which break the LOS links. According to the ITU recommendation report~\cite{ITU}, the probability of a LOS link between a transmitter and a receiver, with heights $h_{\mathrm{TX}}$ and $h_{\mathrm{RX}}$, respectively, is given by~\cite{ITU}:
\begin{equation}\small
P_{\mathrm{LOS}}(r)=\prod_{n=0}^{m}\left[1-\exp\left(-\frac{h_{\mathrm{TX}}-\frac{(n+\frac{1}{2})\left(h_{\mathrm{TX}}-h_{\mathrm{RX}}\right)^{2}}{m+1}}{2\gamma^{2}}\right)\right]
\label{eq:halim}
\end{equation}
where $r$ denotes the horizontal distance between the transmitter and the receiver and $m=\lfloor\frac{r\sqrt{\alpha \beta}}{1000}-1\rfloor$, where $\alpha$, $\beta$ and $\gamma$ are environment related parameters given in Table I in~\cite{Holis}. As we can see from (\ref{eq:halim}), the LOS probability is not a continuous function of the horizontal distance which makes the analysis intractable. As the UE is considered at the ground level, we use the approximation proposed in~\cite{Hourani} to simplify the LOS probability on the access link between a UAV located at a distance $r$ and the UE by a modified Sigmoid function given as:
\begin{equation}\small
\kappa_{u_l}(r)=\frac{1}{1+a \exp (-b[\frac{180}{\pi}\arctan(h_u/\sqrt{r^{2}-h_{u}^{2}})-a])},
\label{eq:LOS_probability}
\end{equation}
where $\sqrt{r^{2}-h_{u}^{2}}$ is the Euclidean horizontal distance separating the projections of the UAV and the UE and $a$ and $b$ are constant values that depend on the environment and are given in Table I of~\cite{Bor}. The NLOS probability is $\kappa_{u_n}=1-\kappa_{u_l}$.

The LOS probability model in (\ref{eq:LOS_probability}) used for the UAV-UE channel was initially developed and approximated for UEs positioned at low height levels from the ground. This model is unsuitable for the BS-UAV aerial channel as both the BSs and UAVs are positioned above the ground level. We use the approximation proposed in~\cite{Cherif} to simplify (\ref{eq:halim}) as a more tractable exponential function, representing the LOS probability between a UAV and a BS separated by a horizontal distance $r$ on the backhaul link as:
\begin{equation}\small
\kappa_{b_l}(r)=-c\exp\left(-d \arctan \left(\frac{|h_u-h_g|}{r}\right)\right)+e,
\label{eq:LOS_probability2}
\end{equation}
where $r$ is the horizontal distance separating the projections of the BS and the UAV, $c$, $d$ and $e$ are parameters that depend on the environment and height of terrestrial BSs and are given in Table II of~\cite{Cherif}. The NLOS probability on the backhaul link is given as $\kappa_{b_n}=1-\kappa_{b_l}$.

The aerial fading channel is characterized using the combination of two components: (a) a large-scale fading modeled using a distance-dependent path-loss and (b) a small-scale Nakagami-$m$ fading modeled using a gamma-distributed random variable. We consider different path-loss exponents ($\eta_{l}$ for LOS and $\eta_{n}$ for NLOS) and fading parameters for the LOS and the NLOS links ($m_{l}$ for LOS and $m_n$ for NLOS). The power received at the reference UE from the $j$-th UAV located at $\textbf{y}_{j}$ is $P^{r}_{u_{\zeta},y_{j}}=P_{u}z_{u_{\zeta},y_{j}}^{-\eta_{\zeta}}\Omega_{u_{\zeta},y_{j}}$, where $z_{u_{\zeta},y_{j}}=||\textbf{y}_{i}-\textbf{v}_{0}||$ is the distance separating the UE from the $i$-th UAV, $\Omega_{u_{\zeta},y_{j}}$ is the small-scale fading, $\eta_{\zeta}$ is the path-loss exponent, and $\zeta\in\{l,n\}$ indicates whether the $j$-th UAV has a LOS link or NLOS link with the UE.

UAVs are connected to ground BSs through mmWave backhaul links. Although mmWave channels share basic propagation characteristics like power law path-loss with sub-$6$~GHz channels, they also have critical distinctions. For instance, beamforming is important to compensate for the high propagation loss in mmWave frequencies as well as the consumed power in analog to digital conversion in large antenna arrays. Thus, the BSs and UAVs must steer their antennas to maximize the directionality gain. However, aligning the antennas is subject to beam-steering errors, which affects the transmission on the backhaul link. We approximate the array patterns of the UAVs and the BSs antennas by the model shown in Fig.~\ref{fig:antenna} and given as
	\vspace{-0.1cm}
	\begin{equation}\small
	G_{s}(\varphi)=\begin{cases}
	G_{s}^{(\text{max})}, & |\varphi|\leq\theta_{s} \\
	G_{s}^{(\text{min})}, & |\varphi|>\theta_{s}
	\end{cases}
	\label{eq:gain}
	\vspace{-0.2cm}
	\end{equation}
	where $\varphi\in[-\pi,\pi)$ is the angle of the boresight direction, $G_{s}^{(\text{max})}$, $G_{s}^{(\text{min})}$ and $\theta_{s}$ are the gains of the main and side lobes and the beamwidth for the BSs and UAVs $(s\in\{g,u\})$, respectively.
	
	For the desired link, the directivity gain is $G_{b,0}=G_{g}^{(\text{max})}G_{u}^{(\text{min})}$ in the absence of the beamforming misalignment errors modeled explicitly in Section~\ref{sec:misalignment}. For the interfering BSs, the beams are assumed to be randomly oriented with respect to each other, and the steering angles are distributed uniformly in $[-\pi,\pi)$. Since (\ref{eq:gain}) produces either gains $G_{s}^{(\text{max})}$ and $G_{s}^{(\text{min})}$, $s\in\{g,u\}$ over all possible input angles, the resulting gain distribution of a BS or UAV is given as
	\vspace{-0.1cm}
	\begin{equation}\small
	f_{G_{s,I}}(g)=\frac{\theta_{s}}{2\pi}\delta\left(g-G_{s}^{(\text{max})}\right)+\left(1-\frac{\theta_{s}}{2\pi}\right)\delta\left(g-G_{s}^{(\text{min})}\right)
	\label{eq:fsI}
	\vspace{-0.1cm}
	\end{equation}
	where $\delta(\cdot)$ is the Dirac delta function, and $s\in\{g,u\}$.
	
	Accordingly, the directivity gain of an interfering link is a discrete random variable denoted as $G_{b,I}$, with a probability mass function (PMF) that can be formulated as $f_{G_{b,I}}(g)=\left(f_{G_{g,I}}\otimes f_{G_{u,I}}\right)(g)$, where $f_{G_{g,I}}(\cdot)$ and $f_{G_{u,I}}(\cdot)$ are the PMFs of the antenna gains of the interfering BS and the corresponding UAV. Finally, based on (\ref{eq:gain}) and (\ref{eq:fsI}), the directivity gain of the interfering links takes the values $G_{k}$ with probability $p_{k}$ ($k\in\{1,2,3,4\}$) defined in Table~\ref{table:antenna}, where $c_{g}=\frac{\theta_{g}}{2\pi}$ and  $c_{u}=\frac{\theta_{u}}{2\pi}$. $\theta_{g}$ and $\theta_{u}$ are the beamwidth of the BS and the UAV antennas.
\begin{table}[t]
	\centering
	\caption{Probability mass function of $G_{b,I}$ and $G_{b,0}$.}
	\vspace{-0.3cm}
	\begin{tabular}{c|c|c|c|c}
		\hline
		\textbf{k} & \textbf{1} & \textbf{2} & \textbf{3} & \textbf{4} \\\hline
		$G_{k}$ & $G_{g}^{(\mathrm{max})}G_{u}^{(\mathrm{max})}$&$G_{g}^{(\mathrm{max})}G_{u}^{(\mathrm{min})}$ & $G_{g}^{(\mathrm{min})}G_{u}^{(\mathrm{max})}$ & $G_{g}^{(\mathrm{min})}G_{u}^{(\mathrm{min})}$\\\hline
		$p_{k}$ & $c_{g} c_{u}$ & $c_{g}(1-c_{u})$& $(1-c_{g})c_{u}$ & $(1-c_{g})(1-c_{u})$ \\\hline
		$p_{k,0}$ & $F_{\lvert\varepsilon_{g}\rvert}(\frac{\theta_{g}}{2})F_{\lvert\varepsilon_{u}\rvert}(\frac{\theta_{u}}{2})$ & $F_{\lvert\varepsilon_{g}\rvert}(\frac{\theta_{g}}{2})\bar{F}_{\lvert\varepsilon_{u}\rvert}(\frac{\theta_{u}}{2})$ & $\bar{F}_{\lvert\varepsilon_{g}\rvert}(\frac{\theta_{g}}{2})F_{\lvert\varepsilon_{u}\rvert}(\frac{\theta_{u}}{2})$ & $\bar{F}_{\lvert\varepsilon_{g}\rvert}(\frac{\theta_{g}}{2})\bar{F}_{\lvert\varepsilon_{u}\rvert}(\frac{\theta_{u}}{2})$\small\\\hline
	\end{tabular}
	\label{table:antenna}
	\vspace{-0.9cm}
\end{table}

The received power from the $i$-th BS located at $\textbf{x}_{i}$ at the $j$-th UAV located at $\textbf{y}_{i}$ is given as $P^{r}_{b,x_{i},y_{j}}=P_{b}G_{b}C_{\xi}\left(s_{b_{\xi},x_{i},y_{j}}^{2}+\Delta_{h}^{2}\right)^{-\eta_{\xi}/2}\Omega_{\xi,x_{i},y_{j}}$, where $P_{b}$ is the BS transmit power on the backhaul link, $\Delta_{h}=|h_{u}-h_{g}|$ is the difference between the BSs and the UAVs heights, $\Omega_{b_{\xi},x_{i}, y_{j}}$ is the small-scale fading, $C_{\xi}$ is the path-loss intercept and $\eta_{\xi}$ is the path-loss exponent, where $\xi\in\{l,n\}$ indicates whether the $i$-th BS has a LOS link or a NLOS link with the $j$-th UAV.
\vspace{-0.7cm}
\subsection{Association Strategy and Performance Metrics}
\vspace{-0.2cm}
This paper studies the performance of a hybrid aerial-terrestrial network where both BSs and UAVs are used to serve UEs. That is, each UE connects to the BS or the UAV that offers the maximum average received signal strength. Note that the UE can either connect to a LOS or a NLOS UAV. The coverage probability, defined as the probability that the signal-to-interference ratio ($\mathrm{SIR}$) exceeds a threshold $\tau_a$, is the main performance metric. Each UAV selects its backhauling BS based on the minimum path-loss association rule. Hence, the BS that provides the backhaul is the best in terms of signal strength. Each UAV and its backhauling BS steer their beams to align their transmissions in order to strengthen the intended link and mitigate interference. Such beam steering along with the high bandwidth and severe path-loss of mmWave frequencies relieve the interference dominance and make the backhaul links sensitive to the ambient noise. Thus, it is more adequate to assess the backhaul links through the signal-to-interference-and-noise-ratio ($\mathrm{SINR}$) rather than the $\mathrm{SIR}$. A backhaul transmission is considered successful if and only if the $\mathrm{SINR}$ exceeds a threshold $\tau_b$, which defines the backhaul probability.

We differentiate between two schemes, namely, backhaul unaware and backhaul aware transmission schemes. In the first scheme, the serving UAV transmits directly to the UE regardless of whether the backhaul link is successful or not. Thus, no further post processing is required when the UAV receives packets from the BS and forwards them to the UE. The aerial coverage in the unaware scheme requires two conditions:
(i) $\mathrm{SIR}>\tau_a$: The received $\mathrm{SIR}$ at the UE from its serving UAV must exceed a threshold $\tau_a$, (ii) $\mathrm{SINR}>\tau_b$: The received $\mathrm{SINR}$ at the serving UAV from the BS to which it connects for backhaul support must exceed a threshold $\tau_b$.

In the backhaul aware scenario, the serving UAV is aware of the backhaul status and checks the integrity of the packets before transmitting them to the UE. This will impose a processing burden on the UAV side but will improve the probability of coverage and limits the aerial interference to the subset of UAVs with successful backhaul links only. An active transmission in the aware scheme implies that the condition ($\mathrm{SINR}>\tau_b$) is already satisfied. Otherwise, the UE halts operation and goes to a service failure status in this specific time slot. Thus, the backhaul aware transmission success is only subject to the $\mathrm{SIR}$ condition ($\mathrm{SIR}>\tau_a$) of the access link.
\vspace{-0.6cm}
\section{Backhaul unaware transmission}\label{sec:unaware}
\vspace{-0.15cm}
In this section, we derive the coverage probability for backhaul unaware transmission. In this scenario, the association rule is only dependent on the quality of the UE-UAV/BS connection. However, a UAV can serve the UE only if it has a successful backhaul link with a ground BS.
\vspace{-0.6cm}
\subsection{Association Probabilities}
\vspace{-0.1cm}
To get coverage, the UE can associate to either a terrestrial BS or a UAV. The set of UAVs is divided into two sub-processes according to the LOS/NLOS status of the access links. According to the association rule, each UE connects to the BS or the UAV that offers the maximum average received power. The serving BS/UAV is not necessarily the nearest to the UE due to the difference in path-loss parameters and transmit powers. However, within a particular set, the path-loss parameters and transmit powers are the same for all links. Therefore, for a specific set, the closest BS/UAV provides a larger average received power than that provided by any other in this set. Thus, the UE is served by the closest BS, LOS UAV or NLOS UAV. The association rule implies an exclusion region on the locations of the nearest interfering device and therefore on the locations of all the other interfering devices in each set for each association type. Specifically, when the UE associates with a terrestrial BS located at a horizontal distance $x_{g}$, all the LOS UAVs are further than $E_{g_{l}}(x_{g})$. Similarly, all the NLOS UAVs are further than $E_{g_{n}}(x_{g})$.
\begin{figure}[t]
	\centering
	\begin{minipage}[t]{0.47\linewidth}
		\includegraphics[width=\linewidth,height=0.2\textheight]{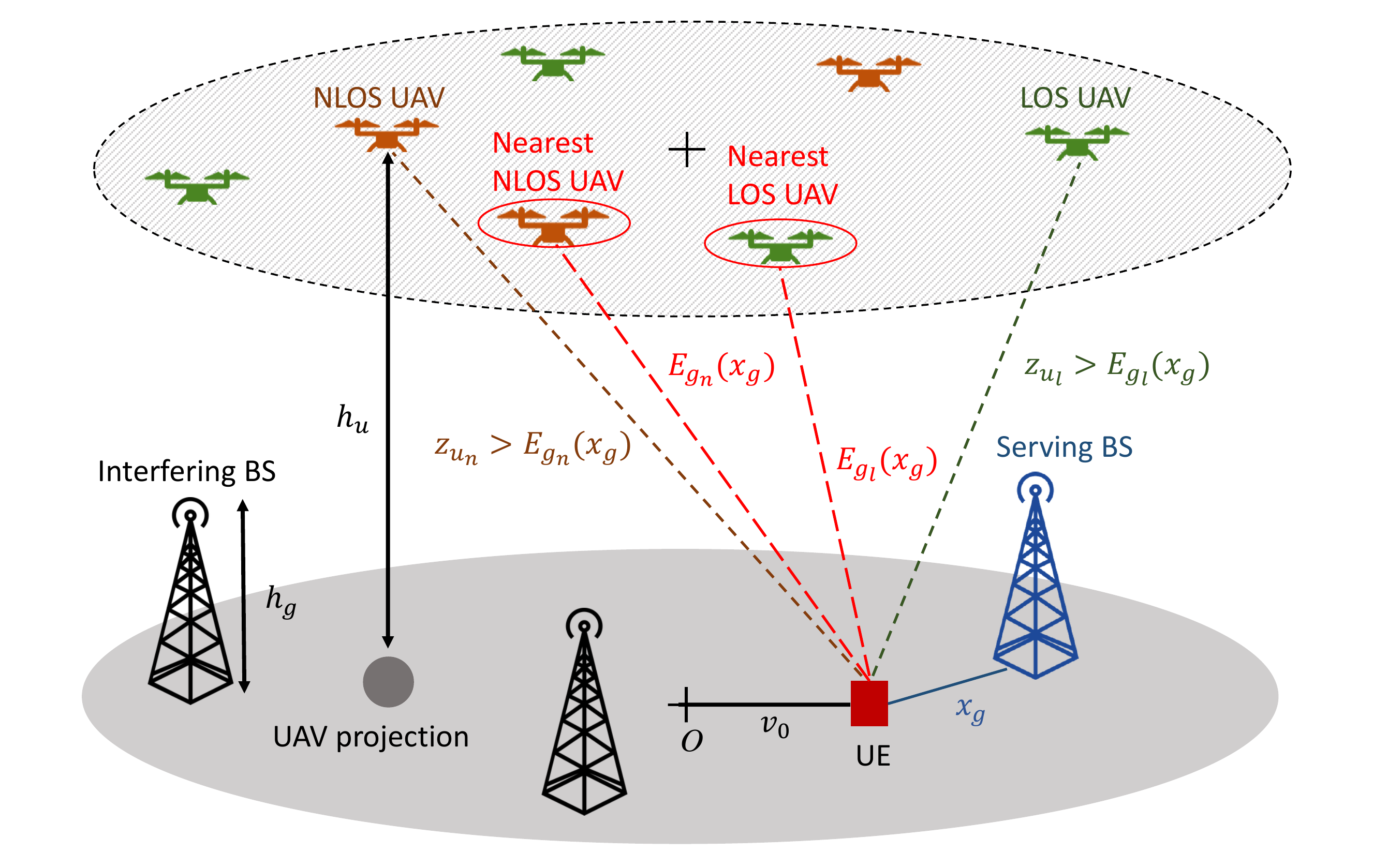}
		\vspace{-1.1cm}
		\caption{Exclusion regions $E_{g_{l}}(x_{g})$ and $E_{g_{n}}(x_{g})$ on the LOS and NLOS UAVs.}
		\label{fig:exclusion_region}
	\end{minipage}
	\begin{minipage}[t]{0.47\linewidth}
		\includegraphics[width=\linewidth,height=0.2\textheight]{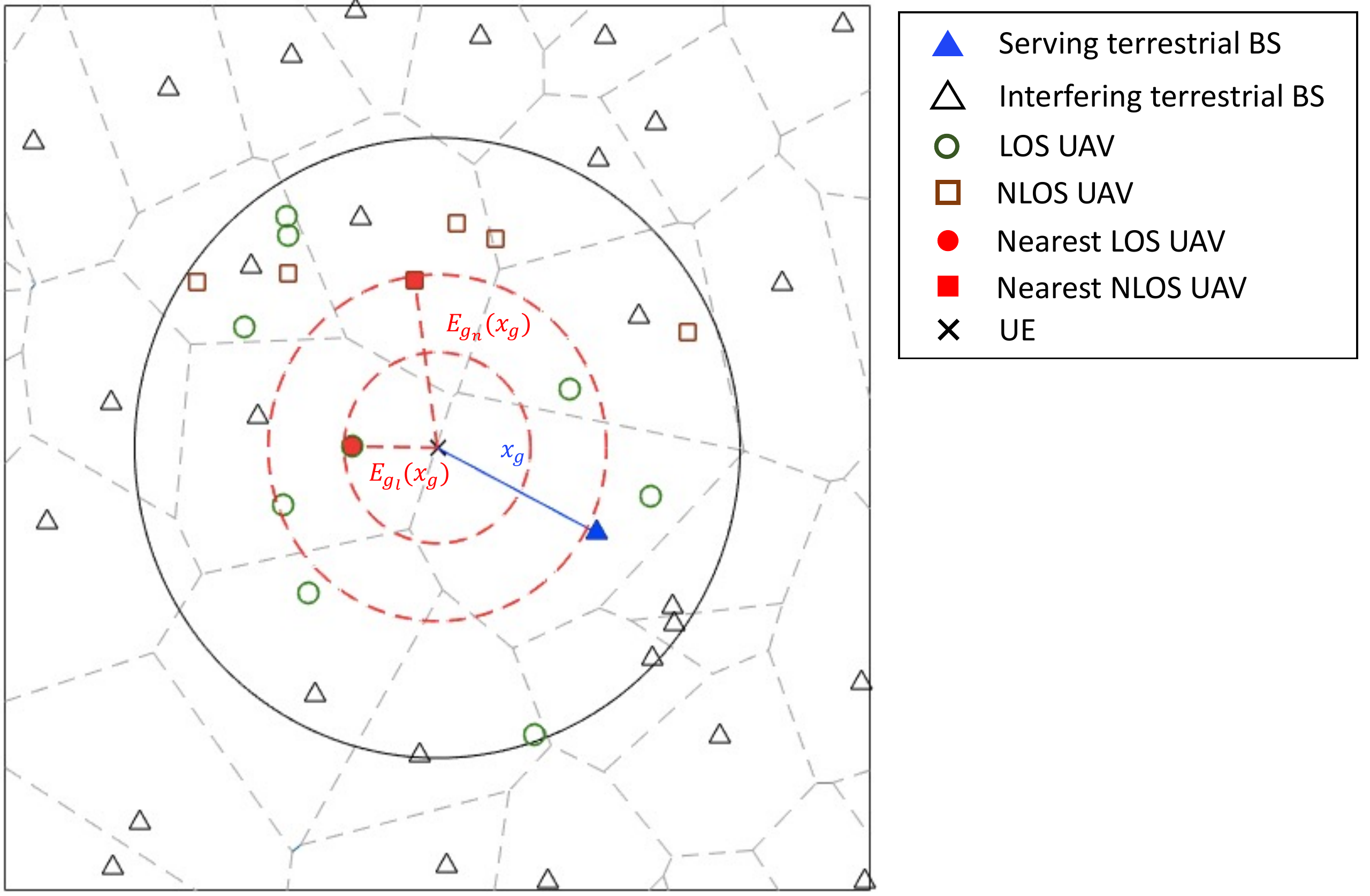}
		\vspace{-1.1cm}
		\caption{Top view snapshot of BSs, LOS/NLOS UAVs and projections of exclusions regions.}
		\label{fig:exclusion_region_horizontal}
	\end{minipage}
	\vspace{-0.8cm}
\end{figure}
Fig.~\ref{fig:exclusion_region} visualizes the exclusion regions on the LOS and NLOS UAVs when the UE associates to a terrestrial BS. Fig.~\ref{fig:exclusion_region_horizontal} presents a top view snapshot on the network showing the exclusion regions on the LOS and NLOS UAVs locations as red dashed circles. Similarly, the exclusion regions $E_{u_l}(x_{u_l})$ and $E_{ln}(x_{u_l})$ are created based on the respective locations of the terrestrial BSs and the NLOS UAVs when the UE associates with a LOS UAV where $x_{u_{l}}$ is the distance to the serving UAV. Finally, the exclusions regions $E_{u_n}(x_{u_n})$ and $E_{nl}(x_{u_n})$ are created based on the respective locations of the BSs and the LOS UAVs, when the UE associates to a NLOS UAV with $x_{u_n}$ being the distance to the serving NLOS UAV. The respective exclusion regions expressions of $E_{g_l}(x)$, $E_{g_n}(x)$, $E_{u_l}(x)$, $E_{u_n}(x)$, $E_{ln}(x)$, and $E_{nl}(x)$ are given as:
\begin{equation}\small
E_{g_\zeta}(x)=\left(\frac{P_{u}}{P_{g}}\right)^{\frac{1}{\eta_{\zeta}}}\left(x^2+h_{g}^2\right)^{\frac{\eta_g}{2\eta_{\zeta}}}, \: E_{u_\zeta}(x)=\sqrt{\left(\frac{P_{g}}{P_{u}}\right)^{\frac{2}{\eta_g}}x^{\frac{2\eta{\zeta}}{\eta_{g}}}-h_{g}^2}, \: E_{\zeta\bar{\zeta}}(x)=x^{\frac{\eta_{\zeta}}{\eta_{\bar{\zeta}}}},
\label{eq:Eg}
\end{equation}
where $\zeta \in \{l,n\}$, $\bar{\zeta}=n$ if $\zeta=l$, and $\bar{\zeta}=l$ if $\zeta=n$.

The association probabilities are defined as the probabilities that the UE associates to either a LOS UAV, a NLOS UAV, or a BS, which are characterized in the following lemma.
\vspace{-0.4cm}
\begin{lemma}
	\emph{Denote by $A_{u_l}$, $A_{u_n}$ and $A_g$ the probabilities that the UE at $\textbf{v}_0$ is served by a LOS UAV, a NLOS UAV or a ground BS, respectively. $A_{u_l}$, $A_{u_n}$ and $A_{g}$ are then given as follows
		\vspace{-0.1cm}
		\begin{equation}\small
		A_{u_\zeta}=N_u\int_{h_u}^{w_{p}}f_{W}(r)\kappa_{u_\zeta}(r)\exp\left(-\pi\lambda_g E^2_{u_\zeta}(r)\right)\left(\int_{r}^{w_{p}}f_{W}(w)\kappa_{u_\zeta}(w)\mathrm{d}w+\int_{E_{\zeta\bar{\zeta}}(r)}^{w_{p}}f_{W}(w)\kappa_{u_{\bar{\zeta}}}(w)\mathrm{d}w\right)^{N_u-1}\mathrm{d}r,
		\label{eq:Aa}
		\end{equation}
		\vspace{-0.8cm}
		\begin{equation}\small
		A_{g}=2\pi\lambda_g\int_{0}^{E_{u}}r \exp\left(-\pi\lambda_g r^2\right)\left(\int_{E_{g_l}(r)}^{w_{p}}f_{W}(w)\kappa_{u_l}(w)\mathrm{d}w+\int_{E_{g_n}(r)}^{w_{p}}f_{W}(w)\kappa_{u_n}(w)\mathrm{d}w\right)^{N_{u}}\mathrm{d}r,
		\label{eq:Ag}
		\end{equation}
		where
		$\zeta \in \{l,n\}$, $\bar{\zeta}=n$ if $\zeta=l$, and $\bar{\zeta}=l$ if $\zeta=n$. $E_u=\mathrm{max}\left(E_{u_l}(w_p), E_{u_n}(w_p)\right)$ and $f_{W}(\cdot)$ is the probability density function (PDF) of the distance $w$ from an arbitrary UAV to the reference UE and is given in Lemma~$2$ of~\cite{Chetlur} as
		\vspace{-0.3cm}
		\begin{equation}\small
		f_{W}(w)=\begin{cases}
		f_{W_{1}}(w)=\frac{2w}{r_{u}^2}, & h_u\leq w\leq w_m\\
		f_{W_{2}}(w)=\frac{2w}{\pi r_{u}^2}\arccos\left(\frac{w^2+v_{0}^2-d^2}{2v_0\sqrt{w^2-h_{u}^2}}\right), & w_m\leq w \leq w_p
		\end{cases}
		\label{eq:fw}
		\end{equation}
		where $r_u$ is the radius of the disk of UAVs, $w_m=\sqrt{(r_u-v_0)^2+h_u^2}$ and $w_p=\sqrt{(r_u+v_0)^2+h_u^2}$.}
	\begin{IEEEproof}
		\emph{See Appendix~\ref{app:association}.}
	\end{IEEEproof}
	\label{lemma:coverage_association}
	\vspace{-0.2cm}
\end{lemma}
Note that the upper limit of the outer integral in (\ref{eq:Ag}) is not $\infty$ and is equal to $E_{u}=\mathrm{max}\left(E_{u_l}(w_p), E_{u_n}(w_p)\right)$, where $E_{u_l}(\cdot)$ and $E_{u_n}(\cdot)$ are given in (\ref{eq:Eg}). Since $w_p$ is the farthest distance between any UAV and the UE, the signal received from the nearest BS is lower than that of any LOS UAV or NLOS UAV if this BS is located further than $E_{u_l}(w_p)$ or $E_{u_n}(w_p)$.
\vspace{-0.5cm}
\subsection{UE-BS Conditional Coverage Probability}
The conditional coverage probability is the probability that the received $\mathrm{SIR}$ is higher than the threshold $\tau_a$ given the association status. When the UE associates to a BS, this probability is expressed as $P_{cov,g}=\mathbb{P}\left[\mathrm{SIR}\geq \tau_a | s=g \right]$ where $\mathbb{P}\left[s=g\right]=A_g$ is the BS association probability. Due to the universal frequency reuse, the aggregate interference $I_{agg,g}$ includes all non-serving BSs (denoted as $\hat{I}_{g}$) and all UAVs (denoted as $I_{u}$). $P_{cov,g}$  is given in Lemma~\ref{lemma:pcovg}.
\vspace{-0.2cm}
\begin{lemma}
	\emph{The conditional coverage probability $P_{cov,g}$ given that the UE connects to a BS is
		\begin{equation}\small
		P_{cov,g}=\int_{0}^{E_u}\mathcal{L}_{\hat{I}_g}(s_1)\mathcal{L}_{I_u}(s_1)f_{X_g}(x_g)\mathrm{d}x_{g},
		\end{equation}
		where $s_{1}=\frac{\tau_a(x_g^2+h_g^2)^{\frac{\eta_g}{2}}}{P_{g}}$, and $E_u=\mathrm{max}\left(E_{u_l}(w_p), E_{u_n}(w_p)\right)$. $\mathcal{L}_{\hat{I}_g}(s_{1})$ and $\mathcal{L}_{I_u}(s_{1})$ are the Laplace transforms of the aggregate interference of the interfering BSs and all the UAVs. $f_{X_g}(x_g)$ is the PDF of the conditional distance $x_g$ to the serving BS.}
	\begin{IEEEproof}
		\emph{The conditional coverage probability $P_{cov,g}$ is calculated as
			\begin{equation}\small
			\begin{aligned}
			&P_{cov,g}=\mathbb{P}\left[\mathrm{SIR}\geq \tau_a | s=g \right]=\mathbb{P}\left[\frac{P_{g}(x_{g}^2+h_{g}^2)^{-\eta_g/2}\Omega_{g,0}}{I_{agg,g}}\geq \tau_a | s=g \right]\\
			&\stackrel{(a)}{=}\mathbb{E}_{X_g}\left[\mathbb{E}_{I_{agg,g}}\left[\exp\left(-\frac{\tau_a (\hat{I}_g+I_u)}{P_{g}(x_{g}^2+h_{g}^2)^{-\eta_g/2}}\right)\right]\right]\stackrel{(b)}{=}\int_{0}^{E_u}\mathcal{L}_{\hat{I}_g}(s_1)\mathcal{L}_{I_u}(s_1)f_{X_g}(x_g)\mathrm{d}x_{g},
			\end{aligned}
			\label{eq:pcovgproof}
			\end{equation}
			where $s_1=\frac{\tau_a(x_g^2+h_g^2)^{\frac{\eta_g}{2}}}{P_{g}}$, (a) follows from the exponential distribution of $\Omega_{g,0}$ and from the expression of the interference $I_{agg,g}=\hat{I}_{g}+I_u$, and (b) follows from the independence of $\hat{I}_g$ and $I_{u}$ and from the Laplace transform definition. The upper limit of the integral in (\ref{eq:pcovgproof}) $E_{u}=\mathrm{max}\left(E_{u_l}(w_p), E_{u_n}(w_p)\right)$ is the maximum distance from the UE to its serving BS considering the exclusion regions of the LOS and NLOS UAVs. $E_{u_l}(\cdot)$ and $E_{u_n}(\cdot)$ are presented in (\ref{eq:Eg}).}
	\end{IEEEproof}
	\label{lemma:pcovg}
	\vspace{-0.1cm}
\end{lemma}
To obtain the expression of $P_{cov,g}$, Laplace transforms of interference terms and the PDF of the distance to serving BS must be computed. Lemma~\ref{lemma:UE_BS_laplace} and Lemma~\ref{lemma:fxg} present these results.
\vspace{-0.4cm}
\begin{lemma}
	\emph{The Laplace transform of the interference $\hat{I}_g$ of all BSs except the serving BS is
		\begin{equation}\small
		\mathcal{L}_{\hat{I}_{g}}(s_{1})=\exp\left[\frac{-2\pi\lambda_g s_1 P_{g}\left(x_{g}^2+h_{g}^2\right)}{\left(\eta_{g}-2\right)\left(\left(x_{g}^2+h_{g}^2\right)^{\frac{\eta_{g}}{2}}+s_1 P_{g}\right)}{_{2}}F_{1}\left(1,1;2-\frac{2}{\eta_{g}};\frac{1}{1+\frac{\left(x_{g}^2+h_{g}^2\right)^{\frac{\eta_{g}}{2}}}{s_1 P_{g}}}\right)\right].
		\label{eq:Ligb}
		\end{equation}
		where ${_{2}}F_{1}$ denotes the Gauss hyper geometric function.
		The Laplace transform of the interference $I_u$ from the UAVs when the UE associates to a BS is
		\begin{equation}\small
		\begin{aligned}
		\mathcal{L}_{I_{u}}(s_{1})&=\left[\frac{1}{\int_{E_{g_l}(x_{g})}^{w_{p}}f_{W}(w)\kappa_{u_l}(w)\mathrm{d}w+\int_{E_{g_n}(x_{g})}^{w_{p}}f_{W}(w)\kappa_{u_n}(w)\mathrm{d}w}\left(\int_{E_{g_l}(x_{g})}^{w_{p}}\left(1+\frac{s_{1}P_{u}v^{-\eta_{l}}}{m_{l}}\right)^{-m_{l}}\right.\right.\\
		&\times f_{W}(v)\kappa_{u_l}(v)\mathrm{d}v+\int_{E_{g_n}(x_{g})}^{w_{p}}\left(1+\frac{s_{1}P_{u}v^{-\eta_{n}}}{m_{n}}\right)^{-m_{n}}f_{W}(v)\kappa_{u_n}(v)\mathrm{d}v\bigg)\bigg]^{N_u},
		\end{aligned}
		\label{eq:Lia}
		\end{equation}
		where $E_{g_l}(x_g)$ and $E_{g_n}(x_g)$ are the minimum distances at which the LOS and NLOS UAVs are placed when the UE associates to a BS at $x_g$. $f_{W}(\cdot)$ and $\kappa_{u_l}(\cdot)$ are given in (\ref{eq:fw}) and (\ref{eq:LOS_probability}).
	}
	\begin{IEEEproof}
		\emph{See Appendix~\ref{app:laplace}}
	\end{IEEEproof}	
	\label{lemma:UE_BS_laplace}
	\vspace{-0.2cm}
\end{lemma}
The PDF of the conditional distance from the UE to the serving BS $x_g$ is provided in Lemma~\ref{lemma:fxg}.
\vspace{-1cm}
\begin{lemma}
	\emph{The PDF of the distance $x_g$ from the UE to its serving BS is given as
		\begin{equation}\small
		f_{X_{g}}(x_g)=\frac{1}{A_{g}}2\pi\lambda_g x_g \exp\left(-\pi\lambda_g x_{g}^2\right)\left(\int_{E_{g_l}(x_g)}^{w_p}f_{W}(w)\kappa_{u_l}(w)\mathrm{d}w + \int_{E_{g_n}(x_g)}^{w_p}f_{W}(w)\kappa_{u_n}(w)\mathrm{d}w\right)^{N_u},
		\label{eq:fxg}
		\end{equation}
		where $E_{g_l}(\cdot)$ and $E_{g_n}(\cdot)$ are given in (\ref{eq:Eg}). $A_{g}$ and $f_W(\cdot)$ are given in Lemma~\ref{lemma:coverage_association}.
	}
	\begin{IEEEproof}
		\emph{See Appendix~\ref{app:distance}.}
	\end{IEEEproof}
	\label{lemma:fxg}
\end{lemma}
\vspace{-1cm}
\subsection{UE-UAV Conditional Coverage Probability}
For a UAV to serve a UE, its backhaul link should be successful. The conditional coverage probability given that the UE associates to a LOS/NLOS UAV is the joint probability of two events and can be expressed as $P_{cov,u_{\zeta}}=\mathbb{P}\left[\mathrm{SIR}\geq \tau_a,\mathrm{SINR}\geq\tau_b\right | s=u_{\zeta}]$, where the first term corresponds to the UAV-UE link and the second term to the BS-UAV backhaul. Since the UE and the UAV connect to different BSs, and for mathematical tractability, we approximate $P_{cov,u_{\zeta}}\approx \mathbb{P}\left[\mathrm{SIR}\geq \tau_a | s=u_{\zeta}\right]\times\mathbb{P}\left[\mathrm{SINR}\geq\tau_b | s=u_{\zeta} \right]$. Such approximation is further validated numerically in Section~\ref{sec:results}. Note that, due to the different propagation environments considered for access and backhaul, and the different used probabilistic LOS models, the backhaul probability is independent of the UE association and the access coverage.
\subsubsection{Backhaul Probability}
We start by providing an expression for the backhaul probability $S(\tau_b)=\mathbb{P}\left[\mathrm{SINR}\geq\tau_b\right]$. The analysis is performed for a reference UAV located at height $h_u$ and at the center of the disk $D_u$. The set of backhaul-enabled BSs is divided into two sub-processes: the LOS BSs set $\Phi_{g_l}$ with density $\lambda_b\kappa_{b_l}(r)$ and the NLOS BSs set $\Phi_{g_n}$ with density $\lambda_b \kappa_{b_n}(r)$, where $r$ is the UAV-BS distance and $\kappa_{b_l}(\cdot)$ is given in (\ref{eq:LOS_probability2}). Since the UAV connects to the BS that provides the strongest signal strength, the serving BS can only be either the nearest BS in $\Phi_{g_l}$ or in $\Phi_{g_n}$. Lemma~\ref{lemma:backhaul_association} provides the probabilities of connecting to a LOS or a NLOS BS.
\vspace{-0.3cm}
\begin{lemma}
	\emph{The probability $A_{b_\xi}$ that a UAV connects to a BS with LOS/NLOS backhaul link is
		\begin{equation}\small
		A_{b_\xi}=\int_{0}^{\infty}e^{-2\pi\lambda_b\int_{0}^{E_{b_\xi}(x)}\kappa_{b_{\bar{\xi}}}(t)t\mathrm{d}t}f_{s_{b_\xi}}(x)\mathrm{d}x,
		\label{eq:Aln}
		\vspace{-0.2cm}
		\end{equation}
		where $\bar{\xi}=n$ if $\xi=l$ and $\bar{\xi}=l$ if $\xi=n$. $E_{b_\xi}(x)=\sqrt{\left(\frac{C_{\bar{\xi}}}{C_{\xi}}\right)^{\frac{2}{\eta_{\bar{\xi}}}}(x^2+\Delta_h^2)^{\frac{\eta_{\xi}}{\eta_{\bar{\xi}}}}-\Delta_h^2}$, $\Delta_h=|h_u-h_g|$ and $f_{s_{b_\xi}}(x)=2\pi\lambda_b x \kappa_{b_\xi}(x)e^{-2\pi\lambda_b\int_{0}^{x}\kappa_{b_\xi}(r)r\mathrm{d}r}$. }
	\begin{IEEEproof}
		\emph{See Appendix~\ref{app:LOS_association}.}
	\end{IEEEproof}
	\label{lemma:backhaul_association}
	\vspace{-0.3cm}
\end{lemma}
The PDF of the distances $x_{b_l}$ and $x_{b_n}$ from the UAV to the serving BS in $\Phi_{g_l}$ and $\Phi_{g_n}$ are provided in Lemma~\ref{lemma:backhaul_distance}.
\vspace{-0.3cm}
\begin{lemma}	
	\emph{Given that a UAV connects to a BS in $\Phi_{g_\xi}$ to get backhaul, the PDF of the distance $x_{b_\xi}$ to the serving BS is
		\vspace{-0.5cm}
		\begin{equation}\small
		f_{X_{b_\xi}}(x)=\frac{f_{s_{b_\xi}}(x)}{A_{b_\xi}}e^{-2\pi\lambda_b\int_{0}^{E_{b_\xi}(x)}\kappa_{b_{\bar{\xi}}}(t)t\mathrm{d}t},
		\label{eq:fsLN}
		\end{equation}
		where $\xi \in \{l,n\}$ indicates if the UAV connects to a LOS or a NLOS BS for backhaul support. $\bar{\xi}=n$ if $\xi=l$ and $\bar{\xi}=l$ if $\xi=n$. $A_{b_\xi}$, $f_{s_{b_\xi}}(\cdot)$ and $E_{b_\xi}(\cdot)$ are given in Lemma~\ref{lemma:backhaul_association}.
	}
	\begin{IEEEproof}
		\emph{See Appendix~\ref{app:backhaul_distance}.}
	\end{IEEEproof}
	\label{lemma:backhaul_distance}
\end{lemma}
\vspace{-0.3cm}
Finally, the backhaul probability is presented in Theorem~\ref{theorem:backhaul_probability}.
\vspace{-0.4cm}
\begin{theorem}[Backhaul Probability]
	\emph{The backhaul probability $S(\tau_b)$ can be derived as
		\vspace{-0.2cm}
		\begin{equation}\small
		S(\tau_b)=A_{b_l}S_{l}(\tau_b)+A_{b_n}S_{n}(\tau_b),
		\label{eq:Stb}
		\vspace{-0.3cm}
		\end{equation}
		where $S_l(\tau_b)$ and $S_n(\tau_b)$ are the conditional backhaul probabilities given that the UAV is connected to a LOS/NLOS BS. $A_{b_l}$ and $A_{b_n}$ are the association probabilities. $S_{\xi}(\tau_b)$ is given as
		\vspace{-0.4cm}
		\begin{equation}\small
		\begin{aligned}
		&S_{\xi}(\tau_b)\approx\sum_{q=1}^{m_\xi}(-1)^{q+1} {m_\mathrm{\zeta}\choose q}\int_{0}^{\infty}e^{-\frac{q \gamma_\xi\left(x^2+\Delta_h^2\right)^{\frac{\eta_{\xi}}{2}}\tau_b\sigma_{b}^2}{P_{b}C_{\xi}G_{b,0}}-Q_{\xi}(q,x)-V_{\xi}(q,x)}f_{X_{b_\xi}}(x)\mathrm{d}x,
		\end{aligned}
		\label{eq:SLN}
		\end{equation}
		where $f_{X_{b_\xi}}(x)$ is given in (\ref{eq:fsLN}) and
		\vspace{-0.2cm}
		\begin{equation}\small
		\begin{aligned}
		Q_{\xi}(q,x)&=2\pi\lambda_b\sum_{k=1}^{4}p_k\int_{x}^{\infty}H\left(m_\xi,\frac{q\gamma_\xi\bar{G}_k\tau_b(x^2+\Delta_h^2)^{\frac{\eta_{\xi}}{2}}}{m_{\xi}(t^2+\Delta_h^2)^{\frac{\eta_{\xi}}{2}}}\right)\times \kappa_{b_{\xi}}(t)t\mathrm{d}t,
		\end{aligned}
		\end{equation}
		\vspace{-0.2cm}
		\begin{equation}\small
		\begin{aligned}
		V_{\xi}(q,x)&=2\pi\lambda_b\sum_{k=1}^{4}p_k\int_{E_{b_\xi}(x)}^{\infty}H\left(m_{\bar{\xi}},\frac{q C_{\bar{\xi}}\gamma_{\xi}\bar{G}_k\tau_b(x^2+\Delta_h^2)^{\frac{\eta_{\xi}}{2}}}{m_{\bar{\xi}}C_{\xi}(t^2+\Delta_h^2)^{\frac{\eta_{\bar{\xi}}}{2}}}\right)\times \kappa_{b_{\bar{\xi}}}(t)t\mathrm{d}t,
		\end{aligned}
		\end{equation}
		where $H(m,x)=1-1/(1+x)^m$, For $\xi \in \{l,n\}$, $\bar{\xi}=n$ if $\xi=l$ and $\bar{\xi}=l$ if $\xi=n$, $m_\xi$ and $\gamma_{\xi}=m_{\xi}(m_{\xi}!)^{-\frac{1}{m_{\xi}}}$ are the Nakagami-m small scale fading parameters; for $k\in\{1,2,3,4\}$, $\bar{G}_k=\frac{G_k}{G_{b,0}}$, $G_k$ and $p_k$ are defined in Table~\ref{table:antenna}.}
	\begin{IEEEproof}
		\emph{See Appendix~\ref{app:backhaul_probability}.}
	\end{IEEEproof}
	\label{theorem:backhaul_probability}
\end{theorem}
\vspace{-0.4cm}
\subsubsection{Beamforming Alignment Errors}\label{sec:misalignment}
To optimize the backhaul connectivity between the UAV and the BS, the antennas must be perfectly aligned. However, perfect alignment in mmWave systems is challenging due to the required high directivity that leads to a narrow beamwidth of the main lobe. In the absence of beam-steering errors, the maximum directivity gain offered by directional beamforming is $G_{g}^{(\text{max})} G_{u}^{(\text{min})}$, where $G_{g}^{(\text{max})}$ and $G_{u}^{(\text{min})}$ are the gains of the main lobes of the BS and the UAV antennas, respectively. To account for the beamforming alignment error of the desired link, a beam steering error model similar to~\cite{Wildman} is considered. We denote by $\varphi_{s}^{*}$, $s \in \{g,u\}$, the angles corresponding to error-free beam-steering and by $\varepsilon_s$ the additive beam-steering errors. $\varepsilon_g$ and $\varepsilon_u$ are assumed to be randomly distributed, independent of each other and to have a symmetric distribution around $\varphi_{g}^{*}$ and $\varphi_{u}^{*}$, respectively. Let $F_{|\varepsilon_{s}|}(x) = \mathbb{P}(|\varepsilon_{s}|\leq x)$ be the cumulative distribution function (CDF) of $|\varepsilon_{\mathrm{s}}|$ for $s \in \{g,u\}$, and let $\bar{F}_{|\varepsilon_{s}|}(x)=\left(1-F_{|\varepsilon_{s}|}(x)\right)$. Then, the PMF of the directivity gain of the desired link $G_{b,0}$ is $f_{G_{b,0}}(g)=\left(f_{G_{g,0}}\otimes f_{G_{u,0}}\right)(g)$, where $f_{G_{g,0}}(.)$ and $f_{G_{u,0}}(.)$ are the PMFs of the directivity gains of the serving BS and the corresponding UAV, respectively and $\otimes$ is the convolution operator. $f_{G_{s,0}}(g)$ can be written as:
	\vspace{-0.1cm}
	\begin{equation}\small
	f_{G_{s,0}}(g)=F_{|\varepsilon_{\mathrm{s}}|}\left(\frac{\theta_s}{2}\right)\delta\left(g-G_{s}^{(\text{max})}\right)+\bar{F}_{|\varepsilon_{s}|}\left(\frac{\theta_s}{2}\right)\delta\left(g-G_{s}^{(\text{min})}\right)
	\vspace{-0.2cm}
	\end{equation}
	where $\delta(\cdot)$ is the Dirac delta function, and $G_{s}^{(\text{max})}$, $G_{s}^{(\text{min})}$ and $\theta_{s}$ are the respective gains of the main and side lobes and the beamwidth of the BS and UAV antennas, for $s \in \{g,u\}$, as illustrated earlier in (\ref{eq:gain}). Similarly to the directivity gain of the interfering links, the directivity gain of the desired link $G_{b,0}$ is a discrete random variable that takes the values $G_{k}$ with probability $p_{k,0}$, where $G_{k}$ and $p_{k_0}$ are defined in Table~\ref{table:antenna}. The backhaul probability can be expressed as
	\begin{equation}\small
	\begin{aligned}
	&S(\tau_b)=\mathbb{E}_{G_{b,0}}\left[S(\tau_b,G_{b,0})\right]=\int_{0}^{\infty}S(\tau_b,g)f_{G_{b,0}}(g)\mathrm{d}g\\
	&=F_{|\varepsilon_{g}|}(\theta_{g}/2)F_{|\varepsilon_{u}|}(\theta_{u}/2)S(\tau_b,G_{g}^{(\text{max})}G_{u}^{(\text{max})})+F_{|\varepsilon_{g}|}(\theta_{g}/2)\bar{F}_{|\varepsilon_{u}|}(\theta_{u}/2)S(\tau_b,G_{g}^{(\text{max})}G_{u}^{(\text{min})})\\
	&+\bar{F}_{|\varepsilon_{g}|}(\theta_{g}/2)F_{|\varepsilon_{u}|}(\theta_{u}/2)S(\tau_b,G_{g}^{(\text{min})}G_{u}^{(\text{max})})+\bar{F}_{|\varepsilon_{g}|}(\theta_{g}/2)\bar{F}_{|\varepsilon_{u}|}(\theta_{u}/2)S(\tau_b,G_{g}^{(\text{min})}G_{u}^{(\text{min})}).
	\end{aligned}\label{eq:misalignment}
	\end{equation}
\subsubsection{Conditional Coverage Probability and Distance Distribution}
The UE can connect to a LOS or a NLOS UAV depending on the association rule. The coverage probability $P_{cov,u_{\zeta}}$ when the UE associates to a LOS UAV ($\zeta=l$) or a NLOS UAV ($\zeta=n$) is given in Lemma~\ref{lemma:pcova}.
\vspace{-0.3cm}
\begin{lemma}
	\emph{The conditional coverage probability $P_{cov,u_{\zeta}}$ given that UE connects to a $\zeta$-UAV is
		\begin{equation}\small
		P_{cov,u_{\zeta}}=S(\tau_b) \int_{h_u}^{w_p}\sum_{k=0}^{m_{\zeta}-1}\frac{(-s_{2_{\zeta}})^k}{k!}\left[\frac{\partial^k}{\partial s_{2_{\zeta}}^{k}}\mathcal{L}_{I_{g_{\zeta}}}(s_{2_{\zeta}})\mathcal{L}_{\hat{I}_{u_{\zeta}}}(s_{2_{\zeta}})\right] f_{Y_{u_{\zeta}}}(y_{u_{\zeta}})\mathrm{d}y_{u_{\zeta}},
		\label{eq:pcova}
		\end{equation}
		where $s_{2_{\zeta}}=\frac{m_{\zeta}\tau_a y_{u_{\zeta}}^{\eta_{\zeta}}}{P_{u}}$ and $f_{Y_{u_{\zeta}}}(y_{u_{\zeta}})$ is the PDF of the conditional distance to the serving $\zeta$-UAV with $\zeta\in\{l,n\}$. $\mathcal{L}_{\hat{I}_{u_{\zeta}}}(s_{2_{\zeta}})$ and $\mathcal{L}_{I_{g_{\zeta}}}(s_{2_{\zeta}})$ are the Laplace transforms of the aggregate interference of all the UAVs except the serving UAV $\hat{I}_{u_{\zeta}}$ and of all the BSs $I_{g_{\zeta}}$.}
	\begin{IEEEproof}
		\emph{See Appendix~\ref{app:pcova}.}
	\end{IEEEproof}
	\label{lemma:pcova}
	\vspace{-0.3cm}
\end{lemma}
As noted in (\ref{eq:pcova}), the Laplace transforms of the interference terms $\mathcal{L}_{I_{g_{\zeta}}}(s_{2_{\zeta}})$ and $\mathcal{L}_{\hat{I}_{u_{\zeta}}}(s_{2_{\zeta}})$ and the PDF of the conditional distance to the serving UAV $f_{Y_{u_{\zeta}}}(y_{u_{\zeta}})$ must be computed to obtain the final expression of $P_{cov,u_{\zeta}}$. Lemma~\ref{lemma:UE_UAV_laplace} presents these Laplace transforms as follows.
\vspace{-0.3cm}
\begin{lemma}
	\emph{The Laplace transform of the interference $I_{g_{\zeta}}$ of all BSs when the UE associates to a $\zeta$-UAV with $\zeta\in\{l,n\}$ is given as
		\begin{equation}\small
		\mathcal{L}_{I_{g_{\zeta}}}(s_{2_{\zeta}})=\exp\left[\frac{-2\pi\lambda_g s_{2_{\zeta}} P_{g}\left(E_{u_{\zeta}}^{2}(y_{u_{\zeta}})+h_{g}^2\right)}{\left(\eta_{g}-2\right)\left(\left(E_{u_{\zeta}}^{2}(y_{u_{\zeta}})+h_{g}^2\right)^{\frac{\eta_{g}}{2}}+s_{2_{\zeta}} P_{g}\right)}{_{2}}F_{1}\left(1,1;2-\frac{2}{\eta_{g}};\frac{1}{1+\frac{\left(E_{u_{\zeta}}^{2}(y_{u_{\zeta}})+h_{g}^2\right)^{\frac{\eta_{g}}{2}}}{s_{2_{\zeta}} P_{g}}}\right)\right].
		\label{eq:laplace_Ig}
		\end{equation}
		where $E_{u_{\zeta}}(y_{u_{\zeta}})$ is given in (\ref{eq:Eg}) and ${_{2}}F_{1}(\cdot)$ is the Gauss hyper geometric function. The Laplace transform of the interference $\hat{I}_{u_{\zeta}}$ from all UAVs except the serving UAV is given as
		\begin{equation}\small
		\begin{aligned}
		\mathcal{L}_{\hat{I}_{u_{\zeta}}}(s_{2_{\zeta}})&=\left[\frac{1}{\int_{y_{u_{\zeta}}}^{w_{p}}f_{W}(w)\kappa_{u_{\zeta}}(w)\mathrm{d}w+\int_{E_{\zeta\bar{\zeta}}\left(y_{u_{\zeta}}\right)}^{w_{p}}f_{W}(w)\kappa_{u_{\bar{\zeta}}}(w)\mathrm{d}w}\left(\int_{y_{u_{\zeta}}}^{w_{p}}\left(1+\frac{s_{2_{\zeta}}P_{u}v^{-\eta_{\zeta}}}{m_{\zeta}}\right)^{-m_{\zeta}}\right.\right.\\
		&\times f_{W}(v)\kappa_{u_\zeta}(v)\mathrm{d}v+\int_{E_{\zeta\bar{\zeta}}(y_{u_{\zeta}})}^{w_{p}}\left(1+\frac{s_{2_{\zeta}}P_{u}v^{-\eta_{\bar{\zeta}}}}{m_{\bar{\zeta}}}\right)^{-m_{\bar{\zeta}}}f_{W}(v)\kappa_{u_{\bar{\zeta}}}(v)\mathrm{d}v\bigg)\bigg]^{N_u-1}.
		\end{aligned}
		\end{equation}}
	\vspace{-0.4cm}
	\begin{IEEEproof}
		\emph{When the UE associates to a $\zeta$-UAV, the BSs are located further than $E_{u_{\zeta}}(y_{u_{\zeta}})$, the interfering $\zeta$-UAV are located further than $y_{u_{\zeta}}$ and the $\bar{\zeta}$-UAV exist beyond $E_{\zeta\bar{\zeta}}(y_{u_{\zeta}})$, where $y_{u_{\zeta}}$ is the distance from the UE to the serving UAV. Thus, $\mathcal{L}_{I_{g_{\zeta}}}(s_{2_{\zeta}})$ is obtained by following a similar approach to Lemma~\ref{lemma:UE_BS_laplace} while replacing the lower limit of the integral in (\ref{eq:prooflemma3}) by $E_{u_{\zeta}}(y_{u_{\zeta}})$. The interference from all UAVs except the serving UAV can be expressed as $\sum_{y_i\in\Phi_{u}\backslash y_{u_{\zeta}}}I_{u,y_j}$, where $I_{u,y_j}$ is the interference from the $j$-th UAV and can be obtained by following a similar approach to Lemma~\ref{lemma:UE_BS_laplace} as $\mathcal{L}_{\hat{I}_{u,y_{j}}}(s_{2_{\zeta}})=\left(\mathbb{E}_{I_{u,y_j}}\left[\exp\left(-s_{2_\zeta} I_{u,y_j}\right)\right]\right)^{N_u-1}$. The final expression of $\mathcal{L}_{\hat{I}_{u_\zeta}}(s_{2_{\zeta}})$ is obtained by replacing the lower limits of the integrals by $y_{u_{\zeta}}$ and $E_{\zeta\bar{\zeta}}(y_{u_{\zeta}})$ and by plugging $f_{V_{u_l}}(v,y_{u_\zeta})$ and $f_{V_{u_n}}(v,E_{\zeta\bar{\zeta}}(y_{u_{\zeta}}))$ in $\mathbb{E}_{I_{u,y_j}}\left[\exp\left(-s_{2_{\zeta}} I_{u,y_j}\right)\right]$ in (\ref{eq:Euai}).}
	\end{IEEEproof}	
	\label{lemma:UE_UAV_laplace}
\end{lemma}
\vspace{-0.2cm}
Next, we derive the PDF of the distance from the UE to the serving UAV $y_{u_{\zeta}}$ in Lemma~\ref{lemma:fxazeta}.
\vspace{-0.3cm}
\begin{lemma}
	\emph{The PDF of the distance $y_{u_{\zeta}}$ to the serving $\zeta$-UAV ($\zeta \in \{l,n\}$) is given as
		\begin{equation}\small
		\begin{aligned}
		f_{Y_{u_{\zeta}}}(y_{u_{\zeta}})&=\frac{N_u}{A_{u_{\zeta}}}f_{W}(y_{u_{\zeta}})\kappa_{u_\zeta}(y_{u_{\zeta}})\exp\left(-\pi\lambda_g E_{u_{\zeta}}(y_{u_{\zeta}})^2\right)\left(\int_{y_{u_{\zeta}}}^{w_p}f_{W}(w)\kappa_{u_\zeta}(w)\mathrm{d}w\right.\\
		&\left.+\int_{E_{\zeta\bar{\zeta}}(y_{u_{\zeta}})}^{w_p}f_{W}(w)\kappa_{u_{\bar{\zeta}}}(w)\mathrm{d}w\right)^{N_u-1},
		\end{aligned}
		\label{eq:fxa}
		\end{equation}
		where $f_W(\cdot)$ and $A_{u_\zeta}$ are given in Lemma~\ref{lemma:coverage_association}. $E_{u_{\zeta}}(\cdot)$ and $E_{\zeta\bar{\zeta}}(\cdot)$ are given in (\ref{eq:Eg}).}
	\begin{IEEEproof}
		\emph{This proof follows a similar approach to Lemma~\ref{lemma:fxg}.}
	\end{IEEEproof}
	\label{lemma:fxazeta}
\end{lemma}
\vspace{-1cm}
\subsection{Overall Coverage Probability}
\vspace{-0.2cm}
After deriving the association probabilities, the conditional distance distributions and the conditional coverage probabilities, the overall coverage probability is obtained through the law of total probability as
$P_{cov}=A_{u_l} P_{cov,u_l}+A_{u_n} P_{cov,u_n}+A_g P_{cov,g}$.
\vspace{-0.5cm}
\section{Backhaul Aware Transmission}\label{sec:aware}
\vspace{-0.1cm}
For the backhaul aware transmission scenario, the UE activity is dependent on the BS-UAV instantaneous backhaul link in case of aerial coverage. That is, the UAV is aware of the backhaul status as it checks the integrity of packets received from the BS before transmitting them to the UE. If the backhaul $\mathrm{SINR}$ satisfies the threshold, the UAV can serve the UE successfully. When the backhaul is down, the UAV refrains its transmission to the UE. The UE does not change its association,\footnote{Changing association based on instantaneous link status may lead to undesirable excessive handovers known as the ping pong effect.} but rather goes to a service failure event in this time slot. As result, the backhaul aware transmission scheme has the following two main advantages, i) the established DL transmissions are subject to the access link $\mathrm{SIR}$ only, since UAVs have the ability to refrain from transmitting corrupted packets, and ii) the interference is relieved because only successfully backhauled UAVs contribute to it, thereby improving the overall coverage probability. Such advantages come at the cost of extra processing at the UAVs because they check the validity of the packets before transmitting to UE. To characterize the coverage probability of the backhaul aware scheme, we follow the same steps of Section~\ref{sec:unaware}. In particular, we start by deriving the transmission probabilities, then we find the conditional coverage probabilities, then we obtain the distance distributions, and finally we provide the unified expression for the coverage probability.
\vspace{-1.4cm}
\subsection{Transmission Probabilities}\label{sec:aware_association}
\vspace{-0.2cm}
In the context of backhaul aware transmission, we define the transmission probability as the probability that the UE is being served by a BS or a successfully backhauled LOS/NLOS UAV. Note that in the case where the UAV-UE access link is better than the BS-UE link and the UAV backhaul link is not successful, the UE is considered in a service failure state and fails to connect. Hence, the service failure event here refers to the event where the UE is not served in the DL in a time slot not as the event that DL is established and a decoding error occurs. It is important to note that an established DL connection is different from UE to UAV association, where the latter is independent from the backhaul link state. When the UE is served by a terrestrial BS, the transmission probability is expressed as $\tilde{A}_g=\mathbb{P}[s=g]$. Note that the transmission probability when the UE connects to a BS is the same as the association probability to a BS for the unaware transmission scenario and is presented in (\ref{eq:Ag}). For a UAV to transmit to the UE, the backhaul link of this UAV should be successful. The UAV transmission probability is the joint probability of two events and can be expressed as $\tilde{A}_{u_{\zeta}}=\mathbb{P}[s=u_{\zeta}, \mathrm{SINR}>\tau_b]$, where the first term corresponds to the UE-UAV association rule $\left(\zeta \in\{l,n\}\right)$ and the second term to the BS-UAV backhaul link. The independence assumption is also considered and validated in Section~\ref{sec:results}. Thus, the UAV transmission probability can be expressed as $\tilde{A}_{u_{\zeta}}=\mathbb{P}[s=u_{\zeta}]\times\mathbb{P}[\mathrm{SINR}>\tau_b]$, where $\mathbb{P}[s=u_{\zeta}]=A_{u_{\zeta}}$ is the UAV association probability for the backhaul unaware scenario derived in Lemma~\ref{lemma:coverage_association} and $\mathbb{P}[\mathrm{SINR}>\tau_b]=S(\tau_b)$ is the backhaul probability given in (\ref{eq:Stb}). The service failure probability is defined as $\tilde{A}_{f}=\mathbb{P}[s=u_l , \mathrm{SINR}<\tau_b \| s=u_n, \mathrm{SINR}<\tau_b]$ and derived as
\vspace{-0.2cm}
\begin{equation}
\begin{aligned}
\tilde{A}_{f} &=\mathbb{P}\left[s=u_l \right]\mathbb{P}\left[\mathrm{SINR} < \tau_b \right]+\mathbb{P}\left[s=u_n \right]\mathbb{P}\left[\mathrm{SINR} < \tau_b \right]=A_{u_l} \left(1-S(\tau_b)\right)+A_{u_n} \left(1-S(\tau_b)\right),
\end{aligned}
\vspace{-0.5cm}
\end{equation}
where $A_{u_l}$ and $A_{u_n}$ are derived in Lemma~\ref{lemma:coverage_association} and $S(\tau_b)$ in Theorem~\ref{theorem:backhaul_probability}.
\vspace{-0.4cm}
\subsection{Conditional and Overall Coverage Probabilities}\label{sec:aware_conditional}
\vspace{-0.2cm}
The conditional coverage probability given that the UE is served by a BS is expressed as $\tilde{P}_{cov,g}=\mathbb{P}[\mathrm{SIR}\geq\tau_a|s=g]$. Due to the dependency of the UAV transmission on the quality of the backhaul link, the aggregate interference for the backhaul aware scenario $\tilde{I}_{agg,g}$ includes the interference from only the UAVs with successful backhaul links denoted as $I_{\tilde{u}}$ and from all BSs except the serving BS denoted as $\hat{I}_{g}$. An expression for $\tilde{P}_{cov,g}$ is given in Lemma~\ref{lemma:pcovglemma}.
\vspace{-0.3cm}
\begin{lemma}
	\emph{The conditional coverage probability given that the UE is served by a BS for the backhaul aware scenario is given as
		\begin{equation}\small
		\tilde{P}_{cov,g}=\int_{0}^{E_u}\mathcal{L}_{\hat{I}_g}(s_1)\mathcal{L}_{I_{\tilde{u}}}(s_1)f_{X_g}(x_g)\mathrm{d}x_{g},
		\label{eq:pcovg_aware}
		\end{equation}
		where $s_{1}=\frac{\tau_a(x_g^2+h_g^2)^{\frac{\eta_g}{2}}}{P_{g}}$, $E_u = \mathrm{max}\left(E_{u_l}(w_p),E_{u_n}(w_{p})\right)$, $\mathcal{L}_{\hat{I}_g}(s_{1})$ is given in (\ref{eq:Ligb}) and $f_{X_g}(x_g)$ in (\ref{eq:fxg}). $\mathcal{L}_{I_{\tilde{u}}}(s_1)$ is the Laplace transform of the aggregate interference from the UAVs that have successful backhaul links with ground BSs.}
	\begin{IEEEproof}
		\emph{This proof follows a similar approach to Lemma~\ref{lemma:pcovg}.
		}
	\end{IEEEproof}
	\label{lemma:pcovglemma}
	\vspace{-0.3cm}
\end{lemma}
A lower bound for the Laplace transform $\mathcal{L}_{I_{\tilde{u}}}(s_1)$ is derived in Lemma~\ref{lemma:laplace_aware}.
\vspace{-0.3cm}
\begin{lemma}
	\emph{The Laplace transform of the aggregate interference $I_{\tilde{u}}$ from all the UAVs that have a successful backhaul link when the UE is served by a BS can be lower bounded by
		\begin{equation}\small
		\begin{aligned}
		\mathcal{L}_{I_{\tilde{u}}}(s_{1})&\geq\left[\frac{1}{\int_{E_{g_l}(x_{g})}^{w_{p}}f_{W}(w)\kappa_{u_l}(w)\mathrm{d}w+\int_{E_{g_n}(x_{g})}^{w_{p}}f_{W}(w)\kappa_{u_n}(w)\mathrm{d}w}\left(\int_{E_{g_l}(x_{g})}^{w_{p}}\left(1+\frac{s_{1}P_{u}v^{-\eta_l}}{m_l}\right)^{-m_l}\right.\right.\\
		&\times f_{W}(v)\kappa_{u_l}(v)\mathrm{d}v+\int_{E_{g_n}(x_{g})}^{w_{p}}\left(1+\frac{s_{1}P_{u}v^{-\eta_n}}{m_n}\right)^{-m_n} f_{W}(v)\kappa_{u_n}(v)\mathrm{d}v\bigg)\bigg]^{N_{u}\times S(\tau_b)},
		\end{aligned}
		\label{eq:Lia_aware}
		\end{equation}
		where $E_{g_l}(\cdot)$ and $E_{g_n}(\cdot)$ are given in (\ref{eq:Eg}), $f_{W}(\cdot)$ and $w_p$ are given in Lemma~\ref{lemma:coverage_association}. }
	\begin{IEEEproof}
		\emph{We denote $\tilde{N_u}$ the number of UAVs that have successful backhaul links. $I_{\tilde{u}}$ is given as $\sum_{y_{j\in\tilde{\Phi_{u}}}}I_{u,y_j}$ where $I_{u,y_j}$ denotes the interference from the $j$-th UAV and $\tilde{\Phi_{u}}$ is the set of UAVs with successful backhaul links. The Laplace transform of $I_{\tilde{u}}$ can be calculated as
			\begin{equation}\small
			\begin{aligned}
			&\mathcal{L}_{I_{\tilde{u}}}(s_{1})=\mathbb{E}_{I_{\tilde{u}}}\left[e^{-s_{1}I_{\tilde{u}}}\right]=\mathbb{E}_{I_{\tilde{u}}}\left[\exp\left(-s_{1}\sum_{j=1}^{\tilde{N_u}}I_{u,y_j}\right)\right]
			\stackrel{(a)}{=}\mathbb{E}_{\tilde{N_u}}\left[\prod_{j=1}^{\tilde{N_u}} \mathbb{E}_{I_{u,y_j}}\left[\exp\left(-s_{1}I_{u,y_j}\right)\right]\right]\\
			&\stackrel{(b)}{=}\mathbb{E}_{\tilde{N_u}}\left[\left(\mathbb{E}_{I_{u,y_j}}\left[\exp\left(-s_1 I_{u,y_j}\right)\right]\right)^{\tilde{N_u}}\right]\stackrel{(c)}{\geq}\left(\mathbb{E}_{I_{u,y_j}}\left[\exp\left(-s_1 I_{u,y_j}\right)\right]\right)^{\mathbb{E}\left[\tilde{N_u}\right]},
			\end{aligned}
			\end{equation}
			where (a) follows from the independent and identically distributed (iid) fading gains and from their independence of the interferers distances in the expression of the interference $I_{u,y_j}$. (b) follows from the iid distribution of the interferers UAVs. (c) follows from the Jensen's inequality that gives a lower bound for $\mathcal{L}_{I_{\tilde{u}}}(s_{1})$. The number of UAVs that have successful backhaul links $\tilde{N_u}$ follows a binomial distribution with backhaul probability $S(\tau_b)$ and mean $\mathbb{E}[\tilde{N_u}]=N_u S(\tau_b)$. The remaining proof follows the procedure adopted in Lemma~\ref{lemma:UE_BS_laplace}.}
	\end{IEEEproof}
	\label{lemma:laplace_aware}
	\vspace{-0.3cm}
\end{lemma}
Similar to the UE-BS conditional coverage probability, the conditional coverage probability given that the UE is served by a $\zeta$-UAV is expressed as $\tilde{P}_{cov,u_{\zeta}}=\mathbb{P}[\mathrm{SIR}\geq\tau_a|s=u_{\zeta}]$. The aggregate interference $\tilde{I}_{agg,u_{\zeta}}$ includes the interference from the UAVs with successful backhaul links except the serving UAV denoted as $\hat{I}_{\tilde{u_{\zeta}}}$ and from all BSs denoted as $I_{g_{\zeta}}$. An expression for $\tilde{P}_{cov,u_{\zeta}}$ is given in the following lemma.
\vspace{-0.3cm}
\begin{lemma}
	\emph{The coverage probability $P_{cov,u_{\zeta}}$ given that the UE is served by a $\zeta$-UAV is
		\begin{equation}\small
		\tilde{P}_{cov,u_{\zeta}}= \int_{h_u}^{w_p}\sum_{k=0}^{m_{\zeta}-1}\frac{(-s_{2_{\zeta}})^k}{k!}\left[\frac{\partial^k}{\partial s_{2_{\zeta}}^{k}}\mathcal{L}_{I_{g_{\zeta}}}(s_{2_{\zeta}})\mathcal{L}_{\hat{I}_{\tilde{u_{\zeta}}}}(s_{2_{\zeta}})\right] f_{Y_{u_{\zeta}}}(y_{u_{\zeta}})\mathrm{d}y_{u_{\zeta}},
		\label{eq:pcova_aware}
		\end{equation}
		where $\zeta\in\{l,n\}$, $s_{2_{\zeta}}=\frac{m_{u_{\zeta}}\tau_a y_{u_{\zeta}}^{\eta_{u_{\zeta}}}}{P_{u}}$. $f_{Y_{u_{\zeta}}}(y_{u_{\zeta}})$ is given in (\ref{eq:fxa}) and $\mathcal{L}_{I_{g_{\zeta}}}(s_{2_{\zeta}})$ in (\ref{eq:laplace_Ig}). $\mathcal{L}_{\hat{I}_{\tilde{u_{\zeta}}}}(s_{2_{\zeta}})$ is the Laplace transform of the aggregate interference of all the UAVs that have successful backhaul links except the serving UAV and is given as
		\begin{equation}\small
		\begin{aligned}
		\mathcal{L}_{\hat{I}_{\tilde{u_{\zeta}}}}(s_{2_{\zeta}})&\geq \left(\frac{1}{\int_{y_{u_{\zeta}}}^{w_{p}}f_{W}(w)\kappa_{u_\zeta}(w)\mathrm{d}w+\int_{E_{\zeta\bar{\zeta}}(y_{u_{\zeta}})}^{w_{p}}f_{W}(w)\kappa_{u_{\bar{\zeta}}}(w)\mathrm{d}w}\left(\int_{y_{u_{\zeta}}}^{w_{p}}\left(1+\frac{s_{2_{\zeta}}P_{u}v^{-\eta_{\zeta}}}{m_{\zeta}}\right)^{-m_{\zeta}}\right.\right.\\
		&\times f_{W}(v)\kappa_{u_\zeta}(v)\mathrm{d}v+\int_{E_{\zeta\bar{\zeta}}(y_{u_{\zeta}})}^{w_{p}}\left(1+\frac{s_{2_{\zeta}}P_{u}v^{-\eta_{\bar{\zeta}}}}{m_{\bar{\zeta}}}\right)^{-m_{\bar{\zeta}}}f_{W}(v)\kappa_{u_{\bar{\zeta}}}(v)\mathrm{d}v\bigg)\bigg)^{N_u S(\tau_b)-1}.
		\end{aligned}
		\end{equation}
		\vspace{-0.5cm}
	}
	\begin{IEEEproof}
		\emph{The proof of this lemma follows a similar approach to Lemma~\ref{lemma:UE_BS_laplace} and Lemma~\ref{lemma:laplace_aware}.}
	\end{IEEEproof}
	\label{lemma:pcova_aware}
	\vspace{-0.2cm}
\end{lemma}
The overall coverage probability for the backhaul aware transmission scenario can be derived through the law of total probability as $\tilde{P}_{cov}=\tilde{A}_{u_l} \tilde{P}_{cov,u_l}+\tilde{A}_{u_n} \tilde{P}_{cov,u_n}+\tilde{A}_g \tilde{P}_{cov,g}$, where $\tilde{A}_{u_l}$, $\tilde{A}_{u_n}$ and $\tilde{A}_g$ are given in Section~\ref{sec:aware_association} and $\tilde{P}_{cov,u_l}$, $\tilde{P}_{cov,u_n}$ and $\tilde{P}_{cov,g}$ are given in Section~\ref{sec:aware_conditional}.
\vspace{-0.5cm}
\section{Numerical results and Discussions}\label{sec:results}
In this section, we verify the analytical analysis against Monte-Carlo simulations and highlight the impact of different system parameters. Unless it is stated explicitly otherwise, we assume that the UE is at the origin ($v_0=0$) and we employ the simulation parameters listed in Table~\ref{table:parameters}. Assuming dense urban environment, we set the environmental parameters $a = 0.136$ and $b = 11.95$ as in~\cite{Saad} to determine the LOS probability on the UAV-UE access links. The dense urban environment parameters of the LOS probability model adopted for the BS-UAV backhaul links are given in~\cite{Cherif} as $h_{g}=25$~m, $c=1$, $d=0.106$ and $e=1$.
\begin{table*}[t]
	\centering
	\caption{Simulation Parameters.}
	\vspace{-0.3cm}
	\begin{tabular}{c|c||c|c||c|c}
		\hline\hline
		\textbf{Parameter} & \textbf{Value} & \textbf{Parameter} & \textbf{Value}  & \textbf{Parameter} & \textbf{Value} \\\hline
		($P_{g}$, $P_{b}$, $P_{u}$) & ($20$, $10$, $1$)~W & ($\eta_{g}$, $\eta_l$, $\eta_n$) & ($4$ , $2.5$, $4$) & ($\theta_{g}, \theta_{u}$) & ($20^{\circ}$, $20^{\circ}$) \\\hline
		$r_{c}$ & $1000$~m & ($m_l$, $m_n$) &($3$, $2$) & ($a$, $b$) & ($11.95$, $0.136$) \\\hline
		($h_{g}$, $h_{u}$) & ($25$, $100$)~m & ($C_l$, $C_n$) & ($-69.8$, $-69.8$)~dB & ($c$, $d$, $e$) & ($1$, $0.106$, $1$) \\\hline
		$\lambda_{g}$ & $10$~BS/$\mathrm{km}^2$ &($G_{g}^{(\text{max})}$, $G_{g}^{(\text{min})}$) & ($18$, $-2$)~dB & $\sigma_{b}^2$ & $4\cdot10^{-11}$~W \\\hline
		$N_{u}$, $\delta_{b}$ & $10$, $1$ & ($G_{u}^{(\text{max})}$, $G_{u}^{(\text{min})}$) & ($18$, $-2$)~dB & ($\tau_a$, $\tau_b$) & ($0$, $10$)~dB \\\hline\hline
	\end{tabular}
	\label{table:parameters}
	\vspace{-0.5cm}
\end{table*}
      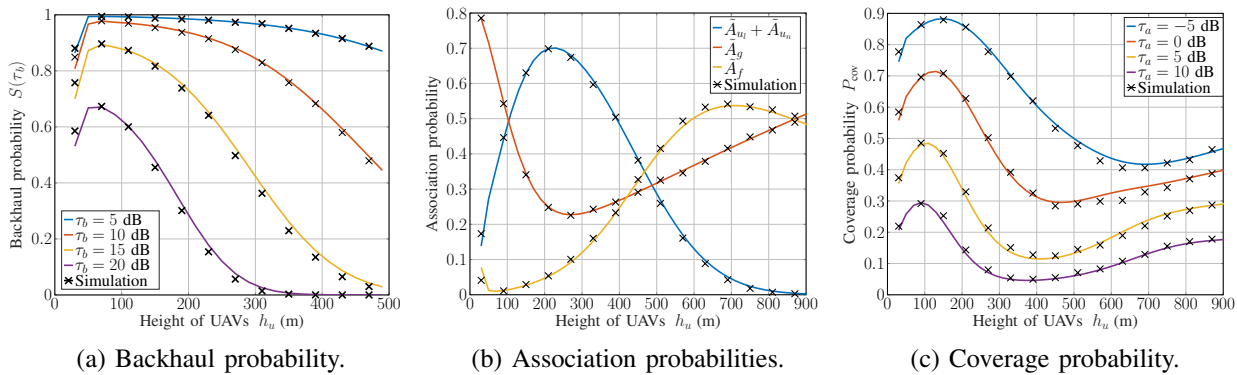
\begin{figure*}[t]
   	\begin{subfigure}[t]{0.33\textwidth}
   	\resizebox{0.97\columnwidth}{!}{\begin{tikzpicture}[thick,scale=1, every node/.style={scale=1.3},font=\Huge]
   			\input{Figures/Fig5a.tex}
   			\end{tikzpicture}}
		\vspace{-0.2cm}
		\caption{Backhaul probability.}
		\label{fig:backhaul_ha}
   	\end{subfigure}
   	~
   	\begin{subfigure}[t]{0.33\textwidth}
   		\resizebox{0.97\columnwidth}{!}{\begin{tikzpicture}[thick,scale=1, every node/.style={scale=1.3},font=\Huge]
   			\input{Figures/Fig5b.tex}
   			\end{tikzpicture}}
		\vspace{-0.2cm}
		\caption{Association probabilities.}
		\label{fig:association}
   	\end{subfigure}
   	~
   	\begin{subfigure}[t]{0.33\textwidth}
   		\resizebox{0.97\columnwidth}{!}{\begin{tikzpicture}[thick,scale=1, every node/.style={scale=1.3},font=\Huge]
   			\input{Figures/Fig5c.tex}
   			\end{tikzpicture}}
	\vspace{-0.2cm}
	\caption{Coverage probability.}
	\label{fig:aware_ha_beta}
   	\end{subfigure}
	\vspace{-0.4cm}
	\caption{Validation of analytical results and impact of UAVs height for backhaul aware.}
	\vspace{-1cm}
   \end{figure*}

\vspace{-0.8cm}
\subsection{Numerical results}
\vspace{-0.25cm}
We analyze the impact of the UAV height on the capability of the UAV to get a good backhaul link in Fig.~\ref{fig:backhaul_ha}. The solid lines represent the analytical results and the markers represent the simulation results. It is noted that the analytical results match perfectly with the simulations, which validates the proposed model. When a UAV hovers at higher altitude, its distance to the serving BS increases, which causes a degradation in the received signal power. However, a higher altitude means a higher LOS probability. For very low altitudes above the terrestrial BSs, the LOS probability is the dominant factor, and hence, the backhaul probability improves. As the UAV increases its altitude, the path-loss dominates and the backhaul probability degrades. An optimal height can be noted in Fig.~\ref{fig:backhaul_ha}, which maximizes the backhaul probability.

In Fig.~\ref{fig:association}, we show the aerial and terrestrial association probabilities in addition to the service failure probability for the backhaul aware scenario as function of the height of the UAVs. In the low heights region, the UE tends to associate more with UAVs and less with terrestrial BSs as the height of these UAVs increases. This is due to the fact that, as the height increases, more UAVs come within LOS conditions with respect to the UE. However, with a further increase in height, the UAV association probability starts to decrease because of the significant path-loss caused by the increasing distances between the UE and the UAVs. At the same time, the backhaul probability degrades which causes the increase of the service failure probability that corresponds to a good UAV-UE access connection with bad backhauling. The existence of UAVs in this case is harming the performance of the UE as backhauling is not guaranteed at high heights. With a further increase in height, the service failure probability starts to decrease because of the significant path-loss encountered on the UAV-UE access links. In this high height regime, the role of UAV becomes minor and the UE sticks to terrestrial connections.

Fig.~\ref{fig:aware_ha_beta} demonstrates how the increase of the UAVs height affects the coverage probability for different $\mathrm{SIR}$ thresholds in the backhaul aware scenario. For low heights, the coverage probability increases as the UAV height increases. This is due to the fact that, as the height increases, more UAVs will have successful backhaul links. For higher altitudes, the backhaul link deteriorates and the distances from the UE to its serving device and interfering UAVs increase, and so the $\mathrm{SIR}$ and the coverage probability decrease. This corresponds to the increase in the service failure probability as seen in Fig.~\ref{fig:association} which reflects the existence of some UEs with good UAV-UE access links and bad backhaul links. With a further increase in height, the coverage probability starts to increase again until a certain limit after which it stabilizes. For these high heights, there is no point of using UAVs and all the UEs are served with the terrestrial network. Fig.~\ref{fig:aware_ha_beta} reveals that the optimal UAV height is related to the value of the $\mathrm{SIR}$ threshold $\tau_a$. For higher thresholds, the UAVs must be deployed at lower height to maximize the coverage probability.

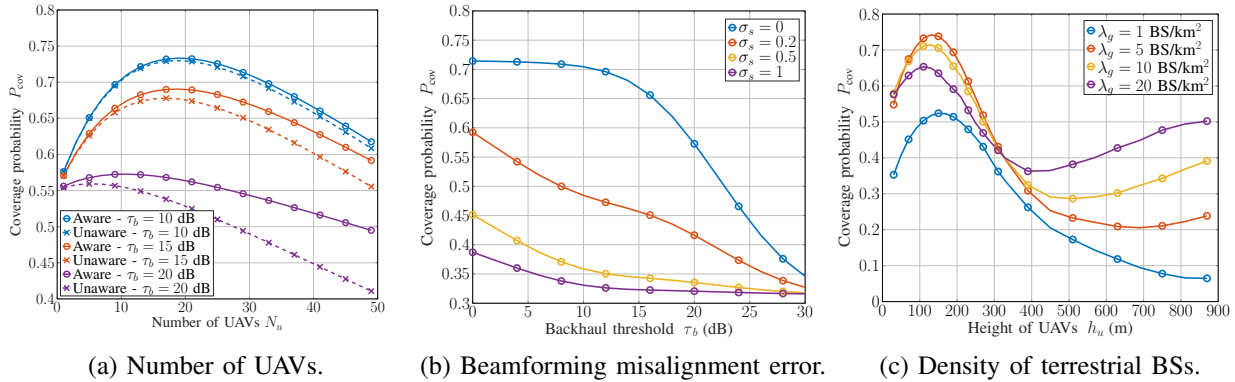
\begin{figure*}[t]
	\begin{subfigure}[t]{0.33\textwidth}
		\resizebox{0.97\columnwidth}{!}{\begin{tikzpicture}[thick,scale=1, every node/.style={scale=1.3},font=\Huge]
			\input{Figures/Fig6a.tex}
			\end{tikzpicture}}
		\vspace{-0.2cm}
		\caption{Number of UAVs.}
		\label{fig:coverage_aware_unaware100}
	\end{subfigure}
	~
	\begin{subfigure}[t]{0.33\textwidth}
		\resizebox{0.97\columnwidth}{!}{\begin{tikzpicture}[thick,scale=1, every node/.style={scale=1.3},font=\Huge]
			\input{Figures/Fig6b.tex}
			\end{tikzpicture}}
		\vspace{-0.2cm}
		\caption{Beamforming misalignment error.}
		\label{fig:misalignment_error}
	\end{subfigure}
	~
	\begin{subfigure}[t]{0.33\textwidth}
		\resizebox{0.97\columnwidth}{!}{\begin{tikzpicture}[thick,scale=1, every node/.style={scale=1.3},font=\Huge]
			\input{Figures/Fig6c.tex}
			\end{tikzpicture}}
		\vspace{-0.2cm}
		\caption{Density of terrestrial BSs.}	
		\label{fig:coverage_lambda}
	\end{subfigure}
	\vspace{-0.4cm}
	\caption{Coverage probability as function of number of UAV (for backhaul unaware and aware transmissions), beamforming misalignment error and density of BSs.}
	\vspace{-1cm}
\end{figure*}
In Fig.~\ref{fig:coverage_aware_unaware100}, we illustrate how the number of UAVs impacts the coverage probability for $\tau_a=0$~dB and $h_u=100$~m for different backhaul thresholds~$\tau_b$. Solid and dashed lines present the results of the backhaul aware and unaware transmission scenarios, respectively. It is clearly seen that increasing the number of UAVs first improves the coverage probability but then deteriorates it for both scenarios. This is because adding more UAVs has both a positive and a negative impact. First, a higher UAVs number increases the average received power from the serving UAV since the distance between the UE and the UAV is lower and the LOS probability is higher. However, the added UAVs impose higher interference levels. Due to the bounded region in which the UAVs are distributed and since the UE-UAV distance is always $\geq h_u$ and the LOS probability $\leq 1$, the power of the received signal reaches a maximum when a sufficient number of UAVs is deployed. On the other side, the interference remains increasing when more UAVs are deployed and the negative effect becomes the dominant factor. Fig.~\ref{fig:coverage_aware_unaware100} also reveals that there exists an optimal number of UAVs that must be deployed to optimize the UE service. Furthermore, Fig.~\ref{fig:coverage_aware_unaware100} shows that the awareness of the backhaul link improves the coverage probability compared to the unaware scenario. This is due to the exclusion of the UAVs that do not have a successful backhaul link from the interfering UAVs. As the number of UAVs in the network increases, the gap between the aware and unaware coverage results becomes larger. It is worth to note here that, even though a service failure event is considered for backhaul aware transmission, its performance remains superior to the unaware scenario for different network parameters.

To evaluate the impact of the beamforming misalignment error on the backhaul link, we introduce the alignment error in the backhaul probability expression in (\ref{eq:misalignment}). We assume that the additive beam steering errors $\varepsilon_{g}$ and $\varepsilon_{u}$ for both BSs and UAVs follow a Gaussian distribution with zero mean and variances $\sigma_{g}^2$ and $\sigma_{u}^2$, respectively. Thus, $\lvert \varepsilon_{s}\rvert$ follows a half-normal distribution with $F_{\lvert\varepsilon_{s}\rvert}=\mathrm{erf}\left(\frac{x}{\sqrt{2}\sigma_{s}}\right)$ and $\bar{F}_{\lvert\varepsilon_{s}\rvert}(x)=1-F_{\lvert\varepsilon_{s}\rvert}(x)$ where $s\in\{g,u\}$ and $\mathrm{erf}(\cdot)$ denotes the error function. Fig.~\ref{fig:misalignment_error} presents the coverage probability as function of the threshold $\tau_b$ for different values of the misalignment error variance ($\sigma_{u}=\sigma_{g}$). We can clearly notice that the quality of the backhaul link has a direct impact on the UE experience. When the BS and UAV antennas are not aligned on the backhaul link, the coverage probability deteriorates. Thus, overlooking the backhaul effect can lead to misleading coverage insights. The obtained results validate the importance of our work compared to the works that consider guaranteed backhaul links.

Fig.~\ref{fig:coverage_lambda} shows the coverage probability $P_{cov}$ as function of $\lambda_{g}$ and the $h_{u}$ considering backhaul aware transmission. When the UAVs hover at low altitudes, increasing the number of BSs decreases $P_{cov}$. Higher densities of BSs result in a higher backhaul probability since the distance separating the UAV from its serving BS decreases, which increases the average received signal power and consequentially enhances the LOS probability. At the same time, at lower heights, the UE tends to associate more to UAVs, and since backhaul aware transmission is considered, more successfully backhauled UAVs means high interference levels and lower coverage probability. At the optimal height, the UEs are striking a balance between good backhauling and association with UAVs. As the height further increases, the coverage probability starts decreasing and reaches a break even point after which adding more terrestrial BSs improves the coverage probability. At this break even point, connecting to a UAV or a terrestrial BS would give the same coverage performance. This analysis is valid until a certain value of the density of BSs, below which insufficient backhauling is encountered. This is clearly noticed for low BSs densities as the UAV has less BSs to align beam with, and initiate backhaul connection. For high heights, the impact of the aerial network is minimal and the UE associates with the terrestrial network. Thus, adding more BSs would decrease the distance separating the UE from its serving BS, which increases the average received signal power and consequentially the coverage probability.

\begin{figure*}[t]
	\begin{subfigure}[t]{0.33\textwidth}
		\resizebox{0.97\columnwidth}{!}{\begin{tikzpicture}[thick,scale=1, every node/.style={scale=1.3},font=\Huge]
			\input{Figures/Fig7a.tex}
			\end{tikzpicture}}
		\vspace{-0.2cm}
		\caption{Backhaul probability.}
		\label{fig:backhaul_fraction}
	\end{subfigure}
	~
	\begin{subfigure}[t]{0.33\textwidth}
		\resizebox{0.97\columnwidth}{!}{\begin{tikzpicture}[thick,scale=1, every node/.style={scale=1.3},font=\Huge]
			\input{Figures/Fig7b.tex}
			\end{tikzpicture}}
		\vspace{-0.2cm}
		\caption{Coverage probability.}
		\label{fig:coverage_fraction}
	\end{subfigure}
	~
	\begin{subfigure}[t]{0.33\textwidth}
		\resizebox{0.97\columnwidth}{!}{\begin{tikzpicture}[thick,scale=1, every node/.style={scale=1.3},font=\Huge]
			\input{Figures/Fig7c.tex}
			\end{tikzpicture}}
		\vspace{-0.2cm}
		\caption{Aware vs. Instantaneous.}	
		\label{fig:outage_no_outage}
	\end{subfigure}
	\vspace{-0.4cm}
	\caption{Backhaul and coverage probabilities as function of the fraction of BSs with backhaul and comparison between backhaul aware transmission and backhaul aware instantaneous association.}
	\vspace{-1cm}
\end{figure*}
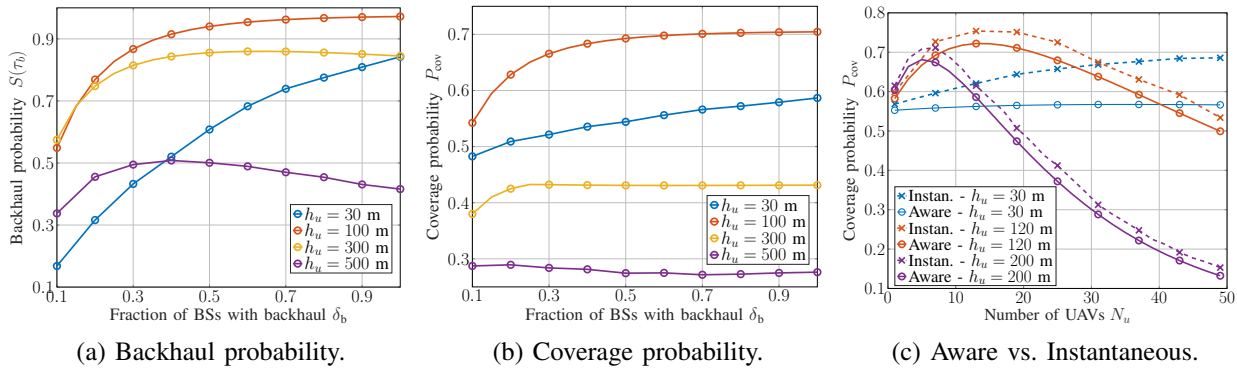


Fig.~\ref{fig:backhaul_fraction} and Fig.~\ref{fig:coverage_fraction} show the impact of having a fraction~$\delta_{b}\leq1$ of backhaul-enabled BSs on both the backhaul and the coverage probabilities, respectively. For low heights, adding more backhaul-enabled BSs increases both the backhaul and the coverage probabilities. This is due to the fact that, as more BSs provide backhauling, the distance separating the UAV from its serving BS decreases, which increases the average received power and consequentially enhances the backhaul probability. As this probability increases, and since at low heights, the UE associates more to UAVs, the coverage probability increases. Thus, at low heights, all the terrestrial BSs must be able to provide backhaul capabilities to optimize the coverage. At higher heights, the backhaul and coverage probabilities increase with the increase of $\delta_{b}$ until a certain limit after which the two probabilities stabilize. Thus, adding more backhaul-enabled BSs will not bring any benefit and only a fraction is needed to achieve the optimal coverage. As the UAVs height further increases, the backhaul probability starts to deteriorate with the increase of $\delta_{b}$. This is caused by the increased interference levels added to the high path loss and LOS probability encountered on the backhaul links. The coverage probability, however, is not affected with the increase of $\delta_{b}$ since at high heights, the UE tends to associate more to the terrestrial network. The obtained results suggest that, a good coverage probability can be achieved when a fraction of the terrestrial BSs are backhaul-enabled. The fraction of backhaul-enabled BSs that optimizes the coverage of UEs is highly dependent on the height of the UAVs.

Finally, we compare the proposed backhaul aware transmission scheme with the instantaneous backhaul aware association scenario. In the latter scenario, when the UE-UAV access link quality is better than the UE-BS link but the UAV does not have a successful backhaul link, the UE is not considered in a service failure state. On the contrary, the UE changes its association to the best terrestrial BS alternative. The instantaneous backhaul aware association scenario models the case when UAVs are used to complement existing networks by providing additional capacity to hotspots areas and network coverage in hard to reach areas. In such scenario, when the UAV connection does not have any added value, the UE operates through the terrestrial connection. The backhaul aware transmission scheme is more applied in unexpected natural disasters. In such scenarios, the existing networks can be damaged. Thus, if the connection to UAV fails and the UE-BS access link quality is bad, the UE cannot connect to the network.

Fig.~\ref{fig:outage_no_outage} presents the comparison between backhaul aware transmission and the simulation results of the instantaneous backhaul aware transmission scheme as a function of the number and the height of UAVs. Note here that the analytical derivations can be conducted by thinning the BPP of UAVs according to the backhaul probability and then conducting the backhaul unaware association analysis. It is clearly seen that, as the number of UAVs increases, the coverage probability of backhaul aware instantaneous association exceeds that of backhaul aware transmission. The gap gets smaller with the increase of the UAVs height. Although the coverage probability is better for backhaul aware association, this scheme suffers from excessive handovers (i.e., ping pong effect) between UAV association and BS association. Such behavior highlights the validity of the proposed aware transmission scheme, especially for high UAVs altitudes.
\vspace{-0.8cm}
\subsection{Discussions and Design Insights}
	\vspace{-0.3cm}
In this section, we investigate the impacts of different system parameters to illustrate generic design guidelines and recommendations. Our results reveal that the height and number of assisting UAVs should be carefully adjusted to achieve optimal system performance. Furthermore, the characteristics of the terrestrial network must be taken into consideration while increasing the number of deployed UAVs. An interesting observation found in Fig.~\ref{fig:aware_ha_beta} is that an optimal height exists for the UAVs at which the coverage probability is maximized. This result is due to the applied probabilistic LOS/NLOS propagation model, implying that transmitting signals from higher altitude can benefit from increased possibility of LOS links. The optimal height is the point at which the UEs strike a balance between benefiting from the improved quality of received signals due to LOS and maintaining a good backhaul link between the UAVs and the terrestrial network. A trade-off can also be observed in Fig.~\ref{fig:coverage_aware_unaware100} regarding the number of deployed UAVs, which demonstrates that dense UAV deployment improves the UE performance if the UAVs are hovering in the low altitudes region and the backhaul links are guaranteed. For high heights, sparse UAV deployment is necessary to lower the interference.

Based on the interpretation of Fig.~\ref{fig:coverage_lambda}, we can notice that, for dense terrestrial BSs deployments, assisting the network with UAVs has a limited contribution on the coverage of UEs. The obtained results in Fig.~\ref{fig:misalignment_error} reveal the importance of considering the backhauling quality of the deployed UAVs as a key parameter in the design of the hybrid network. The advantages of assisting the terrestrial network with aerial BSs would not appear if these BSs are not supported with sufficient backhauling. In addition, adding UAVs with bad backhaul links would harm the UEs performance. Fig.~\ref{fig:coverage_fraction} offers the network operator the possibility of upgrading only a fraction of the existing BS network to provide backhauling for UAVs and optimized coverage for UEs; thereby reducing the needed infrastructure cost. Finally, supporting the UAVs with sufficient processing capabilities to be aware of the backhaul link quality would further improve the coverage.
\vspace{-0.6cm}
\section{Conclusion}\label{sec:conclusion}
\vspace{-0.2cm}
This paper uses stochastic geometry to assess the performance of mmWave backhauling for UAVs in a hybrid aerial-terrestrial cellular network considering key system parameters, e.g., UAVs height and number, and beamforming alignment error. We addressed both the backhaul unaware and the backhaul aware transmission scenarios. After characterizing the backhaul probability, the association and transmission probabilities and the distance distributions, we obtained an expression for the overall coverage probability and validated our results using Monte-Carlo simulations. Our results show that the quality of the backhaul link has a significant impact on the UE experience, and directly affects the deployment of UAVs for assisting terrestrial networks.
\vspace{-1.3cm}
\appendix
\vspace{-0.5cm}
\subsection{Proof of Lemma~\ref{lemma:coverage_association}}\label{app:association}
To derive the association probabilities, we follow a similar approach to~\cite{Wang}. Thus, the UE associates to a LOS UAV at distance $r$ if three conditions are satisfied:
\begin{itemize}
	\item The UAV at distance $r$ is a LOS UAV with a LOS probability $\kappa_{u_l}(r)$.
	\item The remaining $(N_u-1)$ UAVs are either LOS UAVs located further than $r$ or NLOS UAVs located further than $E_{ln}(r)$. The probabilities of these two events are calculated as $\int_{r}^{w_p}f_{W}(w)\kappa_{u_l}(w)\mathrm{d}w$ and $\int_{E_{ln}(r)}^{w_{p}}f_{W}(w)\kappa_{u_n}(w)\mathrm{d}w$, where $f_W(w)$ is the PDF of the distance from an arbitrary UAV to the UE. Since the $(N_u-1)$ UAVs are iid, the probability of this event is $\left(\int_{r}^{w_p}f_{W}(w)\kappa_{u_l}(w)\mathrm{d}w+\int_{E_{ln}(r)}^{w_p}f_{W}(w)\kappa_{u_n}(w)\mathrm{d}w\right)^{N_u-1}$.
	\item The nearest BS is located further than $E_{u_l}(r)$ with a probability of $\exp\left(-\pi\lambda_g E_{u_l}^2(r)\right)$.
\end{itemize}
Since these conditions are independent, the probability that the serving device at distance $r$ is a LOS UAV is $\kappa_{u_l}(r)\left(\int_{r}^{w_p}f_{W}(w)\kappa_{u_l}(w)\mathrm{d}w+\int_{E_{ln}(r)}^{w_p}f_{W}(w)\kappa_{u_n}(w)\mathrm{d}w\right)^{N_u-1}\exp\left(-\pi\lambda_g E_{u_l}^{2}(r)\right)$.
There are $N_u$ ways of choosing a UAV from iid $N_u$ UAVs. Thus, the probability that the UE associates to a LOS UAV at distance $r$ is calculated as:
\begin{equation}\small
N_u \kappa_{u_l}(r)\left(\int_{r}^{w_p}f_{W}(w)\kappa_{u_l}(w)\mathrm{d}w+\int_{E_{ln}(r)}^{w_p}f_{W}(w)\kappa_{u_n}(w)\mathrm{d}w\right)^{N_u-1}\exp\left(-\pi\lambda_g E_{u_l}^{2}(r)\right)
\vspace{-0.1cm}
\end{equation}
Finally, $A_{u_l}$ is obtained by integrating over the feasible region $h_u\leq r \leq w_p$. The association probabilities to a NLOS UAV and to a BS $A_{u_n}$ and $A_g$ can be obtained through similar derivation.
\vspace{-1.4cm}
\subsection{Proof of Lemma~\ref{lemma:UE_BS_laplace}}\label{app:laplace}
\vspace{-0.2cm}
The aggregate interference of the interfering BSs is given as $\hat{I}_{g}=\sum\limits_{x_{i}\in \phi_g \backslash x_g}P_{g}(s_{g,x_{i}}^2+h_{g}^2)^{-\eta_{g}/2}\Omega_{g,x_{i}}$. Thus, $\mathcal{L}_{\hat{I}_g}(s_{1})$ can be derived as
\begin{equation}\small
\begin{aligned}
&\mathcal{L}_{\hat{I}_g}(s_{1})=\mathbb{E}_{\phi_g}\left[e^{-\sum_{x_{i}\in \phi_g \backslash x_g}s_{1}P_{g}(s_{g,x_{i}}^2+h_{g}^2)^{-\frac{\eta_{g}}{2}}\Omega_{g,x_{i}}}\right]\stackrel{(a)}{=}\mathbb{E}_{\phi_g}\left[\prod_{x_{i}\in \phi_g \backslash x_g}^{}\mathbb{E}_{\Omega_{g,x_{i}}}\left[ e^{-s_{1}P_{g}(s_{g,x_{i}}^2+h_{g}^2)^{-\frac{\eta_{g}}{2}}\Omega_{g,x_{i}}}\right]\right]\\
&\stackrel{(b)}{=}\mathbb{E}_{\phi_g}\left[\prod_{x_{i}\in \phi_g \backslash x_g}^{}\frac{1}{1+s_{1}P_{g}\left(s_{g,x_i}^2+h_{g}^2\right)^{-\frac{\eta_{g}}{2}}}\right]\stackrel{(c)}{=}\exp\left[-2\pi\lambda_g\int_{x_g}^{\infty}\left(1-\frac{1}{1+s_{1}P_{g}(z^2+h_g^2)^{-\frac{\eta_{g}}{2}}}\right)z\mathrm{d}z\right],
\end{aligned}
\label{eq:prooflemma3}
\vspace{-0.2cm}
\end{equation}
where (a) follows from the iid distribution of the fading gain and its independence of $\phi_g$, (b) from the exponential distribution of $\Omega_{g,x_{i}}$ and (c) from the probability generation functional (PGFL) of $\phi_g$~\cite{Sultan} and from replacing $s_{g,x_{i}}$ with $z$. Using (p.$315$) from~\cite{integral} and the Euler's hyper geometric transformation, a closed form expression is found for (\ref{eq:prooflemma3}) as presented in (\ref{eq:Ligb}).

To derive the Laplace transform of the interference from all UAVs $I_u=\sum\limits_{y_{j}\in\Phi_{u}}I_{u,y_{j}}$, where $I_{u,y_{j}}$ is the interference from the $j$-th UAV, we use a similar approach to~\cite{Wang} as follows
\vspace{-0.5cm}
\begin{equation}\small
\begin{aligned}
\mathcal{L}_{I_u}(s_{1})=\mathbb{E}_{I_u}\left[\exp\left(-s_{1}\sum_{j=1}^{N_u}I_{u,y_{j}}\right)\right]\stackrel{(a)}{=}\prod_{j=1}^{N_u} \mathbb{E}_{I_{u,y_{j}}}\left[\exp\left(-s_{1}I_{u,y_{j}}\right)\right]=\left(\mathbb{E}_{I_{u,y_{j}}}\left[ \exp\left(-s_{1}I_{u,y_{j}}\right)\right]\right)^{N_u},
\end{aligned}
\label{eq:LIa_proof}
\end{equation}
where (a) follows from the iid distribution of the fading gains and from their independence of the interferers distances in the interference expression. Since the UE associates to a BS, for any of the $N_u$ interfering UAVs, it is either a LOS UAV located further than $E_{g_l}(x_{g})$ or a NLOS UAV further than $E_{g_n}(x_{g})$. The probabilities of these two events are  $\frac{\int_{E_{g_l}(x_{g})}^{w_p}f_{W}(w)\kappa_{u_l}(w)\mathrm{d}w}{\int_{E_{g_l}(x_{g})}^{w_p}f_{W}(w)\kappa_{u_l}(w)\mathrm{d}w+\int_{E_{g_n}(x_{g})}^{w_p}f_{W}(w)\kappa_{u_n}(w)\mathrm{d}w}$ and $\frac{\int_{E_{g_n}(x_{g})}^{w_p}f_{W}(w)\kappa_{u_n}(w)\mathrm{d}w}{\int_{E_{g_l}(x_{g})}^{w_p}f_{W}(w)\kappa_{u_l}(w)\mathrm{d}w+\int_{E_{g_n}(x_{g})}^{w_p}f_{W}(w)\kappa_{u_n}(w)\mathrm{d}w}$, respectively, where $f_W(w)$ is given in (\ref{eq:fw}).
\vspace{0.3cm}
Thus, $\mathbb{E}_{I_{u,y_{j}}}\left[\exp\left(-s_{1}I_{u,y_{j}}\right)\right]$ can be calculated as:
\begin{equation}\small
\begin{aligned}
&\mathbb{E}_{I_{u,y_{j}}}\left[ \exp\left(-s_{1}I_{u,y_{j}}\right)\right]
=\frac{\int_{E_{g_l}(x_{g})}^{w_p}f_{W}(w)\kappa_{u_l}(w)\mathrm{d}w}{\int_{E_{g_l}(x_{g})}^{w_p}f_{W}(w)\kappa_{u_l}(w)\mathrm{d}w+\int_{E_{g_n}(x_{g})}^{w_p}f_{W}(w)\kappa_{u_n}(w)\mathrm{d}w}\mathbb{E}_{v_{u_l,y_j}}\left[\exp\left(-s_{1}P_{u}\Omega_{u_l,y_j}v_{u_l,y_j}^{-\eta_{l}}\right)\right]\\
&+\frac{\int_{E_{g_n}(x_{g})}^{w_p}f_{W}(w)\kappa_{u_n}(w)\mathrm{d}w}{\int_{E_{g_l}(x_{g})}^{w_p}f_{W}(w)\kappa_{u_l}(w)\mathrm{d}w+\int_{E_{g_n}(x_{g})}^{w_p}f_{W}(w)\kappa_{u_n}(w)\mathrm{d}w}\mathbb{E}_{v_{u_n,y_j}}\left[\exp\left(-s_{1}P_{u}\Omega_{u_n,y_j}v_{u_n,y_j}^{-\eta_n}\right)\right].
\end{aligned}
\label{eq:Euai}
\vspace{-0.3cm}
\end{equation}
where $v_{u_l,y_j}$ and $v_{u_n,y_j}$ are the distances of the UE to the $j$-th interfering LOS UAV and the $j$-th interfering NLOS UAV. By omitting $y_j$ we obtain
\begin{equation}\small
\begin{aligned}
\mathbb{E}_{v_{u_l}}\left[ \exp\left(-s_{1}P_{u}\Omega_{u_l}v_{u_l}^{-\eta_l}\right)\right]
&\stackrel{(a)}{=}\mathbb{E}_{v_{u_l}}\left[\left(1+\frac{s_1P_{u}v_{u_l}^{-\eta_l}}{m_l}\right)^{-m_l}\right]
=\int_{E_{g_l}(x_{g})}^{w_{p}}f_{V_{u_l}}\left(v, E_{g_l}(x_g)\right)\left(1+\frac{s_1P_u {v}^{-\eta_l}}{m_l}\right)^{-m_l}\mathrm{d}v
\end{aligned}
\label{eq:Eual}
\vspace{-0.3cm}
\end{equation}
where (a) follows from the moment generating functional (MGF) of the fading gain $\Omega_{u_l}$ that follows a gamma distribution and from the iid distribution of the interferers distances. All LOS UAVs are further than $E_{g_l}(x_{g})$ from the UE. $f_{V_{u_l}}(v,x)$ is the distribution of the distance from the interfering LOS UAV to the reference UE given in Lemma~$4$ in~\cite{Wang} as $f_{V_{u_l}}(v,x)=\frac{f_{W}(v)\kappa_{u_l}(v)}{\int_{x}^{w_p}f_{W}(w)\kappa_{u_l}(w)\mathrm{d}w}$. Similarly, for a NLOS UAV:
\vspace{-0.2cm}
\begin{equation}\small
\mathbb{E}_{v_{u_n}}\left[\exp\left(-s_{1}P_{u}\Omega_{u_n} v_{u_n}^{-\eta_n}\right)\right]=\int_{E_{g_n}(x_{g})}^{w_{p}}f_{V_{u_n}}\left(v, E_{g_n}(x_g)\right)\left(1+\frac{s_1 P_{u}v^{-\eta_n}}{m_n}\right)^{-m_n}\mathrm{d}v.
\label{eq:Euan}
\end{equation}
$f_{V_{u_n}}\left(v, E_{g_n}(x_g)\right)$ is the distance distribution from the NLOS interfering UAV to the reference UE given as $f_{V_{u_n}}(v,x)=\frac{f_{W}(v)\kappa_{u_n}(v)}{\int_{x}^{w_p}f_{W}(w)\kappa_{u_n}(w)\mathrm{d}w}$. By plugging (\ref{eq:Eual}), (\ref{eq:Euan}), $f_{V_{u_l}}\left(v, E_{g_l}(x_g)\right)$ and $f_{V_{u_n}}\left(v, E_{g_n}(x_g)\right)$ in (\ref{eq:Euai}) and (\ref{eq:LIa_proof}), we can get the expression in (\ref{eq:Lia}).
\vspace{-0.4cm}
\subsection{Proof of Lemma~\ref{lemma:fxg}}\label{app:distance}
\vspace{-0.2cm}
The event $X_{g}>x_{g}$ is equivalent to the event $R_{g}>x_g$ given that the UE associates with a ground BS, where $R_g$ is the horizontal distance separating the nearest BS from the reference UE. The complementary cumulative distribution function (CCDF) of $X_g$ is given as
\begin{equation}\small
\begin{aligned}
&\bar{F}_{X_{g}}(x_g)=\mathbb{P}[R_{g}>x_g\mid s=g]=\frac{\mathbb{P}[R_g>x_g,s=g]}{\mathbb{P}[s=g]}\\
&\stackrel{(a)}{=}\frac{1}{A_{g}}\int_{x_{g}}^{\infty}f_{R_{g}}(r)\left(\int_{E_{g_l}(r)}^{w_p}f_{W}(w)\kappa_{u_l}(w)\mathrm{d}w+\int_{E_{g_n}(r)}^{w_p}f_{W}(w)\kappa_{u_n}(w)\mathrm{d}w\right)^{N_u}\mathrm{d}r
\end{aligned}
\label{eq:Fxgb}
\end{equation}
where (a) follows from the iid distribution of $N_{u}$ UAVs and the exclusion regions $E_{g_l}(\cdot)$ and $E_{g_n}(\cdot)$ given in (\ref{eq:Eg}) on the locations of the LOS and NLOS UAVs, respectively. $f_{R_g}(r)=2\pi\lambda_g r \exp\left(-\pi\lambda_g r^2\right)$ is the PDF of the horizontal distance separating the nearest BS from the reference UE~\cite{Moltchanov}. The CDF of $x_g$ is $F_{X_{g}}(x_g)=1-\bar{F}_{X_{g}}(x_g)$ and the PDF is given as
\begin{equation}\small
f_{X_{g}}(x_{g})=\frac{\mathrm{d}F_{X_{g}(x_{g})}}{\mathrm{d}x_{g}}=\frac{1}{A_{g}}f_{R_{g}}(x_{g})\left(\int_{E_{g_l}(x_g)}^{w_p}f_{W}(w)\kappa_{u_l}(w)\mathrm{d}w+\int_{E_{g_n}(x_g)}^{w_p}f_{W}(w)\kappa_{u_n}(w)\mathrm{d}w\right)^{N_u},
\vspace{-0.2cm}
\end{equation}
where $f_{W}(\cdot)$ and $\kappa_{u_l}(\cdot)$ are given in (\ref{eq:fw}) and (\ref{eq:LOS_probability}), respectively. Finally the distance distribution of $x_g$ can be obtained as in (\ref{eq:fxg}).
\vspace{-0.6cm}
\subsection{Proof of Lemma~\ref{lemma:backhaul_association}}\label{app:LOS_association}
\vspace{-0.1cm}
We start by providing the distributions of distances from the reference UAV to the nearest LOS BS in $\phi_{g_l}$ and the nearest NLOS BS in $\phi_{g_n}$. Denoting by $s_{b_l}$ the horizontal distance from the reference UAV to the nearest LOS BS in $\phi_{g_l}$, the CCDF of $s_{b_l}$ can be calculated as
\begin{equation}\small
\begin{aligned}
\bar{F}_{s_{b_l}}(s)&=\mathbb{P}(s_{b_l} > s)= \mathbb{P}(\mathrm{No \; backhaul-enabled \; LOS \; BS \; closer \; than\;} s)=e^{-2\pi\lambda_b\int_{0}^{s}\kappa_{b_l}(r)r\mathrm{d}r}.
\end{aligned}
\end{equation}
Therefore, the CDF is $1-e^{-2\pi\lambda_b\int_{0}^{s}\kappa_{b_l}(r)r\mathrm{d}r}$ and the PDF can be found as
\begin{equation}\small
f_{s_{b_l}}(s)=	2\pi\lambda_b s \kappa_{b_l}(s) e^{-2\pi\lambda_b\int_{0}^{s}\kappa_{b_l}(r)r\mathrm{d}r}
\label{eq:fsl}
\vspace{-0.1cm}
\end{equation}
Similarly, the CDF and the PDF of the horizontal distance $s_{b_n}$ from the UAV to the nearest NLOS BS from $\phi_{g_n}$ are given as $\bar{F}_{s_{b_n}}(s)=e^{-2\pi\lambda_b\int_{0}^{s}\kappa_{b_n}(r)r\mathrm{d}r}$ and $f_{s_{b_n}}(s)= 2\pi\lambda_b s \kappa_{b_n}(s) e^{-2\pi\lambda_b\int_{0}^{s}\kappa_{b_n}(r)r\mathrm{d}r}$, where $\kappa_{b_n}=1-\kappa_{b_l}$ and $\kappa_{b_l}$ is given in (\ref{eq:LOS_probability2}).

The reference UAV connects with a backhaul-enabled LOS BS in $\phi_{g_l}$ to get backhaul support if the nearest LOS BS has smaller path-loss than that of the nearest NLOS BS in $\phi_{g_n}$. Thus, the probability $A_{b_l}$ that the reference UAV is associated with a LOS BS can be derived as follows
\begin{equation}\small
\begin{aligned}
&A_{b_l}=\mathbb{P}\left[C_{l}z_{b_l}^{-\eta_{l}}>C_{n}z_{b_n}^{-\eta_{n}}\right]=\int_{0}^{\infty}\mathbb{P}\left[C_{l}(s_{b_l}^2+\Delta_h^2)^{-\frac{\eta_{l}}{2}}> C_{n}\left(s_{b_n}^2+\Delta_h^2\right)^{-\frac{\eta_{n}}{2}}\right]f_{s_{b_l}}(s_{b_l})\mathrm{d}s_{b_l}\\
&=\int_{0}^{\infty}\mathbb{P}\left[s_{b_n}>E_{b_l}(s_{b_l})\right]f_{s_{b_l}}(s_{b_l})\mathrm{d}s_{b_l}=\int_{0}^{\infty}\bar{F}_{s_{b_n}}(E_{b_l}(s_{b_l}))f_{s_{b_l}}(s_{b_l})\mathrm{d}s_{b_l}\\
\end{aligned}
\vspace{-0.1cm}
\end{equation}
where $z_{b_l}$, $z_{b_n}$, $s_{b_l}$, $s_{b_n}$ are the actual and horizontal distances from the UAV to the nearest LOS and NLOS BSs and $E_{b_l}(s)=\sqrt{\left(\frac{C_n}{C_l}\right)^{\frac{2}{\eta_n}}({s}^2+\Delta_h^2)^{\frac{\eta_l}{\eta_n}}-\Delta_h^2}$. By replacing $s_{b_l}$ with $x$, $A_{b_l}$ is obtained as in (\ref{eq:Aln}). $A_{b_n}$  is obtained with the same procedure.
\vspace{-0.6cm}
\subsection{Proof of Lemma~\ref{lemma:backhaul_distance}}\label{app:backhaul_distance}
\vspace{-0.2cm}
Denote $X_{b_l}$ as the horizontal distance between the reference UAV and its serving LOS BS. Since the event $X_{b_l}>x$ is the event of $s_{b_l}>x$ given that the reference UAV connects to a LOS BS to get backhaul support, the probability of $X_{b_l}>x$ can be given as
\begin{equation}\small
\mathbb{P}[X_{b_l}>x]=\mathbb{P}[s_{b_l}>x\mid s=b_l]=\frac{\mathbb{P}[s_{b_l}>x,s=b_l]}{\mathbb{P}[s=b_l]}
\label{eq:cond_backhaul_distance_proof1}
\end{equation}
where $\mathbb{P}[s=b_l]=A_{b_l}$ is the probability that the UE associates to a LOS BS. The joint probability of $s_{b_l}>x$ and $s=b_l$ is
\begin{equation}\small
\begin{aligned}
&\mathbb{P}[s_{b_l}>x,s=b_l]=\mathbb{P}\left[s_{b_l}>x,C_l(s_{b_l}^2+\Delta_h^2)^{-\frac{\eta_l}{2}}>C_n(s_{b_n}^2+\Delta_h^2)^{-\frac{\eta_n}{2}}\right]\\
&=\int_{x}^{\infty}\mathbb{P}[s_{b_n}>E_{b_l}(s_{b_l})]f_{s_{b_l}}(s_{b_l})\mathrm{d}s_{b_l}=\int_{x}^{\infty}\bar{F}_{s_{b_n}}(E_{b_l}(s_{b_l}))f_{s_{b_l}}(s_{b_l})\mathrm{d}s_{b_l}.
\end{aligned}
\label{eq:cond_backhaul_distance_proof2}
\end{equation}
Plugging (\ref{eq:cond_backhaul_distance_proof2}) in (\ref{eq:cond_backhaul_distance_proof1}) gives
$\mathbb{P}[X_{b_l}>x]=\frac{1}{A_{b_l}}\int_{x}^{\infty}\bar{F}_{s_{b_n}}(E_{b_l}(s_{b_l}))f_{s_{b_l}}(s_{b_l})\mathrm{d}s_{b_l}.$ The CDF of $X_{b_l}$ is $F_{X_{b_l}}(x)=1-\mathbb{P}[X_{b_l}>x]$ and the PDF is given as
\begin{equation}\small
f_{X_{b_l}}(x)=\frac{\mathrm{d}F_{X_{b_l}}(x)}{\mathrm{d}x}=\frac{1}{A_{b_l}}\bar{F}_{s_{b_n}}(E_{b_l}(x))f_{s_{b_l}}(x)\label{eq:cond_backhaul_distance_proof3}
\end{equation}
By plugging $\bar{F}_{s_{b_n}}(\cdot)$ and $f_{s_{b_l}}(\cdot)$ in (\ref{eq:cond_backhaul_distance_proof3}), the PDF of the horizontal distance to the serving LOS BS is given as in (\ref{eq:fsLN}) in Lemma~\ref{lemma:backhaul_distance}. Following the same procedure, the PDF of the horizontal distance between the UAV and its serving NLOS BS is determined and presented in (\ref{eq:fsLN}).
\vspace{-0.7cm}
\subsection{Proof of Theorem~\ref{theorem:backhaul_probability}}\label{app:backhaul_probability}
Given that the reference UAV is connected to a BS in $\phi_{g_l}$, and that the desired link has a length of $x_{b_{l}}=x$, by Slivnyak’s Theorem, the conditional backhaul probability is given as
\begin{equation}\small
S_l(\tau_b)=\mathbb{P}[\mathrm{SINR}>\tau_b]=\int_{0}^{\infty}\mathbb{P}\left[\Omega_{b_l,0}>\frac{\tau_b(\sigma_{b}^{2}+I_b)}{P_{b}G_{b,0}C_l(x^2+\Delta_h^2)^{-\frac{\eta_l}{2}}}\right]f_{X_{b_l}}(x)\mathrm{d}x
\label{eq:theorem1_proof_1}
\end{equation}
where $I_b=I_{b_l}+I_{b_n}$ is the total interference from the LOS and the NLOS backhaul-enabled BSs, $\sigma_{b}^{2}$ is the noise power and $G_{b,0}$ is the maximum antennas gain. Noting that $\Omega_{b_l,0}$ is a normalized gamma random variable with parameter $m_l$, we have the following approximation
	\begin{equation}\small
	\begin{aligned}
	&\mathbb{P}\left[\Omega_{b_l,0}>\frac{\tau_b(\sigma_{b}^{2}+I_b)}{P_{b}G_{b,0}C_l(x^2+\Delta_h^2)^{-\frac{\eta_l}{2}}}\right]=1-\mathbb{P}\left[\Omega_{b_l,0}<\frac{\tau_{b}\left(\sigma_{b}^{2}+I_{b}\right)}{P_{b}G_{b,0}C_{l}\left(x^2+\Delta h^{2}\right)^{-\frac{\eta_{l}}{2}}}\right]\\
	&\stackrel{(a)}{\approx}1-\mathbb{E}_{\phi_g}\left[\left(1-e^{-\frac{\gamma_l \tau_b (\sigma_{b}^2+I_b)}{P_{b}G_{b,0}C_l (x^2+\Delta_h^2)^{-\frac{\eta_l}{2}}}}\right)^{m_{l}}\right]
	\stackrel{(b)}{\approx}\sum_{q=1}^{m_l}(-1)^{q+1}{m_l\choose q}\mathbb{E}_{\phi_g}\left[e^{-\frac{q\gamma_l \tau_b (\sigma_{b}^{2}+I_b)(x^2+\Delta_h^2)^{\frac{\eta_l}{2}}}{P_{b}G_{b,0}C_l}}\right]\\
	&\stackrel{(c)}{=}\sum_{q=1}^{m_l}(-1)^{q+1}{m_l\choose q}e^{-q\mu_l\tau_b\sigma_{b}^2}\mathcal{L}_{I_{b_l}}\left(q\mu_l\tau_b\right)\mathcal{L}_{I_{b_n}}\left(q\mu_l\tau_b\right)\\
	\end{aligned}
	\label{eq:theorem1_proof_2}
	\vspace{-0.1cm}
	\end{equation}
	where $\gamma_{l}=m_{l}(m_{l}!)^{-\frac{1}{m_{l}}}$, $(a)$ follows from the upper bound of the CDF of a gamma random variable deduced from Theorem~$1$ of~\cite{Alzer}, $(b)$ follows from the binomial theorem and the assumption that $m_{l}$ is an integer. (c) follows from denoting $\mu_{l}=\frac{\gamma_{l} (x^2+\Delta_h^2)^{\frac{\eta_{l}}{2}}}{P_{b}G_{b,0}C_{l}}$, and from denoting the Laplace functionals of the interference of the LOS and NLOS BSs as $\mathcal{L}_{I_{b_{l}}}(s)=\mathbb{E}[e^{-sI_{b_l}}]$ and $\mathcal{L}_{I_{b_{n}}}(s)=\mathbb{E}[e^{-sI_{b_{n}}}]$, respectively, and the fact that $\phi_{g_{l}}$ and $\phi_{g_{n}}$ are independent.

Given that the backhaul link is LOS of length $x_{b_l}$, based on the association rule, all the LOS interfering BSs are further than $x_{b_{l}}$, and all NLOS BSs are further than $E_{b_{l}}(x_{b_{l}})$. The Laplace transform $\mathcal{L}_{I_{b_l}}(t)$ is derived  as
\begin{equation}\small
\begin{aligned}
&\mathcal{L}_{I_{b_{l}}}(t)=\mathbb{E}\left[e^{-tI_{b_{l}}}\right]=\mathbb{E}\left[e^{-t\sum_{x_{i}\in \phi_{g_{l}}\backslash x_{b_{l}}}^{}P_{b}\Omega_{b_{l},x_{i}}G_{b,I}C_l\left(s_{b_{l},x_{i}}^2+\Delta_h^2\right)^{-\frac{\eta_{l}}{2}}}\right]\\
&\stackrel{(a)}{=}\exp\left[-2\pi\int_{x_{b_{l}}}^{\infty}\left(1-\mathbb{E}_{\Omega_{b_l},G_{b}}\left[e^{-tP_{b}\Omega_{l}G_{b}C_l\left(r^2+\Delta_h^2\right)^{-\frac{\eta_l}{2}}}\right]\right)\right. \lambda_b \kappa_{b_l}(r)r\mathrm{d}r\Bigg].
\end{aligned}\label{eq:theorem1_proof_3}
\end{equation}
where (a) follows from the PGFL of a PPP~\cite{Sultan} and from omitting the indices $x_{i}$ and $I$. The expectation term in (\ref{eq:theorem1_proof_3}) is given as
\begin{equation}\small
\begin{aligned}
&\mathbb{E}_{\Omega_{b_l},G_{b}}\left[e^{-tP_{b}\Omega_{b_l}G_{b}C_l\left(r^2+\Delta_h^2\right)^{-\frac{\eta_l}{2}}}\right]\\&\stackrel{(a)}{=}\sum_{k=1}^{4}p_{k}\mathbb{E}_{\Omega_{b_l}}\left[e^{-tP_{b}\Omega_{b_l}G_{k}C_l\left(r^2+\Delta_h^2\right)^{-\frac{\eta_l}{2}}}\right]\stackrel{(b)}{=}\sum_{k=1}^{4}p_{k}\left(1+\frac{tP_{b}G_{k}C_l\left(r^2+\Delta_h^2\right)^{-\frac{\eta_l}{2}}}{m_{l}}\right)^{-m_l}
\end{aligned}
\vspace{-0.1cm}
\end{equation}
where (a) follows from the discrete directivity gain in the interference channels, and (b) follows from computing the Laplace transform of $\Omega_{b_l}$ which follows a gamma distribution. Similarly, for the NLOS interfering links the Laplace transform $\mathcal{L}_{I_{b_n}}(t)$ is given as
\vspace{-0.1cm}
\begin{equation}\small
\mathcal{L}_{I_{b_n}}(t)=\exp\left(-2\pi\int_{E_{b_l}(x_{b_l})}^{\infty}\left(1-\mathbb{E}_{\Omega_{b_n},G_{b}}\left[e^{-tP_{b}\Omega_{b_n}G_{b}C_n\left(r^2+\Delta_h^2\right)^{-\frac{\eta_n}{2}}}\right]\right)\right. \lambda_b\kappa_{b_n}(r)r\mathrm{d}r\Bigg)
\label{eq:theorem1_proof_4}
\vspace{-0.5cm}
\end{equation}
where
\begin{equation}\small
\mathbb{E}_{\Omega_{b_n},G_{b}}\left[e^{-tP_{b}\Omega_{b_n}G_{b}C_n\left(r^2+\Delta_h^2\right)^{-\frac{\eta_n}{2}}}\right]=\sum_{k=1}^{4}p_{k}\left(1+\frac{tP_{b}G_{k}C_n\left(r^2+\Delta_h^2\right)^{-\frac{\eta_n}{2}}}{m_{n}}\right)^{-m_n}
\vspace{-0.3cm}
\end{equation}
Finally, by plugging (\ref{eq:theorem1_proof_2}), (\ref{eq:theorem1_proof_3}) and (\ref{eq:theorem1_proof_4}) in (\ref{eq:theorem1_proof_1}) and by replacing $\mu_l$ by $\frac{\gamma_l \left(x^2+\Delta_h^2\right)^{\frac{\eta_l}{2}}}{P_{b}G_{b,0}C_l}$ and $x_{b_{l}}$ by $x$, we get $S_{l}(\tau_b)$ as in (\ref{eq:SLN}) given that the UAV is served by a LOS BS. The same procedure is followed to obtain $S_{n}(\tau_b)$. Here, all NLOS interferers are farther than $x_{b_n}$ and all LOS interferers are farther than $E_{b_n}(x_{b_n})$. The detailed proof is omitted here and $S_{n}(\tau_b)$ is given in (\ref{eq:SLN}). Finally, by the law of total probability, the backhaul probability can be derived as in~(\ref{eq:Stb}).
\vspace{-0.5cm}
\subsection{Proof of Lemma~\ref{lemma:pcova}}\label{app:pcova}
The conditional coverage probability $P_{cov,u_l}$ is defined as the coverage probability given that the UE associates to a LOS UAV at distance $y_{u_l}$ and is given as
\begin{equation}\small
\begin{aligned}
&P_{cov,u_l}=\mathbb{P}\left[\mathrm{SIR}\geq \tau_a,\mathrm{SINR}\geq\tau_b\right | s=u_l]\stackrel{(a)}{\approx}\mathbb{E}_{Y_{u_l}}\left[\mathbb{E}_{I_{agg,u_l}}\left[\mathbb{P}\left(\Omega_{u_l,0}\geq\frac{\tau_a I_{agg,u_l}}{P_{u}y_{u_l}^{-\eta_l}}\right)\right]\right]\times S(\tau_b) \\
&\stackrel{(b)}{=}\mathbb{E}_{Y_{u_l}}\left[\mathbb{E}_{I_{agg,u_l}}\left[\sum_{k=0}^{m_l-1}\frac{(\hat{I}_{u_l}+I_{g_l})^k}{k!}\left(\frac{m_{l}\tau_a y_{u_l}^{\eta_l}}{P_{u}}\right)^k\exp\left(-\left(\frac{m_l \tau_a y_{u_l}^{\eta_l}}{P_{u}}\right)\left(\hat{I}_{u_l}+I_{g_l}\right)\right)\right]\right] \times S(\tau_b)\\
&\stackrel{(c)}{=}S(\tau_b) \int_{h_u}^{w_p}\sum_{k=0}^{m_l-1}\frac{(-s_{2_l})^k}{k!}\left[\frac{\partial^k}{\partial s_{2_l}^{k}}\mathcal{L}_{I_g}(s_{2_l})\mathcal{L}_{\hat{I}_{u_l}}(s_{2_l})\right]f_{Y_{u_l}}(y_{u_l})\mathrm{d}y_{u_l},
\end{aligned}
\end{equation}
where $s_{2_l}=\frac{m_l\tau_a y_{u_l}^{\eta_l}}{P_{u}}$ and $S(\tau_b)$ is the backhaul probability given in (\ref{eq:Stb}), (a) follows from the independence assumption, (b) from the CCDF of the UE-UAV channel power gain $\Omega_{u_l,0}$ and from the expression of the interference $I_{agg,u_l}=\hat{I}_{u_l}+I_{g_l}$. Finally, (c) follows from the independence of $\hat{I}_{u_l}$ and $I_{g_l}$, where $\hat{I}_{u_l}$ and $I_{g_l}$ are the interference expressions from all the UAVs except the serving UAV and from all the BSs. $P_{cov,u_n}$ can be derived following a similar approach.
\vspace{-0.5cm}
\bibliography{IEEEabrv,references}
\bibliographystyle{ieeetr}
\end{document}

%% file: Figures/defs_tikzpgf.tex
\usepackage{tikz}
\usepackage{pgfplotstable}
\usepackage{rotate}

\definecolor{rubblue}{cmyk}{1,0.5,0,0.6}
\definecolor{rubgreen}{cmyk}{0.5,0,1,0}
\definecolor{rubgray}{cmyk}{0.03,0.03,0.03,0.1}

\usepgfplotslibrary{units}

\usetikzlibrary{%
patterns,%
calc,%
fit,%
arrows,%
plotmarks,%
shadows,%
chains,%
shapes%
}

\tikzset{>=latex'} 
\tikzstyle{every picture}+=[remember picture] 
\pgfdeclarelayer{background}
\pgfdeclarelayer{foreground}
\pgfsetlayers{background,main,foreground}

\tikzstyle{blueblock}=[draw=rubblue, rectangle, thick, drop shadow, minimum width=20mm, minimum height=8mm,fill=rubblue!20, text width=20mm, text centered]
\tikzstyle{bluebox}=[draw=rubblue, rectangle, thick, drop shadow, minimum width=8mm, minimum height=8mm,fill=rubblue!20, text width=8mm, text centered]
\tikzstyle{greenblock}=[draw=rubgreen, rectangle, thick, drop shadow, minimum width=20mm, minimum height=8mm,fill=rubgreen!20, text width=20mm, text centered]
\tikzstyle{dot} = [draw, circle, minimum size=0.2pt,scale=0.3,fill=black,black]
\tikzstyle{smalldot} = [draw, circle, minimum size=0.1pt,scale=0.2,fill=black,black]
\tikzstyle{reddot}  =[draw,circle,minimum size=0.2pt,scale=0.8,fill=red,thin]
\tikzstyle{greendot}  =[draw,circle,minimum size=0.2pt,scale=0.8,fill=Green,thin]
\tikzstyle{bluedot}  =[draw,circle,minimum size=0.2pt,scale=0.8,fill=blue,thin]
\tikzstyle{whitedot}=[draw,circle,minimum size=0.2pt,scale=0.8,fill=white,thin]
\tikzstyle{blackdot} = [draw, circle, minimum size=0.2pt,scale=0.7,fill=black,black]
\tikzstyle{sum} = [drop shadow, draw=rubblue, thick, fill=rubblue!20, circle]
\tikzstyle{relay} = [blueblock, minimum width=5mm, minimum height=20mm, text width=5mm, rounded corners=2pt]
\tikzstyle{relay2} = [blueblock, minimum width=5mm, minimum height=15mm, text width=5mm, rounded corners=2pt]
\tikzstyle{relay3} = [blueblock, minimum width=5mm, minimum height=25mm, text width=5mm, rounded corners=2pt]
\tikzstyle{relay4} = [blueblock, minimum width=5mm, minimum height=10mm, text width=5mm, rounded corners=2pt]
\tikzstyle{relay5} = [blueblock, minimum width=5mm, minimum height=50mm, text width=5mm, rounded corners=2pt]
\tikzstyle{relay6} = [blueblock, minimum width=5mm, minimum height=5mm, text width=5mm, rounded corners=2pt]
\tikzstyle{circgreen} = [draw, circle, inner sep=2pt, fill=rubgreen, drop shadow, thick]
\tikzstyle{circwhite} = [draw, circle, inner sep=2pt, fill=white, drop shadow, thick]
\tikzstyle{circdashed} = [draw, dashed, circle, inner sep=2pt, fill=rubgray, drop shadow, thick]
\tikzstyle{vertbox} = [rectangle, draw=rubblue, thick, rotate=90, text centered, minimum width=16.5mm, minimum height=8mm, text width=16.5mm, inner sep=0pt, fill=rubblue!20, drop shadow]
\tikzstyle{vertboxb} = [rectangle, draw=rubblue, thick, rotate=90, text centered, minimum width=16.5mm, minimum height=8mm, text width=16.5mm, fill=rubblue!20, drop shadow]
\tikzstyle{vertboxshort} = [rectangle, draw=rubblue, thick, rotate=90, text centered, minimum width=10mm, minimum height=8mm, text width=10mm, inner sep=0pt, fill=rubblue!20, drop shadow]
\tikzstyle{smalldotgreen} = [draw=rubgreen, circle, minimum size=0.2pt,scale=0.8,fill=rubgreen!20]
\tikzstyle{antenna} = [regular polygon, regular polygon sides=3, draw, shape border rotate=180, minimum size=0.2pt, scale=0.3]

\tikzstyle{poly} = [regular polygon, regular polygon sides=6, shape aspect=0.5, minimum width=1.5cm, minimum height=0.35cm, draw, dashed]

\definecolor{cff9e00}{RGB}{255,158,0}
\definecolor{c4fff00}{RGB}{79,255,0}
\definecolor{cff0012}{RGB}{255,0,18}
\definecolor{c00c5ff}{RGB}{0,197,255}
\definecolor{c046f00}{RGB}{4,111,0}
\definecolor{c004b9d}{RGB}{0,75,157}

\newlength{\mylen}

%% file: Figures/Fig5a.tex
%
%
\definecolor{mycolor1}{rgb}{0.00000,0.44700,0.74100}%
\definecolor{mycolor2}{rgb}{0.85000,0.32500,0.09800}%
\definecolor{mycolor3}{rgb}{0.92900,0.69400,0.12500}%
\definecolor{mycolor4}{rgb}{0.49400,0.18400,0.55600}%
\definecolor{mycolor5}{rgb}{0.74902,0.00000,0.74902}%
\definecolor{mycolor6}{rgb}{0.00000,0.44706,0.74118}%
\definecolor{mycolor7}{rgb}{0.63500,0.07800,0.18400}

\begin{axis}[%
width=8.73in,
height=7.323in,
at={(1.464in,1.169in)},
scale only axis,
xmin=0,
xmax=500,
xtick={0,100,...,500},
xlabel style={font=\Huge, font=\color{white!15!black}},
xlabel={\Huge{$\text{Height of UAVs} \;\; h_u \; \text{(m)}$}},
ymin=0,
ymax=1,
ytick={0,0.2,...,1},
ylabel style={font=\Huge, font=\color{white!15!black}},
ylabel={\Huge{$\text{Backhaul probability} \;\; S(\tau{}_b)$}},
axis background/.style={fill=white},
xmajorgrids,
ymajorgrids,
legend style={at={(0.02,0.02)}, anchor=south west, legend cell align=left, align=left, draw=white!15!black}
]
\addplot [color=mycolor1, line width=3.0pt]
  table[row sep=crcr]{%
30	0.8636274\\
50	0.9938362\\
70	0.9943466\\
90	0.993554\\
110	0.992541\\
130	0.991306\\
150	0.989856\\
170	0.988165\\
190	0.986193\\
210	0.983895\\
230	0.981213\\
250	0.978085\\
270	0.974436\\
290	0.970182\\
310	0.965227\\
330	0.959466\\
350	0.952788\\
370	0.945071\\
390	0.936192\\
410	0.926028\\
430	0.914459\\
450	0.901373\\
470	0.886671\\
490	0.87027\\
};
\addlegendentry{\Huge{$\tau_b = 5$ dB}}

\addplot [color=mycolor2, line width=3.0pt]
  table[row sep=crcr]{%
30	0.807285\\
50	0.9669238\\
70	0.9775598\\
90	0.973446\\
110	0.970203\\
130	0.965348\\
150	0.95913\\
170	0.951448\\
190	0.942069\\
210	0.930682\\
230	0.91693\\
250	0.900433\\
270	0.880812\\
290	0.857727\\
310	0.830905\\
330	0.800184\\
350	0.765536\\
370	0.727098\\
390	0.685176\\
410	0.640248\\
430	0.59294\\
450	0.544003\\
470	0.494263\\
490	0.444582\\
};
\addlegendentry{\Huge{$\tau_b = 10$ dB}}

\addplot [color=mycolor3, line width=3.0pt]
  table[row sep=crcr]{%
30	0.6997438\\
50	0.8722438\\
70	0.8940008\\
90	0.884197\\
110	0.871733\\
130	0.852073\\
150	0.826083\\
170	0.7935\\
190	0.753825\\
210	0.706745\\
230	0.652428\\
250	0.591701\\
270	0.526101\\
290	0.457766\\
310	0.389209\\
330	0.322992\\
350	0.261409\\
370	0.206232\\
390	0.158558\\
410	0.118796\\
430	0.0867424\\
450	0.0617393\\
470	0.0428436\\
490	0.0289935\\
};
\addlegendentry{\Huge{$\tau_b = 15$ dB}}

\addplot [color=mycolor4, line width=3.0pt]
  table[row sep=crcr]{%
30	0.5312664\\
50	0.6675142\\
70	0.6710618\\
90	0.643667\\
110	0.602995\\
130	0.547808\\
150	0.480797\\
170	0.405049\\
190	0.325327\\
210	0.24766\\
230	0.177923\\
250	0.120303\\
270	0.0764574\\
290	0.0456605\\
310	0.0256305\\
330	0.0135289\\
350	0.0067175\\
370	0.00313776\\
390	0.00137843\\
410	0.000569197\\
430	0.000220752\\
450	8.03341e-05\\
470	2.74027e-05\\
490	8.75208e-06\\
};
\addlegendentry{\Huge{$\tau_b = 20$ dB} }

\addplot [color=black, line width=3.0pt, draw=none, mark repeat={2}, mark size=9.0pt, mark=x, mark options={solid, black}]
  table[row sep=crcr]{%
30	0.8808\\
50	0.9944\\
70	0.9944\\
90	0.9927\\
110	0.9924\\
130	0.9921\\
150	0.9882\\
170	0.9868\\
190	0.985\\
210	0.9823\\
230	0.9807\\
250	0.9755\\
270	0.9733\\
290	0.9697\\
310	0.9684\\
330	0.9549\\
350	0.9516\\
370	0.947\\
390	0.9341\\
410	0.9241\\
430	0.9153\\
450	0.9033\\
470	0.8877\\
490	0.8701\\
};

\addplot [color=black, line width=2.0pt, draw=none, mark repeat={2}, mark size=9.0pt, mark=x, mark options={solid, black}]
  table[row sep=crcr]{%
30	0.849\\
50	0.9693\\
70	0.9786\\
90	0.9722\\
110	0.9699\\
130	0.9648\\
150	0.9543\\
170	0.9494\\
190	0.9374\\
210	0.9251\\
230	0.9148\\
250	0.8963\\
270	0.876\\
290	0.8568\\
310	0.8291\\
330	0.7924\\
350	0.7579\\
370	0.7139\\
390	0.683\\
410	0.6451\\
430	0.5798\\
450	0.5419\\
470	0.479\\
490	0.4235\\
};

\addplot [color=black, line width=3.0pt, draw=none, mark repeat={2}, mark size=9.0pt, mark=x, mark options={solid, black}]
  table[row sep=crcr]{%
30	0.7575\\
50	0.8791\\
70	0.8964\\
90	0.8868\\
110	0.8727\\
130	0.8464\\
150	0.8169\\
170	0.7852\\
190	0.7383\\
210	0.6955\\
230	0.6411\\
250	0.5763\\
270	0.4975\\
290	0.4317\\
310	0.3629\\
330	0.2967\\
350	0.2294\\
370	0.1841\\
390	0.1349\\
410	0.0968\\
430	0.065\\
450	0.047\\
470	0.0301\\
490	0.0173\\
};

\addplot [color=black, line width=3.0pt, draw=none, mark repeat={2}, mark size=9.0pt, mark=x, mark options={solid, black}]
  table[row sep=crcr]{%
30	0.5854\\
50	0.6769\\
70	0.6731\\
90	0.639\\
110	0.6001\\
130	0.5352\\
150	0.4554\\
170	0.3898\\
190	0.3013\\
210	0.2311\\
230	0.154\\
250	0.0994\\
270	0.0566\\
290	0.0312\\
310	0.0144\\
330	0.0063\\
350	0.003\\
370	0.0009\\
390	0.0004\\
410	0.0003\\
430	0.0001\\
450	0\\
470	0\\
490	0\\
};
\addlegendentry{\Huge{Simulation}}

\end{axis}

%% file: Figures/Fig5b.tex
%
%
\definecolor{mycolor1}{rgb}{0.00000,0.44700,0.74100}%
\definecolor{mycolor2}{rgb}{0.85000,0.32500,0.09800}%
\definecolor{mycolor3}{rgb}{0.92900,0.69400,0.12500}%
\definecolor{mycolor4}{rgb}{0.49400,0.18400,0.55600}%
\definecolor{mycolor5}{rgb}{0.74902,0.00000,0.74902}%
\definecolor{mycolor6}{rgb}{0.00000,0.44706,0.74118}%
\definecolor{mycolor7}{rgb}{0.63500,0.07800,0.18400}

\begin{axis}[%
width=8.73in,
height=7.323in,
at={(1.464in,1.169in)},
scale only axis,
xmin=0,
xmax=900,
xtick={0,100,...,900},
xlabel style={font=\color{white!15!black}},
xlabel={\Huge{$\text{Height of UAVs} \;\; h_u \; \text{(m)}$}},
ymin=0,
ymax=0.8,
ytick={0,0.1,...,0.8},
ylabel style={font=\color{white!15!black}},
ylabel={\Huge{Association probability}},
axis background/.style={fill=white},
xmajorgrids,
ymajorgrids,
legend style={at={(0.715,0.715)}, anchor=south west, legend cell align=left, align=left, draw=white!15!black}
]
\addplot [color=mycolor1, line width=3.0pt]
  table[row sep=crcr]{%
30	0.1381255\\
50	0.2724548\\
70	0.3580463\\
90	0.44122651\\
110	0.51643824\\
130	0.5795704309\\
150	0.6289516433\\
170	0.66456730426\\
190	0.687384597234\\
210	0.6988590979123\\
230	0.7005860144901\\
250	0.69408000192587\\
270	0.680662000228746\\
290	0.66142500002417\\
310	0.637251000002263\\
330	0.608873000000187\\
350	0.576924000000014\\
370	0.542009000000001\\
390	0.50473\\
410	0.465724\\
430	0.425657\\
450	0.385221\\
470	0.345102\\
490	0.30596\\
510	0.268397\\
530	0.232928\\
550	0.199966\\
570	0.169807\\
590	0.142632\\
610	0.118508\\
630	0.097402\\
650	0.0791962\\
670	0.0637078\\
690	0.0507072\\
710	0.0399369\\
730	0.0311276\\
750	0.0240116\\
770	0.0183329\\
790	0.0138551\\
810	0.0103651\\
830	0.00767625\\
850	0.00562787\\
870	0.00408478\\
890	0.00293511\\
910	0.00208789\\
930	0.00147032\\
950	0.00102501\\
970	0.000707352\\
990	0.000483194\\
};
\addlegendentry{\Huge{$\tilde{A}_{u_{l}} + \tilde{A}_{u_{n}}$} }

\addplot [color=mycolor2, line width=3.0pt]
  table[row sep=crcr]{%
30	0.784324\\
50	0.715428\\
70	0.632169\\
90	0.546737\\
110	0.467701\\
130	0.399625\\
150	0.344247\\
170	0.30152\\
190	0.270346\\
210	0.249091\\
230	0.235944\\
250	0.229206\\
270	0.227233\\
290	0.228863\\
310	0.233063\\
330	0.239081\\
350	0.246384\\
370	0.254561\\
390	0.263359\\
410	0.272589\\
430	0.282125\\
450	0.291879\\
470	0.301786\\
490	0.311797\\
510	0.321894\\
530	0.33203\\
550	0.342197\\
570	0.352382\\
590	0.362534\\
610	0.372677\\
630	0.382807\\
650	0.392879\\
670	0.402913\\
690	0.412883\\
710	0.422792\\
730	0.432633\\
750	0.442402\\
770	0.452089\\
790	0.461687\\
810	0.471201\\
830	0.480627\\
850	0.489956\\
870	0.499185\\
890	0.508319\\
910	0.517348\\
930	0.526269\\
950	0.535092\\
970	0.543797\\
990	0.552399\\
};
\addlegendentry{\Huge{$\tilde{A}_{g}$}}

\addplot [color=mycolor3, line width=3.0pt]
  table[row sep=crcr]{%
30	0.0775521\\
50	0.0121173\\
70	0.00978467\\
90	0.0120359\\
110	0.0158609\\
130	0.0208042\\
150	0.0268006\\
170	0.0339126\\
190	0.0422696\\
210	0.0520517\\
230	0.0634701\\
250	0.0767492\\
270	0.0921046\\
290	0.109712\\
310	0.129685\\
330	0.152043\\
350	0.176697\\
370	0.203432\\
390	0.231913\\
410	0.261688\\
430	0.292218\\
450	0.322902\\
470	0.353113\\
490	0.382238\\
510	0.40971\\
530	0.43504\\
550	0.457837\\
570	0.477822\\
590	0.49483\\
610	0.508806\\
630	0.519795\\
650	0.527923\\
670	0.533382\\
690	0.53641\\
710	0.53727\\
730	0.536239\\
750	0.533589\\
770	0.529581\\
790	0.524456\\
810	0.518431\\
830	0.511697\\
850	0.504416\\
870	0.496728\\
890	0.488747\\
910	0.480565\\
930	0.472257\\
950	0.463884\\
970	0.45549\\
990	0.447112\\
};
\addlegendentry{\Huge{$\tilde{A}_{f}$}}

\addplot [color=black, line width=2.0pt, draw=none, mark repeat={3}, mark size=9.0pt, mark=x, mark options={solid, black}]
  table[row sep=crcr]{%
30	0.1734\\
50	0.2767\\
70	0.3592\\
90	0.4463\\
110	0.513\\
130	0.5773\\
150	0.6303\\
170	0.665\\
190	0.6929\\
210	0.6984\\
230	0.7008\\
250	0.6819\\
270	0.674\\
290	0.6594\\
310	0.6394\\
330	0.5968\\
350	0.578\\
370	0.5462\\
390	0.504\\
410	0.4567\\
430	0.4143\\
450	0.382\\
470	0.3252\\
490	0.295\\
510	0.2593\\
530	0.22\\
550	0.1921\\
570	0.1608\\
590	0.1351\\
610	0.1085\\
630	0.0885\\
650	0.0695\\
670	0.0543\\
690	0.0427\\
710	0.031\\
730	0.0237\\
750	0.018\\
770	0.016\\
790	0.0092\\
810	0.0071\\
830	0.0046\\
850	0.0032\\
870	0.0029\\
890	0.0013\\
};

\addplot [color=black, line width=2.0pt, draw=none, mark repeat={3}, mark size=9.0pt, mark=x, mark options={solid, black}]
  table[row sep=crcr]{%
30	0.7856\\
50	0.7131\\
70	0.6323\\
90	0.5426\\
110	0.4686\\
130	0.4022\\
150	0.3407\\
170	0.2979\\
190	0.2645\\
210	0.2479\\
230	0.2352\\
250	0.2395\\
270	0.2256\\
290	0.2254\\
310	0.231\\
330	0.243\\
350	0.2402\\
370	0.2514\\
390	0.2631\\
410	0.2717\\
430	0.2745\\
450	0.2908\\
470	0.3033\\
490	0.3079\\
510	0.3253\\
530	0.329\\
550	0.3447\\
570	0.3459\\
590	0.3598\\
610	0.3707\\
630	0.3787\\
650	0.3936\\
670	0.4064\\
690	0.4158\\
710	0.4269\\
730	0.4261\\
750	0.4483\\
770	0.4477\\
790	0.4654\\
810	0.4677\\
830	0.4818\\
850	0.4919\\
870	0.5072\\
890	0.5029\\
};

\addplot [color=black, line width=2.0pt, draw=none, mark repeat={3}, mark size=9.0pt, mark=x, mark options={solid, black}]
  table[row sep=crcr]{%
30	0.041\\
50	0.0102\\
70	0.0085\\
90	0.0111\\
110	0.0184\\
130	0.0205\\
150	0.029\\
170	0.0370999999999999\\
190	0.0426000000000001\\
210	0.0537\\
230	0.0640000000000001\\
250	0.0786\\
270	0.1004\\
290	0.1152\\
310	0.1296\\
330	0.1602\\
350	0.1818\\
370	0.2024\\
390	0.2329\\
410	0.2716\\
430	0.3112\\
450	0.3272\\
470	0.3715\\
490	0.3971\\
510	0.4154\\
530	0.451\\
550	0.4632\\
570	0.4933\\
590	0.5051\\
610	0.5208\\
630	0.5328\\
650	0.5369\\
670	0.5393\\
690	0.5415\\
710	0.5421\\
730	0.5502\\
750	0.5337\\
770	0.5363\\
790	0.5254\\
810	0.5252\\
830	0.5136\\
850	0.5049\\
870	0.4899\\
890	0.4958\\
};
\addlegendentry{\Huge{Simulation}}

\end{axis}

%% file: Figures/Fig5c.tex
%
%
\definecolor{mycolor1}{rgb}{0.00000,0.44700,0.74100}%
\definecolor{mycolor2}{rgb}{0.85000,0.32500,0.09800}%
\definecolor{mycolor3}{rgb}{0.92900,0.69400,0.12500}%
\definecolor{mycolor4}{rgb}{0.49400,0.18400,0.55600}%
\definecolor{mycolor5}{rgb}{0.74902,0.00000,0.74902}%
\definecolor{mycolor6}{rgb}{0.00000,0.44706,0.74118}%
\definecolor{mycolor7}{rgb}{0.63500,0.07800,0.18400}

\begin{axis}[%
width=8.73in,
height=7.323in,
at={(1.464in,1.169in)},
scale only axis,
xmin=0,
xmax=900,
xtick={0,100,...,900},
xlabel style={font=\color{white!15!black}},
xlabel={\Huge{$\text{Height of UAVs} \;\; h_u \; \text{(m)}$}},
ymin=0,
ymax=0.9,
ytick={0,0.1,...,0.9},
ylabel style={font=\color{white!15!black}},
ylabel={\Huge{$\text{Coverage probability} \;\; P_{\text{cov}}$}},
axis background/.style={fill=white},
xmajorgrids,
ymajorgrids,
legend style={at={(0.705,0.705)}, anchor=south west, legend cell align=left, align=left, draw=white!15!black}
]
\addplot [color=mycolor1, line width=3.0pt]
  table[row sep=crcr]{%
30	0.745429\\
50	0.820026\\
70	0.842502\\
90	0.860083\\
110	0.873116\\
130	0.880851\\
150	0.882893\\
170	0.879187\\
190	0.869939\\
210	0.85559\\
230	0.836732\\
250	0.814078\\
270	0.788426\\
290	0.760627\\
310	0.731545\\
330	0.701969\\
350	0.672677\\
370	0.644267\\
390	0.61721\\
410	0.591813\\
430	0.568232\\
450	0.546494\\
470	0.526548\\
490	0.508293\\
510	0.491647\\
530	0.476544\\
550	0.462993\\
570	0.451056\\
590	0.440808\\
610	0.432354\\
630	0.425798\\
650	0.421136\\
670	0.418351\\
690	0.417328\\
710	0.417919\\
730	0.419929\\
750	0.423139\\
770	0.427367\\
790	0.432376\\
810	0.438007\\
830	0.4441\\
850	0.450521\\
870	0.457161\\
890	0.46394\\
910	0.470785\\
930	0.477644\\
950	0.484483\\
970	0.491261\\
990	0.497969\\
};
\addlegendentry{\Huge{$\tau_a = -5$ dB}}

\addplot [color=mycolor2, line width=3.0pt]
  table[row sep=crcr]{%
30	0.557597\\
50	0.633383\\
70	0.668873\\
90	0.695264\\
110	0.710854\\
130	0.714275\\
150	0.70536\\
170	0.685051\\
190	0.655172\\
210	0.618136\\
230	0.576589\\
250	0.533258\\
270	0.490442\\
290	0.450113\\
310	0.413697\\
330	0.382093\\
350	0.355727\\
370	0.334637\\
390	0.318592\\
410	0.307177\\
430	0.299883\\
450	0.296153\\
470	0.295417\\
490	0.2971\\
510	0.300627\\
530	0.30544\\
550	0.311009\\
570	0.316863\\
590	0.322614\\
610	0.328027\\
630	0.332992\\
650	0.337535\\
670	0.341791\\
690	0.345924\\
710	0.350109\\
730	0.354472\\
750	0.359077\\
770	0.363933\\
790	0.369003\\
810	0.374233\\
830	0.379553\\
850	0.384897\\
870	0.39021\\
890	0.395455\\
910	0.400597\\
930	0.405617\\
950	0.41051\\
970	0.41526\\
990	0.419874\\
};
\addlegendentry{\Huge{$\tau_a = 0$ dB}}

\addplot [color=mycolor3, line width=3.0pt]
  table[row sep=crcr]{%
30	0.355714\\
50	0.422166\\
70	0.45883\\
90	0.480226\\
110	0.484356\\
130	0.471397\\
150	0.443813\\
170	0.4056\\
190	0.361377\\
210	0.315546\\
230	0.271723\\
250	0.232436\\
270	0.199102\\
290	0.172179\\
310	0.151432\\
330	0.13621\\
350	0.125688\\
370	0.119035\\
390	0.115522\\
410	0.114562\\
430	0.115714\\
450	0.118665\\
470	0.123208\\
490	0.129201\\
510	0.136547\\
530	0.145154\\
550	0.154913\\
570	0.165667\\
590	0.177189\\
610	0.189184\\
630	0.201296\\
650	0.213147\\
670	0.224385\\
690	0.234728\\
710	0.244006\\
730	0.252158\\
750	0.259229\\
770	0.265328\\
790	0.270598\\
810	0.275189\\
830	0.279236\\
850	0.282845\\
870	0.286101\\
890	0.289068\\
910	0.291792\\
930	0.294306\\
950	0.29664\\
970	0.298811\\
990	0.30084\\
};
\addlegendentry{\Huge{$\tau_a = 5$ dB}}

\addplot [color=mycolor4, line width=3.0pt]
  table[row sep=crcr]{%
30	0.206527\\
50	0.258304\\
70	0.285576\\
90	0.294836\\
110	0.286498\\
130	0.26398\\
150	0.232428\\
170	0.197172\\
190	0.162617\\
210	0.131747\\
230	0.106094\\
250	0.086026\\
270	0.0711384\\
290	0.0606367\\
310	0.0536216\\
330	0.0492641\\
350	0.0468867\\
370	0.0459817\\
390	0.0461938\\
410	0.0472894\\
430	0.049125\\
450	0.0516254\\
470	0.0547539\\
490	0.0585067\\
510	0.062898\\
530	0.067949\\
550	0.0736789\\
570	0.0800916\\
590	0.0871633\\
610	0.0948276\\
630	0.102965\\
650	0.111402\\
670	0.119919\\
690	0.128269\\
710	0.136217\\
730	0.143563\\
750	0.150166\\
770	0.155953\\
790	0.160917\\
810	0.165099\\
830	0.168573\\
850	0.171426\\
870	0.17375\\
890	0.175634\\
910	0.177156\\
930	0.178383\\
950	0.179375\\
970	0.180177\\
990	0.180831\\
};
\addlegendentry{\Huge{$\tau_a = 10$ dB}}

\addplot [color=black, line width=2.0pt, draw=none, mark repeat={3}, mark size=9.0pt, mark=x, mark options={solid, black}]
  table[row sep=crcr]{%
30	0.7771\\
50	0.81947304625292\\
70	0.844411799474772\\
90	0.864041297324905\\
110	0.870460059409448\\
130	0.880521114449766\\
150	0.879462888742828\\
170	0.872722343293252\\
190	0.872457864391794\\
210	0.856189941785361\\
230	0.838863545736264\\
250	0.811961489397819\\
270	0.777573270481144\\
290	0.754664260140028\\
310	0.732946252518504\\
330	0.69876901914783\\
350	0.67077696\\
370	0.64556712\\
390	0.62030937\\
410	0.58439184\\
430	0.54656352\\
450	0.53229284\\
470	0.49854918\\
490	0.48336401\\
510	0.4771\\
530	0.452664301903576\\
550	0.448537035506632\\
570	0.428572036522245\\
590	0.42370302435078\\
610	0.413165896801079\\
630	0.406366060879282\\
650	0.405480820377637\\
670	0.404585799320583\\
690	0.40620537637184\\
710	0.408928419786316\\
730	0.404630221750585\\
750	0.421534347272051\\
770	0.420674997955702\\
790	0.43119306951393\\
810	0.432203980652118\\
830	0.438991703716285\\
850	0.448480107571483\\
870	0.464315306659952\\
890	0.459317177418218\\
};

\addplot [color=black, line width=2.0pt, draw=none, mark repeat={3}, mark size=9.0pt, mark=x, mark options={solid, black}]
  table[row sep=crcr]{%
30	0.5845\\
50	0.635478245252012\\
70	0.664637164319706\\
90	0.696234458611057\\
110	0.711059476252668\\
130	0.714884523851684\\
150	0.708071060508644\\
170	0.691266227783738\\
190	0.660194121335765\\
210	0.627820092280169\\
230	0.588899802271847\\
250	0.543922672268908\\
270	0.502411562605042\\
290	0.459361609014704\\
310	0.421133459945984\\
330	0.391795876086826\\
350	0.36382018979796\\
370	0.344055442554511\\
390	0.325235621360272\\
410	0.3102064\\
430	0.29690759\\
450	0.28432399\\
470	0.28359181\\
490	0.27933536\\
510	0.2904\\
530	0.293377873515584\\
550	0.28686091725448\\
570	0.299027870845204\\
590	0.291974757590197\\
610	0.295347573617719\\
630	0.301905963293606\\
650	0.310830755132023\\
670	0.314330166819811\\
690	0.328780360984928\\
710	0.329723690028597\\
730	0.346184664843192\\
750	0.342999781342005\\
770	0.350625297540606\\
790	0.356866480414208\\
810	0.370758112863652\\
830	0.376767300882279\\
850	0.386876032769386\\
870	0.387637127045798\\
890	0.394633887169811\\
};

\addplot [color=black, line width=2.0pt, draw=none, mark repeat={3}, mark size=9.0pt, mark=x, mark options={solid, black}]
  table[row sep=crcr]{%
30	0.3742\\
50	0.415836290373355\\
70	0.460978646994504\\
90	0.48523223409887\\
110	0.488265081975168\\
130	0.482416493958398\\
150	0.452033944179\\
170	0.425274823420781\\
190	0.377963055937476\\
210	0.329384541567152\\
230	0.279544146317686\\
250	0.242424628760132\\
270	0.21385127849925\\
290	0.182924095798319\\
310	0.167833859047619\\
330	0.151007250986296\\
350	0.14136136989798\\
370	0.133903496741674\\
390	0.127906143531353\\
410	0.126347473011301\\
430	0.12662529\\
450	0.12490926\\
470	0.12615378\\
490	0.12598801\\
510	0.145\\
530	0.149847219347215\\
550	0.157299257104754\\
570	0.159091904075868\\
590	0.172641475864891\\
610	0.17917210669934\\
630	0.189520599459289\\
650	0.195625465639722\\
670	0.212023794680321\\
690	0.22035727929015\\
710	0.234941999639591\\
730	0.244675484268606\\
750	0.251750228731041\\
770	0.253042153809637\\
790	0.264976143387676\\
810	0.269657293345098\\
830	0.277920068890399\\
850	0.282980426455344\\
870	0.287064125979785\\
890	0.293873057427386\\
};

\addplot [color=black, line width=2.0pt, draw=none, mark repeat={3}, mark size=9.0pt, mark=x, mark options={solid, black}]
  table[row sep=crcr]{%
30	0.2194\\
50	0.254100276227141\\
70	0.282493460031192\\
90	0.291635913488681\\
110	0.293650687644395\\
130	0.277912759133645\\
150	0.252841504737279\\
170	0.215555162599387\\
190	0.179223069791676\\
210	0.143496981249556\\
230	0.121073246131016\\
250	0.0978935107292188\\
270	0.0796658809281856\\
290	0.0672945931026205\\
310	0.0568113904460892\\
330	0.0538727922944589\\
350	0.0520987219481948\\
370	0.0494801074667467\\
390	0.04966253\\
410	0.05075581\\
430	0.0532806\\
450	0.05477234\\
470	0.05549094\\
490	0.05841594\\
510	0.0707\\
530	0.072317380533718\\
550	0.0781557780521851\\
570	0.0828085953535761\\
590	0.0901437236348125\\
610	0.099509711766303\\
630	0.107277696208206\\
650	0.119356185575814\\
670	0.122644624507596\\
690	0.129214901315553\\
710	0.137612143839195\\
730	0.147838711257218\\
750	0.155517159719531\\
770	0.162252664161402\\
790	0.164002934563005\\
810	0.17001682248233\\
830	0.170226708824729\\
850	0.174857742620768\\
870	0.17806307924753\\
890	0.178888655949367\\
};
\addlegendentry{\Huge{Simulation}}

\end{axis}

%% file: Figures/Fig6a.tex
%
%
\definecolor{mycolor1}{rgb}{0.00000,0.44700,0.74100}%
\definecolor{mycolor2}{rgb}{0.85000,0.32500,0.09800}%
\definecolor{mycolor3}{rgb}{0.92900,0.69400,0.12500}%
\definecolor{mycolor4}{rgb}{0.49400,0.18400,0.55600}%
\definecolor{mycolor5}{rgb}{0.74902,0.00000,0.74902}%
\definecolor{mycolor6}{rgb}{0.00000,0.44706,0.74118}%
\definecolor{mycolor7}{rgb}{0.63500,0.07800,0.18400}

\begin{axis}[%
width=9.2in,
height=8.3in,
at={(1.464in,1.02in)},
scale only axis,
xmin=0,
xmax=50,
xtick={0,10,...,50},
xlabel style={font=\color{white!15!black}},
xlabel={\Huge{$\text{Number of UAVs} \; N_u$}},
ymin=0.4,
ymax=0.8,
ytick={0.4,0.45,...,0.8},
ylabel style={font=\color{white!15!black}},
ylabel={\Huge{$\text{Coverage probability} \;\; P_{\text{cov}}$}},
axis background/.style={fill=white},
xmajorgrids,
ymajorgrids,
legend style={at={(0.01,0.01)}, anchor=south west, legend cell align=left, align=left, draw=white!15!black}
]
\addplot [color=mycolor1, line width=3.0pt, mark repeat={2}, mark size=6.0pt, mark=o, mark options={solid, mycolor1}]
  table[row sep=crcr]{%
1	0.576216\\
3	0.618011\\
5	0.651099\\
7	0.676909\\
9	0.696631\\
11	0.711254\\
13	0.721605\\
15	0.728372\\
17	0.73213\\
19	0.733361\\
21	0.732466\\
23	0.729784\\
25	0.725599\\
27	0.720149\\
29	0.713638\\
31	0.706236\\
33	0.698086\\
35	0.689314\\
37	0.680022\\
39	0.670301\\
41	0.660229\\
43	0.64987\\
45	0.639283\\
47	0.628516\\
49	0.617612\\
};
\addlegendentry{Aware - $\tau_b = 10$ dB}

\addplot[color=mycolor1, dash pattern=on 8pt off 8pt on 8pt off 8pt, line width=3.0pt, mark repeat={2}, mark size=8.0pt, mark=x, mark options={solid, mycolor1}]  table[row sep=crcr]{%
1	0.576085\\
3	0.617595\\
5	0.650368\\
7	0.675838\\
9	0.695199\\
11	0.709446\\
13	0.719407\\
15	0.725776\\
17	0.729128\\
19	0.729948\\
21	0.728641\\
23	0.725546\\
25	0.72095\\
27	0.715092\\
29	0.708176\\
31	0.700375\\
33	0.691834\\
35	0.682676\\
37	0.673008\\
39	0.66292\\
41	0.652491\\
43	0.641785\\
45	0.630862\\
47	0.619771\\
49	0.608555\\
};
\addlegendentry{Unaware - $\tau_b = 10$ dB}

\addplot [color=mycolor2, line width=3.0pt, mark repeat={2}, mark size=6.0pt, mark=o, mark options={solid, mycolor2}]
  table[row sep=crcr]{%
1	0.570809\\
3	0.603355\\
5	0.628997\\
7	0.648876\\
9	0.663943\\
11	0.674983\\
13	0.682654\\
15	0.687501\\
17	0.689979\\
19	0.690468\\
21	0.689285\\
23	0.686695\\
25	0.682922\\
27	0.678151\\
29	0.672541\\
31	0.666225\\
33	0.659315\\
35	0.651906\\
37	0.644079\\
39	0.635905\\
41	0.627441\\
43	0.618739\\
45	0.609844\\
47	0.600792\\
49	0.591618\\
};
\addlegendentry{Aware - $\tau_b = 15$ dB}

\addplot [color=mycolor2, dash pattern=on 8pt off 8pt on 8pt off 8pt, line width=3.0pt, mark repeat={2}, mark size=8.0pt, mark=x, mark options={solid, mycolor2}]
  table[row sep=crcr]{%
1	0.570251\\
3	0.601594\\
5	0.625935\\
7	0.644432\\
9	0.65805\\
11	0.667588\\
13	0.673715\\
15	0.676986\\
17	0.677866\\
19	0.676742\\
21	0.673938\\
23	0.669727\\
25	0.664337\\
27	0.65796\\
29	0.650758\\
31	0.642869\\
33	0.634408\\
35	0.625475\\
37	0.616154\\
39	0.606517\\
41	0.596625\\
43	0.586533\\
45	0.576286\\
47	0.565924\\
49	0.555481\\
};
\addlegendentry{Unaware - $\tau_b = 15$ dB}

\addplot [color=mycolor4, line width=3.0pt, mark repeat={2}, mark size=6.0pt, mark=o, mark options={solid, mycolor4}]  table[row sep=crcr]{%
1	0.556037\\
3	0.563091\\
5	0.567922\\
7	0.570928\\
9	0.572434\\
11	0.57271\\
13	0.571973\\
15	0.570405\\
17	0.568151\\
19	0.565334\\
21	0.562052\\
23	0.558386\\
25	0.554404\\
27	0.55016\\
29	0.545699\\
31	0.541059\\
33	0.536271\\
35	0.53136\\
37	0.526348\\
39	0.521252\\
41	0.516087\\
43	0.510866\\
45	0.5056\\
47	0.500297\\
49	0.494966\\
};
\addlegendentry{Aware - $\tau_b = 20$ dB}

\addplot [color=mycolor4, dash pattern=on 8pt off 8pt on 8pt off 8pt, line width=3.0pt, mark repeat={2}, mark size=8.0pt, mark=x, mark options={solid, mycolor4}]
  table[row sep=crcr]{%
1	0.554349\\
3	0.557984\\
5	0.559344\\
7	0.558838\\
9	0.556801\\
11	0.553506\\
13	0.549181\\
15	0.544012\\
17	0.538153\\
19	0.53173\\
21	0.524848\\
23	0.517594\\
25	0.510041\\
27	0.502248\\
29	0.494266\\
31	0.486137\\
33	0.477897\\
35	0.469576\\
37	0.461199\\
39	0.45279\\
41	0.444367\\
43	0.435946\\
45	0.427541\\
47	0.419166\\
49	0.41083\\
};
\addlegendentry{Unaware - $\tau_b = 20$ dB}

\end{axis}

\begin{axis}[%
width=11.264in,
height=9.181in,
at={(0in,0in)},
scale only axis,
xmin=0,
xmax=1,
ymin=0,
ymax=1,
axis line style={draw=none},
ticks=none,
axis x line*=bottom,
axis y line*=left,
legend style={legend cell align=left, align=left, draw=white!15!black}
]
\end{axis}

%% file: Figures/Fig6b.tex
%
%
\definecolor{mycolor1}{rgb}{0.00000,0.44700,0.74100}%
\definecolor{mycolor2}{rgb}{0.85000,0.32500,0.09800}%
\definecolor{mycolor3}{rgb}{0.92900,0.69400,0.12500}%
\definecolor{mycolor4}{rgb}{0.49400,0.18400,0.55600}%
\definecolor{mycolor5}{rgb}{0.74902,0.00000,0.74902}%
\definecolor{mycolor6}{rgb}{0.00000,0.44706,0.74118}%
\definecolor{mycolor7}{rgb}{0.63500,0.07800,0.18400}

\begin{axis}[%
width=8.73in,
height=7.682in,
at={(1.464in,1.12in)},
scale only axis,
xmin=0,
xmax=30,
xtick={0,5,...,30},
xlabel style={font=\color{white!15!black}},
xlabel={\Huge{$\text{Backhaul threshold} \; \; \tau{}_b \; \text{(dB)}$}},
ymin=0.3,
ymax=0.8,
ytick={0.3,0.35,...,0.8},
ylabel style={font=\color{white!15!black}},
ylabel={\Huge{$\text{Coverage probability} \;\; P_{\text{cov}}$}},
axis background/.style={fill=white},
xmajorgrids,
ymajorgrids,
legend style={at={(0.76,0.76)}, anchor=south west, legend cell align=left, align=left, draw=white!15!black}
]
\addplot [color=mycolor1, line width=3.0pt, mark repeat={2}, mark size=6.0pt, mark=o, mark options={solid, mycolor1}]
  table[row sep=crcr]{%
0	0.714267\\
2	0.713745\\
4	0.712878\\
6	0.711456\\
8	0.709008\\
10	0.704525\\
12	0.696127\\
14	0.680997\\
16	0.656076\\
18	0.619614\\
20	0.572711\\
22	0.519488\\
24	0.465633\\
26	0.416491\\
28	0.375942\\
30	0.346177\\
};
\addlegendentry{$\sigma_s = 0$}

\addplot [color=mycolor2, line width=3.0pt, mark repeat={2}, mark size=6.0pt, mark=o, mark options={solid, mycolor2}]
  table[row sep=crcr]{%
0	0.592392\\
2	0.567381\\
4	0.542203\\
6	0.51918\\
8	0.49979\\
10	0.484499\\
12	0.472666\\
14	0.462354\\
16	0.450704\\
18	0.435436\\
20	0.416306\\
22	0.394918\\
24	0.373575\\
26	0.354342\\
28	0.338634\\
30	0.327191\\
};
\addlegendentry{$\sigma_s = 0.2$}

\addplot [color=mycolor3, line width=3.0pt, mark repeat={2}, mark size=6.0pt, mark=o, mark options={solid, mycolor3}]
  table[row sep=crcr]{%
0	0.451216\\
2	0.42916\\
4	0.407109\\
6	0.387174\\
8	0.370773\\
10	0.358589\\
12	0.350571\\
14	0.345793\\
16	0.342567\\
18	0.33932\\
20	0.335451\\
22	0.331163\\
24	0.32691\\
26	0.323099\\
28	0.32\\
30	0.317752\\
};
\addlegendentry{$\sigma_s = 0.5$}

\addplot [color=mycolor4, line width=3.0pt, mark repeat={2}, mark size=6.0pt, mark=o, mark options={solid, mycolor4}]
  table[row sep=crcr]{%
0	0.387136\\
2	0.373601\\
4	0.360125\\
6	0.347999\\
8	0.338088\\
10	0.33082\\
12	0.326205\\
14	0.323757\\
16	0.322534\\
18	0.321625\\
20	0.320617\\
22	0.319507\\
24	0.318408\\
26	0.317424\\
28	0.316624\\
30	0.316045\\
};
\addlegendentry{$\sigma_s = 1$}

\end{axis}

%% file: Figures/Fig6c.tex
%
%
\definecolor{mycolor1}{rgb}{0.00000,0.44700,0.74100}%
\definecolor{mycolor2}{rgb}{0.85000,0.32500,0.09800}%
\definecolor{mycolor3}{rgb}{0.92900,0.69400,0.12500}%
\definecolor{mycolor4}{rgb}{0.49400,0.18400,0.55600}%
\definecolor{mycolor5}{rgb}{0.74902,0.00000,0.74902}%
\definecolor{mycolor6}{rgb}{0.00000,0.44706,0.74118}%
\definecolor{mycolor7}{rgb}{0.63500,0.07800,0.18400}

\begin{axis}[%
width=8.73in,
height=7.482in,
at={(1.464in,1.01in)},
scale only axis,
xmin=0,
xmax=900,
xtick={0,100,...,900},
xlabel style={font=\color{white!15!black}},
xlabel={\Huge{$\text{Height of UAVs}  \;\; h_u \; \text{(m)}$}},
ymin=0,
ymax=0.8,
ytick={0.0,0.1,...,0.8},
ylabel style={font=\color{white!15!black}},
ylabel={\Huge{$\text{Coverage probability} \;\; P_{\text{cov}}$}},
axis background/.style={fill=white},
xmajorgrids,
ymajorgrids,
legend style={at={(0.6,0.72)}, anchor=south west, legend cell align=left, align=left, draw=white!15!black}
]
\addplot [color=mycolor1, line width=3.0pt, mark repeat={2}, mark size=6.0pt, mark=o, mark options={solid, mycolor1}]  table[row sep=crcr]{%
30	0.3528\\
50	0.407403\\
70	0.451303\\
90	0.481548\\
110	0.503448\\
130	0.517716\\
150	0.524102\\
170	0.522644\\
190	0.513952\\
210	0.499084\\
230	0.479347\\
250	0.456104\\
270	0.430649\\
290	0.395244381297999\\
310	0.36197052360718\\
330	0.334873250754315\\
390	0.262180976220667\\
450	0.205130990251411\\
510	0.17262\\
570	0.142727342996321\\
630	0.118285040526386\\
690	0.0944719008267378\\
750	0.0777983412033301\\
810	0.0661303558710746\\
870	0.0646481364403137\\
};
\addlegendentry{$\lambda_g = 1$ BS/km$^2$}

\addplot [color=mycolor2, line width=3.0pt, mark repeat={2}, mark size=6.0pt, mark=o, mark options={solid, mycolor2}]
  table[row sep=crcr]{%
30	0.548\\
50	0.625239\\
70	0.674601\\
90	0.709851\\
110	0.732687\\
130	0.742132\\
150	0.738051\\
170	0.721322\\
190	0.693627\\
210	0.657690520951363\\
230	0.613105330705977\\
250	0.566979418985362\\
270	0.518585149281491\\
290	0.471798801891026\\
310	0.431075981766296\\
330	0.394310339189189\\
390	0.308559800160128\\
450	0.252859916937081\\
510	0.23262\\
570	0.220368631738058\\
630	0.209024783873469\\
690	0.205426932837247\\
750	0.210673521595536\\
810	0.222418649388489\\
870	0.238501837831804\\
};
\addlegendentry{$\lambda_g = 5$ BS/km$^2$}

\addplot [color=mycolor3, line width=3.0pt, mark repeat={2}, mark size=6.0pt, mark=o, mark options={solid, mycolor3}]
  table[row sep=crcr]{%
30	0.5794\\
50	0.633383\\
70	0.668873\\
90	0.695264\\
110	0.710854\\
130	0.714275\\
150	0.70536\\
170	0.685051\\
190	0.655172\\
210	0.627950750834924\\
230	0.584998494695983\\
250	0.550314588578574\\
270	0.500469197459238\\
290	0.464774895747575\\
310	0.422725040258026\\
330	0.396471845206521\\
390	0.3246474\\
450	0.29086603\\
510	0.2867\\
570	0.291272039036129\\
630	0.301802130777242\\
690	0.323248370841084\\
750	0.342753997498109\\
810	0.36807594995141\\
870	0.391217418144854\\
};
\addlegendentry{$\lambda_g = 10$ BS/km$^2$}

\addplot [color=mycolor4, line width=3.0pt, mark repeat={2}, mark size=6.0pt, mark=o, mark options={solid, mycolor4}]
  table[row sep=crcr]{%
30	0.5764\\
50	0.603929\\
70	0.628918\\
90	0.646187\\
110	0.653393\\
130	0.649621\\
150	0.635355\\
170	0.612153\\
190	0.591241774166318\\
210	0.569114530887194\\
230	0.532450244520136\\
250	0.496315573216251\\
270	0.46912895696709\\
290	0.44354428349635\\
310	0.4209652\\
330	0.39854418\\
390	0.36265302\\
450	0.36461816\\
510	0.3819\\
570	0.400866673526222\\
630	0.427307414234999\\
690	0.448656842737095\\
750	0.477030047516199\\
810	0.493802586633907\\
870	0.50216062\\
};
\addlegendentry{$\lambda_g = 20$ BS/km$^2$}

\end{axis}

%% file: Figures/Fig7a.tex
%
%
\definecolor{mycolor1}{rgb}{0.00000,0.44700,0.74100}%
\definecolor{mycolor2}{rgb}{0.85000,0.32500,0.09800}%
\definecolor{mycolor3}{rgb}{0.92900,0.69400,0.12500}%
\definecolor{mycolor4}{rgb}{0.49400,0.18400,0.55600}%
\definecolor{mycolor5}{rgb}{0.74902,0.00000,0.74902}%
\definecolor{mycolor6}{rgb}{0.00000,0.44706,0.74118}%
\definecolor{mycolor7}{rgb}{0.63500,0.07800,0.18400}

\begin{axis}[%
width=8.6in,
height=7in,
at={(1.017in,0.665in)},
scale only axis,
xmin=0.1,
xmax=1,
xtick={0.1,0.3,...,1},
xlabel style={font=\color{white!15!black}},
xlabel={\Huge{$\text{Fraction of BSs with backhaul }\delta{}_\text{b}$}},
ymin=0.1,
ymax=1,
ytick={0.1,0.3,...,1},
ylabel style={font=\color{white!15!black}},
ylabel={\Huge{$\text{Backhaul probability} \;\; S(\tau_{b})$}},
axis background/.style={fill=white},
xmajorgrids,
ymajorgrids,
legend style={at={(0.68,0.05)}, anchor=south west, legend cell align=left, align=left, draw=white!15!black}
]
\addplot [color=mycolor1, line width=3.0pt, mark size=6.0pt, mark=o, mark options={solid, mycolor1}]
  table[row sep=crcr]{%
0.1	0.1679\\
0.2	0.3161\\
0.3	0.4325\\
0.4	0.5203\\
0.5	0.608\\
0.6	0.6828\\
0.7	0.7391\\
0.8	0.7752\\
0.9	0.809\\
1	0.8423\\
};
\addlegendentry{$h_u = 30$ m}

\addplot [color=mycolor2, line width=3.0pt, mark repeat={2}, mark size=6.0pt, mark=o, mark options={solid, mycolor2}]
  table[row sep=crcr]{%
0.1	0.548579\\
0.15	0.681871\\
0.2	0.768959\\
0.25	0.827274\\
0.3	0.867212\\
0.35	0.895135\\
0.4	0.915031\\
0.45	0.929455\\
0.5	0.940076\\
0.55	0.948008\\
0.6	0.954008\\
0.65	0.958599\\
0.7	0.962145\\
0.75	0.964909\\
0.8	0.967077\\
0.85	0.968788\\
0.9	0.970144\\
0.95	0.971221\\
1	0.972077\\
};
\addlegendentry{$h_u = 100$ m}

\addplot [color=mycolor3, line width=3.0pt, mark repeat={2}, mark size=6.0pt, mark=o, mark options={solid, mycolor3}]
  table[row sep=crcr]{%
0.1	0.574902\\
0.15	0.683517\\
0.2	0.748175\\
0.25	0.788552\\
0.3	0.814625\\
0.35	0.831807\\
0.4	0.843206\\
0.45	0.850692\\
0.5	0.855441\\
0.55	0.858217\\
0.6	0.859533\\
0.65	0.859745\\
0.7	0.859101\\
0.75	0.857784\\
0.8	0.855927\\
0.85	0.853632\\
0.9	0.850974\\
0.95	0.848014\\
1	0.844798\\
};
\addlegendentry{$h_u = 300$ m}

\addplot [color=mycolor4, line width=3.0pt, mark size=6.0pt, mark=o, mark options={solid, mycolor4}]
  table[row sep=crcr]{%
0.1	0.33746\\
0.2	0.45536\\
0.3	0.4948\\
0.4	0.50806\\
0.5	0.50066\\
0.6	0.4891\\
0.7	0.47004\\
0.8	0.45404\\
0.9	0.43076\\
1	0.41588\\
};
\addlegendentry{$h_u = 500$ m}

\end{axis}

%% file: Figures/Fig7b.tex
%
%
\definecolor{mycolor1}{rgb}{0.00000,0.44700,0.74100}%
\definecolor{mycolor2}{rgb}{0.85000,0.32500,0.09800}%
\definecolor{mycolor3}{rgb}{0.92900,0.69400,0.12500}%
\definecolor{mycolor4}{rgb}{0.49400,0.18400,0.55600}%
\definecolor{mycolor5}{rgb}{0.74902,0.00000,0.74902}%
\definecolor{mycolor6}{rgb}{0.00000,0.44706,0.74118}%
\definecolor{mycolor7}{rgb}{0.63500,0.07800,0.18400}

\begin{axis}[%
width=8.6in,
height=7in,
at={(1.017in,0.665in)},
scale only axis,
xmin=0.1,
xmax=1,
xtick={0.1,0.3,...,1},
xlabel style={font=\color{white!15!black}},
xlabel={\Huge{$\text{Fraction of BSs with backhaul }\delta{}_\text{b}$}},
ymin=0.25,
ymax=0.75,
ytick={0.2,0.3,...,0.75},
ylabel style={font=\color{white!15!black}},
ylabel={\Huge{$\text{Coverage probability} \;\; P_{\text{cov}}$}},
axis background/.style={fill=white},
xmajorgrids,
ymajorgrids,
legend style={at={(0.68,0.08)}, anchor=south west, legend cell align=left, align=left, draw=white!15!black}
]
\addplot [color=mycolor1, line width=3.0pt, mark size=6.0pt, mark=o, mark options={solid, mycolor1}]
  table[row sep=crcr]{%
0.1	0.4826\\
0.2	0.509024802299234\\
0.3	0.521431657255888\\
0.4	0.535490819926087\\
0.5	0.54415692545024\\
0.6	0.556037455704324\\
0.7	0.566025006855499\\
0.8	0.5721\\
0.9	0.579022526716003\\
1	0.586647614441143\\
};
\addlegendentry{$h_u = 30$ m}

\addplot [color=mycolor2, line width=3.0pt, mark repeat={2}, mark size=6.0pt, mark=o, mark options={solid, mycolor2}]
  table[row sep=crcr]{%
0.1	0.542445\\
0.15	0.594725\\
0.2	0.628258\\
0.25	0.650434\\
0.3	0.665491\\
0.35	0.675957\\
0.4	0.683382\\
0.45	0.688748\\
0.5	0.692691\\
0.55	0.695631\\
0.6	0.697851\\
0.65	0.699549\\
0.7	0.700859\\
0.75	0.70188\\
0.8	0.70268\\
0.85	0.703312\\
0.9	0.703812\\
0.95	0.704209\\
1	0.704525\\
};
\addlegendentry{$h_u = 100$ m}

\addplot [color=mycolor3, line width=3.0pt, mark repeat={2}, mark size=6.0pt, mark=o, mark options={solid, mycolor3}]
  table[row sep=crcr]{%
0.1	0.38\\
0.15	0.41\\
0.2	0.425\\
0.25	0.432515\\
0.3	0.432175\\
0.35	0.431759\\
0.4	0.431402\\
0.45	0.431133\\
0.5	0.430949\\
0.55	0.430836\\
0.6	0.430782\\
0.65	0.430773\\
0.7	0.4308\\
0.75	0.430854\\
0.8	0.43093\\
0.85	0.431021\\
0.9	0.431123\\
0.95	0.431233\\
1	0.431347\\
};
\addlegendentry{$h_u = 300$ m}

\addplot [color=mycolor4, line width=3.0pt, mark size=6.0pt, mark=o, mark options={solid, mycolor4}]
  table[row sep=crcr]{%
0.1	0.2873\\
0.2	0.289321920556325\\
0.3	0.283840907954668\\
0.4	0.281310347781184\\
0.5	0.274443029697767\\
0.6	0.27495691779941\\
0.7	0.271607849717174\\
0.8	0.272725694299947\\
0.9	0.274908226911392\\
1	0.276252908787614\\
};
\addlegendentry{$h_u = 500$ m}

\end{axis}

%% file: Figures/Fig7c.tex
%
%
\definecolor{mycolor1}{rgb}{0.00000,0.44700,0.74100}%
\definecolor{mycolor2}{rgb}{0.85000,0.32500,0.09800}%
\definecolor{mycolor3}{rgb}{0.92900,0.69400,0.12500}%
\definecolor{mycolor4}{rgb}{0.49400,0.18400,0.55600}%
\definecolor{mycolor5}{rgb}{0.74902,0.00000,0.74902}%
\definecolor{mycolor6}{rgb}{0.00000,0.44706,0.74118}%
\definecolor{mycolor7}{rgb}{0.63500,0.07800,0.18400}

\begin{axis}[%
width=8.73in,
height=7.1in,
at={(1.464in,1.01in)},
scale only axis,
xmin=0,
xmax=50,
xtick={0,10,...,50},
xlabel style={font=\color{white!15!black}},
xlabel={\Huge{$\text{Number of UAVs} \; N_u$}},
ymin=0.1,
ymax=0.8,
ytick={0,0.1,...,0.8},
ylabel style={font=\color{white!15!black}},
ylabel={\Huge{$\text{Coverage probability} \;\; P_{\text{cov}}$}},
axis background/.style={fill=white},
xmajorgrids,
ymajorgrids,
legend style={at={(0.02,0.01)}, anchor=south west, legend cell align=left, align=left, draw=white!10!black}
]
\addplot [color=mycolor1, dash pattern=on 8pt off 8pt on 8pt off 8pt, line width=3.0pt, mark size=8pt, mark=x, mark options={solid, draw=mycolor1}]
  table[row sep=crcr]{%
1	0.5683\\
7	0.596107632662526\\
13	0.620651967012972\\
19	0.643907694582519\\
25	0.657513785905081\\
31	0.668824392899576\\
37	0.676438908646534\\
43	0.684285662122494\\
49	0.685830864031404\\
};
\addlegendentry{Instan. - $h_u = 30$ m}

\addplot [color=mycolor1, line width=2.0pt, mark size=6.0pt, mark repeat={3}, mark=o, mark options={solid, draw=mycolor1}]
  table[row sep=crcr]{%
1	0.552671\\
3	0.55478\\
5	0.55667\\
7	0.558354\\
9	0.559849\\
11	0.561169\\
13	0.562326\\
15	0.563333\\
17	0.564202\\
19	0.564943\\
21	0.565567\\
23	0.566082\\
25	0.566499\\
27	0.566826\\
29	0.56707\\
31	0.567238\\
33	0.567338\\
35	0.567376\\
37	0.567357\\
39	0.567287\\
41	0.567171\\
43	0.567014\\
45	0.56682\\
47	0.566593\\
49	0.566337\\
};
\addlegendentry{Aware - $h_u = 30$ m}

\addplot [color=mycolor2, dash pattern=on 8pt off 8pt on 8pt off 8pt, line width=3.0pt, mark size=8pt, mark=x, mark options={solid, draw=mycolor2}]
  table[row sep=crcr]{%
1	0.5989\\
7	0.726687643537659\\
13	0.753914728716048\\
19	0.751242457803216\\
25	0.725194647508751\\
31	0.67434172\\
37	0.6309106\\
43	0.59190348\\
49	0.53389155\\
};
\addlegendentry{Instan. - $h_u = 120$ m}

\addplot [color=mycolor2, line width=3.0pt, mark size=6.0pt, mark=o, mark repeat={3}, mark options={solid, draw=mycolor2}]
  table[row sep=crcr]{%
1	0.582341\\
3	0.631665\\
5	0.667407\\
7	0.692335\\
9	0.708644\\
11	0.718077\\
13	0.722027\\
15	0.7216\\
17	0.717684\\
19	0.710991\\
21	0.702094\\
23	0.691452\\
25	0.679439\\
27	0.666359\\
29	0.652459\\
31	0.637941\\
33	0.62297\\
35	0.607685\\
37	0.592196\\
39	0.576598\\
41	0.56097\\
43	0.545371\\
45	0.529861\\
47	0.514483\\
49	0.499274\\
};
\addlegendentry{Aware - $h_u = 120$ m}

\addplot [color=mycolor4, dash pattern=on 8pt off 8pt on 8pt off 8pt , line width=3.0pt, mark size=8.0pt, mark repeat={3}, mark=x, mark options={solid, draw=mycolor4}]
  table[row sep=crcr]{%
1	0.6149\\
3	0.684426736336907\\
5	0.708048128510573\\
7	0.711245178124312\\
9	0.689001000692256\\
11	0.656575888067657\\
13	0.615070545892336\\
15	0.582563435141919\\
17	0.55148947935743\\
19	0.507306053356749\\
21	0.475407168013637\\
23	0.438383216959006\\
25	0.412267485641591\\
27	0.379829371257761\\
29	0.347177839383322\\
31	0.312589859631705\\
33	0.289521979137482\\
35	0.268959727933967\\
37	0.247580850468187\\
39	0.224589670460184\\
41	0.211187373733493\\
43	0.191651418156447\\
45	0.177151416763353\\
47	0.166632922044409\\
49	0.152964774278856\\
};
\addlegendentry{Instan. - $h_u = 200$ m}

\addplot [color=mycolor4, line width=3.0pt, mark size=6pt, mark=o, mark repeat={3}, mark options={solid, draw=mycolor4}]
  table[row sep=crcr]{%
1	0.604308\\
3	0.66353\\
5	0.681397\\
7	0.67417\\
9	0.651988\\
11	0.621222\\
13	0.585899\\
15	0.548577\\
17	0.51089\\
19	0.473862\\
21	0.438146\\
23	0.404132\\
25	0.372042\\
27	0.341986\\
29	0.313992\\
31	0.28804\\
33	0.264073\\
35	0.242009\\
37	0.221755\\
39	0.203206\\
41	0.186255\\
43	0.170791\\
45	0.156709\\
47	0.143901\\
49	0.132268\\
};
\addlegendentry{Aware - $h_u = 200$ m}

\end{axis}